\documentclass[a4paper,12pt]{article}

\usepackage{graphicx, amssymb, amstext, amsmath}
\usepackage[left=1.5cm, right=1.5cm, top=2.50cm, bottom=2.00cm]{geometry}
\usepackage{mathrsfs}
\usepackage{bbding}
\usepackage{subfigure}
\usepackage{float}
\usepackage{undertilde}
\usepackage{algorithm} 
\usepackage{algorithmic}
\usepackage{enumitem}
\usepackage{setspace}
\usepackage{color}
\date{\today}

\usepackage{tocstyle}
\usetocstyle{standard}
\settocfeature{raggedhook}{\raggedright}

\renewcommand{\newline}{\medskip}

\newcommand{\der}{\ensuremath{\rm d}}
\newcommand{\vect}[1]{\pmb{#1}}

\newcommand{\divergence}{\ensuremath{\nabla}\cdot}
\newcommand{\grad}{\ensuremath{\nabla}}
\newcommand{\lap}{\ensuremath{\nabla^2}}
\newcommand{\austenite}{A }
\newcommand{\mOne}{M$_1$ }
\newcommand{\mTwo}{M$_2$ }
\newcommand{\mThree}{M$_3$ }
\newcommand{\pooz}{[110] }
\newcommand{\pozo}{[101] }

\newcommand{\lx}{$L_{x_1}$}
\newcommand{\ly}{$L_{x_2}$}
\newcommand{\lz}{$L_{x_3}$}

\pdfoutput=1
\begin{document}

\setlength{\parindent}{0pt}
\newenvironment{nospace}%
{\noindent\ignorespaces}%
{\par\noindent%
\ignorespacesafterend}
\long\def\symbolfootnote[#1]#2{\begingroup%
\def\thefootnote{\fnsymbol{footnote}}\footnote[#1]{#2}\endgroup}
\allowdisplaybreaks
\newpage

\begin{center}                                           
\Large{Effect of Aspect Ratio and Boundary Conditions in \\
Modeling Shape Memory Alloy Nanostructures \\with 3D Coupled Dynamic Phase-Field Theories}
\end{center}
\normalsize

\begin{center}
R. Dhote$^{1,3}$, H. Gomez$^2$, R. Melnik$^3$, J. Zu$^1$

$^1$Mechanical and Industrial Engineering, University 
of Toronto, \\5 King's College Road, Toronto, ON, M5S3G8, Canada\\
$^2$Department of Applied Mathematics, University of A Coru\~{n}a \\ Campus de Elvina, s/n. 15192 A Coru\~{n}a, Spain\\
$^3$ The MS2Discovery Interdisciplinary Research Institute, \\
M$^2$NeT Laboratory, Wilfrid Laurier University, Waterloo, ON,  
N2L3C5, Canada
\end{center}

\section*{Abstract}

The behavior of shape memory alloy (SMA) nanostructures is influenced by strain rate and temperature evolution during dynamic loading. The coupling between temperature, strain and strain rate effects is essential to capture inherent thermo-mechanical behavior in SMAs. In this paper, we  propose a new fully coupled thermo-mechanical 3D phase-field model that accounts for two-way coupling between mechanical (or structural) and thermal physics. The 3D model provides a realistic description of the properties of  SMAs nanostructures. We use the strain-based Ginzburg-Landau potential for cubic-to-tetragonal phase transformations. The variational formulation of the developed model is implemented in the isogeometric analysis framework to overcome numerical challenges. We have observed a complete disappearance of the out-of-plane martensitic variant in a very high aspect ratio SMA domain as well as the presence of three variants in equal portions in a low aspect ratio SMA domain. The sensitive dependence of different boundary conditions on the microstructure morphology has been examined energetically. The tensile tests on a rectangular prism nanowires, using the displacement based loading, demonstrate  the shape memory effect and pseudoelastic behavior. We have also observed that higher strain rates, as well as the lower aspect ratio domains, resulting in high yield stress and  phase transformations occur at higher stress during dynamic axial loading. The simulation results using the developed model are in qualitative agreement with the numerical and experimental results from the literature.

\textbf{Keywords}: Phase-field model, Ginzburg-Landau theory, nonlinear thermo-elasticity, shape memory alloys, nanostructures.  

\section{Introduction}

Shape memory alloys (SMAs) are metallic alloys with distinguished  characteristics like thermoelastic phase transformations, unique shape memory effect and pseudoelastic hysteretic behaviors, excellent corrosion resistance,  biocompatibility along with high strength, strain and power density. SMAs have been widely used as transducers in sensing and actuation applications \cite{Lagoudas,Otsuka}. More recently, they found applications in SMA nanotubes \cite{vzuvzek2012electrochemical}, nanofilms \cite{bhattacharya2005material,koig2010micro,bayer2011carbon}, nanowires \cite{clements2003wireless,san2008superelasticity,Juan2009} for nanoelectromechanical and microelectromechanical systems, biomedical devices, as well as for a number of other applications \cite{benard1998thin,kohl2004shape,miyazaki2009thin,yoneyama2009shape}. In many of the above applications, SMAs are  actuated dynamically at different loading rates under various boundary conditions. Factors like geometry \cite{zakharov2012submicron}, boundary conditions \cite{Juan2009,koig2010micro,phillips2011phase}, loading rates \cite{shaw1995thermomechanical}, have substantial influence on the microstructure evolution and their thermo-mechanical behaviors. However, these factors have not been studied in a fully coupled thermo-mechanical framework for SMAs. Such studies would aid the understanding of phase transformation mechanisms and thermo-mechanical behavior of SMAs for better SMA-based application developments.\\

SMAs have inherent thermo-mechanical properties. In the various experimental studies, the sensitivity of dynamic loading and strong interactions of the thermal and mechanical effects have been emphasized for SMA specimens  \cite{leo1993transient,shaw1995thermomechanical,shaw1997nucleation,gadaj2002temperature,pieczyska2004thermomechanical,pieczyska2006phase,pieczyskaqirt,he2010rate}. The drastic changes in a fundamental mechanical response of SMAs have been attributed to self-heating  and cooling due to exothermic and endothermic processes during latent heat transfer. However, the experimental setups for dynamic loading behavior are complex and time consuming. Arguably, the modeling framework is a complementary tool to study the dynamic behavior, its influence on microstructure evolution and thermo-mechanical response of SMAs. \\

Several modeling approaches have significantly contributed to the understanding of temperature- and stress- induced transformations \cite{Lagoudas,Otsuka}. Extensive discussions of various models have been reviewed in \cite{khandelwal2011models,mamivand2013review}. Here we are interested in phase-field (PF) models, which are used across different length scales, including the nanoscale \cite{Levitas2002a,bouville2008microstructure}. The PF models provide a unified framework that describes stress- and temperature- induced phase transformations, including their dynamics. The characteristic features of transformations have been thoroughly studied, in particular, in the context of nucleation, growth, domain wall movement, merging, and elimination \cite{salje1993phase,provatas2010phase}.  In the literature, there have been extensive studies of 2D SMA PF models under quasistatic loading conditions. The 2D models can qualitatively capture the basic mechanisms during phase transformations, under the assumption of constant strain across the out-of-plane direction. However, the real-life applications often require 3D geometries, where the strain can be accommodated in all three directions. At the same time, very few 3D studies have been reported in the literature for quasistatic and dynamic loading conditions. \\  

An important pioneering work on the 3D PF model was reported by Barsch et al. \cite{barsch1984twin}. They introduced the strain-based order parameters (OPs) model to describe a cubic-to-tetragonal phase transformation consisting of inhomogeneous strain field associated with domain walls, constituent phases and transformation precursors. Later, Jacobs et al. \cite{jacobs2003simulations} incorporated the Rayleigh dissipation and solved the isothermal model to obtain  static domain patterns during temperature induced transformations. The computer simulations on a large domain revealed several microstructure features  observed experimentally. Wang et al. \cite{wang1997three} described a 3D PF model that accounts for the transformation induced elastic strain of martensitic phase transformations. They developed a stochastic field kinetic model and studied the nucleation, growth of martensitic evolution in cubic-to-tetragonal transformations. Later, Artemev et al. \cite{artemev2001three} investigated the effect of external load on phase transformations in a polycrystalline material. Using the transformation strain-related order parameters, Levitas et al. \cite{Levitas2002a,Levitas2002b,Levitas2003} developed a PF model that accounts for temperature dependent thermo-mechanical properties of different phases of SMAs. Subsequently, Idesman et al. \cite{Idesman2008} used the model and demonstrated the effect of inertial forces that can alter the microstructure during evolution. Following the strain-based order parameter modeling frameworks \cite{barsch1984twin,jacobs2003simulations}, Ahluwalia et al. \cite{ahluwalia2004pattern,Ahluwalia2006} reported the only study at that time showing dynamic strain loading characteristics of SMAs. However, the influence of different strain rates on microstructure and mechanical response of FePd samples on temperature- and stress- induced PTs was examined by using exclusively the isothermal model. Other 3D PF models have reported morphological evolution in spinodal decomposition \cite{seol2003computer}, thermoelastic transformations \cite{man2011study} and decomposition of the supersaturated binary solid solution \cite{ni2007transformation}. \\ 

Most of the above 3D models examined the phase transformations under static or quasistatic conditions in a controlled temperature environment with periodic boundary conditions on simple cubic geometries. With the inherent thermo-mechanical coupling and other factors mentioned above, it is essential to model the thermo-mechanical physics and the rate dependent coupling in a single framework. Recently, for the first time, we presented a coupled thermo-mechanical framework for modeling 2D square-to-rectangular phase transformations \cite{Dhote2012} and developed a new computational framework, based on the isogeometric analysis (using non-uniform rational B-spline (NURBS) basis functions) \cite{dhote2014isogeometric}. In this work, we extend this new framework to the 3D model for  cubic-to-tetragonal phase transformations and numerically solve the resulting coupled, strongly non-linear equations by using the isogeometric analysis (IGA). A point to remark is  that the energy functionals used in the 2D \cite{dhote2014isogeometric} and 3D models must be distinct in order to  satisfy the symmetry related to martensitic variants in the energy landscape. This leads to distinct expressions for constitutive relationships and the thermo-mechanical coupling term. The 3D constitutive equations and thermo-mechanical coupling term are not reducible to the 2D constitutive equations and the thermo-mechanical coupling term due to the selection of distinct functionals. Consequently, this leads to different models and distinct numerical implementations for the 2D and 3D models.\\

In this paper, we first provide the derivation of the 3D fully coupled thermo-mechanical model for the cubic-to-tetragonal martensitic transformations in Section \ref{sec:Ch14PF3DGL}. The numerical formulation of the developed model using isogeometric analysis is presented in Section \ref{sec:Ch14NumericalFormulation}. The effect of aspect ratio and boundary conditions on microstructure morphology and energetics is discussed during the temperature induced phase transformations in Section \ref{sec:Ch14tempinduced}. Section \ref{sec:Ch14TT} describes the effect of aspect ratio and strain rate sensitivity on thermo-mechanical behavior of SMAs during stress-induced transformations. Finally, the conclusions are summarized in Section \ref{sec:Ch14Conclusions}.

\section{3D Thermo-Mechanical Phase-Field Model} \label{sec:Ch14PF3DGL}

\begin{figure}[h!]
\centering
\subfigure[]
{
\includegraphics[width=0.28\linewidth]{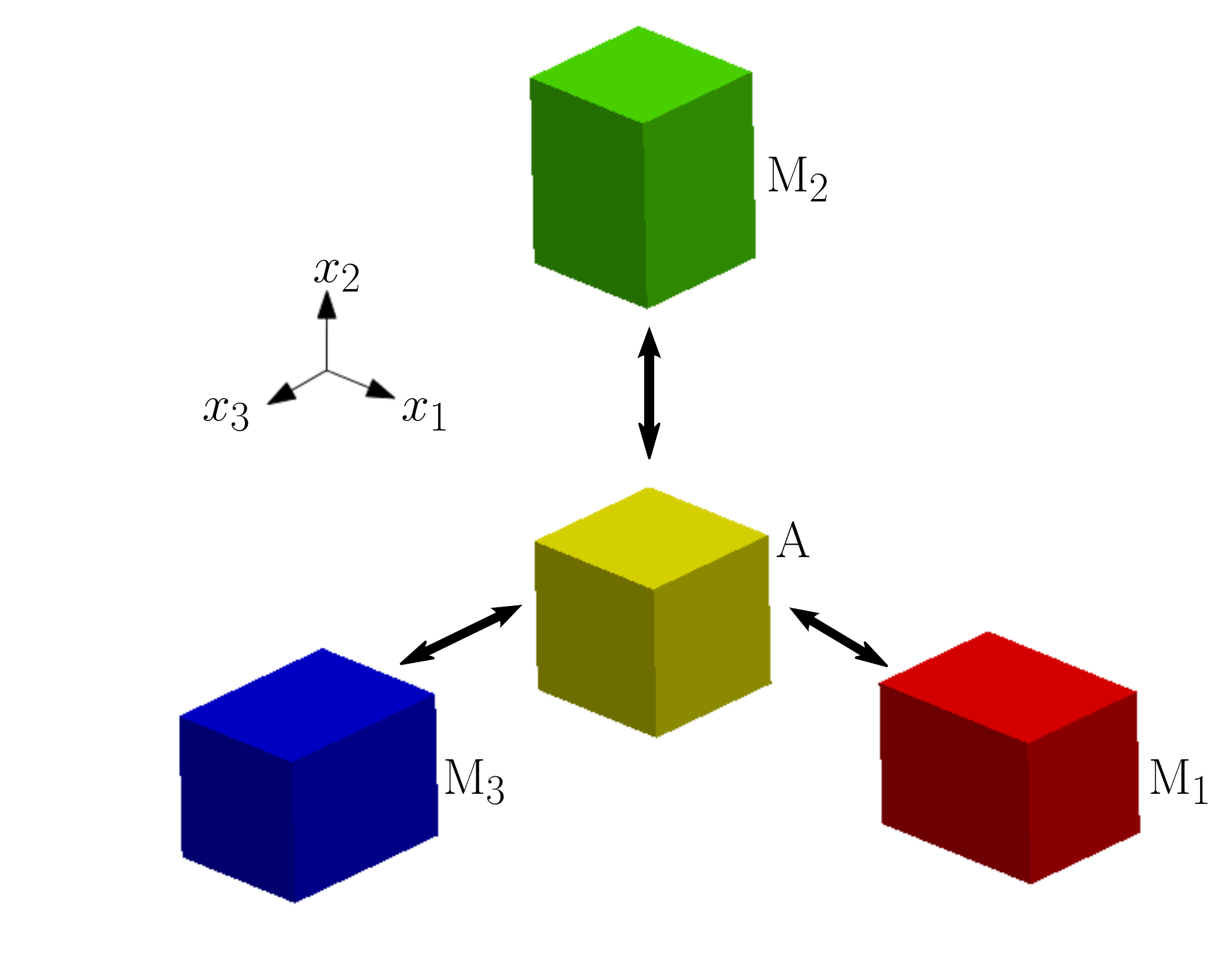}
\label{fig:Ch14Cubic2Tet1}
}
\subfigure[]
{
\includegraphics[width=0.32\textwidth]{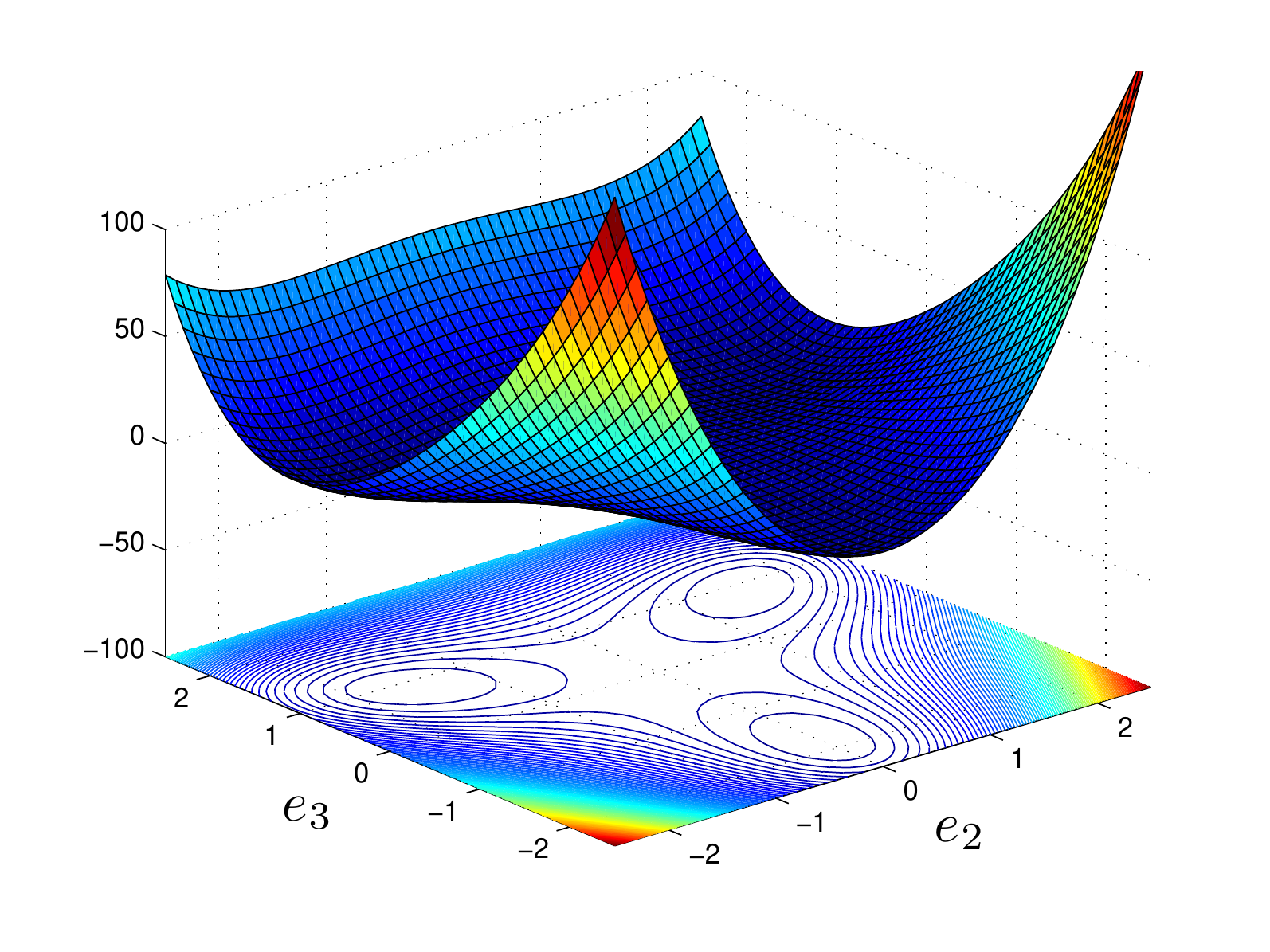}
\label{fig:Ch14ThreeDFunctionalM}
}
\subfigure[]
{
\includegraphics[width=0.32\textwidth]{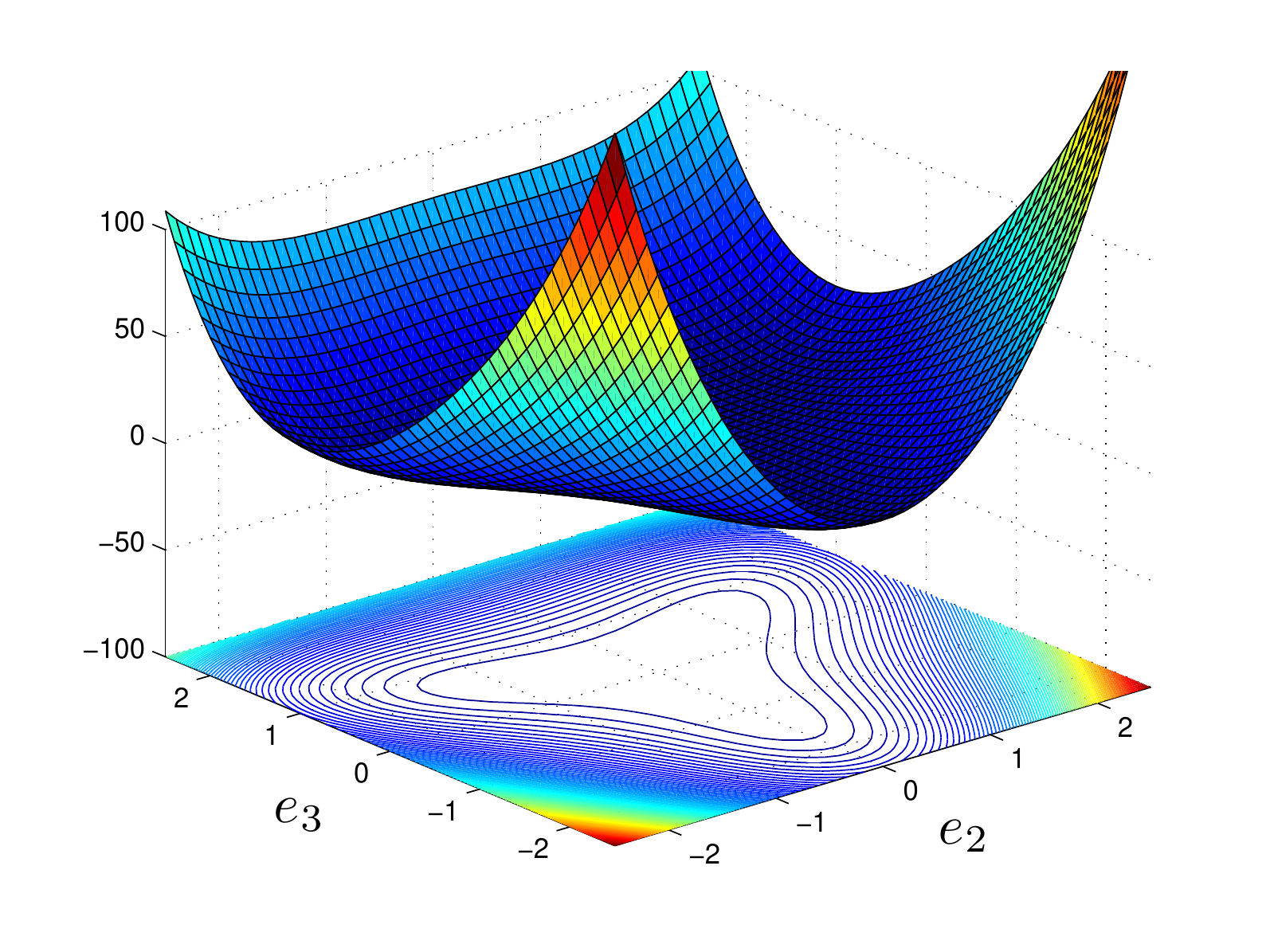}
\label{fig:Ch14ThreeDFunctionalP}
}
\caption{Cubic-to-tetragonal phase transformations (a) schematic of microstructures (austenite (A), martensite variants (M$_1$, M$_2$, M$_3$), and $\mathscr{F}_L$ plot at (b) $ \tau  = -1.2 $ and (c) $ \tau  = 1.2 $.}
\label{fig:Ch14CubicToTetragonal}
\end{figure}

The thermo-mechanical properties of SMAs are functions of their microstructures and loadings. The properties can be described by  a phase-field model using the Ginzburg-Landau approach \cite{mamivand2013review}.  As different parts of the domain may evolve into distinct phases, a continuum description of phases in the model can be carried out by introducing a diffuse interface domain wall between phases as a function of OP(s). By choosing appropriate OPs, all possible phases in a domain can be described. Unlike other approaches \cite{wang1997three,Levitas2002a}, here we describe the cubic-to-tetragonal phase transformations in a PF model with respect to the strain-based (deviatoric) OPs. The cubic-to-tetragonal phase transformation can be  schematically depicted as shown in Fig. \ref{fig:Ch14Cubic2Tet1}, with the cubic austenite (A) phase and the three tetragonal martensite variants (M$_1$, M$_2$, M$_3$) stretched along the three rectilinear directions. The thermo-mechanical behavior of SMAs is established by solving a coupled 3D model described below.

We begin by introducing basic notations for the kinematics of SMAs. Let us declare $\vect u=\{u_1,u_2, u_3\}^T$ the displacement field. We will work on the physical domain $\Omega\subset\mathbb{R}^3$, which is assumed to be an open set parameterized by Cartesian coordinates $\vect x=\{x_1,x_2,x_3\}^T$. We define the strain measures $e_i$, for $i=1,\dots,6$, using the symmetric strain tensor as follows:

\begin{equation}
\label{Ch14kin1}
\left\{\begin{matrix}
e_1 \\
e_2 \\
e_3 \\
e_4 \\
e_5 \\
e_6 \\
\end{matrix}\right\} = \left[\begin{array}{c|c}
\mathbb{D}_3 \hspace{.05cm}& \hspace{.05cm}\mathbb{O}_3 \\ [.1cm]
\hline \\[-.3cm]
\mathbb{O}_3 & \mathbb{I}_3
\end{array}\right] \left\{\begin{matrix}
\epsilon_{11} \\
\epsilon_{22} \\
\epsilon_{33} \\
\epsilon_{23} \\
\epsilon_{13} \\
\epsilon_{12} \\
\end{matrix}\right\}, 
\end{equation}
where $\mathbb{D}_3$, $\mathbb{O}_3$, $\mathbb{I}_3$ are the $3\times3$ constant matrices. In particular,
\begin{equation}
\mathbb{D}_3=\left[\begin{matrix}
1/\sqrt{3} &  1/\sqrt{3} &  1/\sqrt{3} \\
1/\sqrt{2} & -1/\sqrt{2} &  0          \\
1/\sqrt{6} & -1/\sqrt{6} & -2/\sqrt{6}  
\end{matrix}\right],
\end{equation}
while $\mathbb{I}_3$, and $\mathbb{O}_3$ are, respectively, the $3\times3$ identity and zero matrices.

We call $ e_1 $ hydrostatic strain, $ e_2,  e_3 $ deviatoric strains, and $ e_4 $, $ e_5 $, $ e_6 $ shear strains. In Eqs. \eqref{Ch14kin1}, $\vect\epsilon$ denotes the Cauchy-Lagrange infinitesimal strain tensor, whose components are  
$\epsilon_{ij} = \left( u_{i,j} +  u_{j,i} \right)/2,\, i,j\in\{1,2,3\}$, where an inferior comma denotes partial differentiation (e.g., $u_{i,j}=\partial u_i/\partial x_j$). The deviatoric strains $ e_2 $ and $ e_3 $ are selected as the OPs to describe different phases in a domain. 

We use the free-energy  functional initially proposed by Barsch et al. \cite{barsch1984twin} and later modified by Ahluwalia et al. \cite{Ahluwalia2006} to study the cubic-to-tetragonal martensitic transformations in SMAs. The free energy functional $ \mathscr{F} $ for the cubic-to-tetragonal transformations is written as 
%
%
\begin{equation}
\mathscr{F}[\vect u, \theta] = \int_{\Omega} \left[ \mathscr{F}_b + \mathscr{F}_s + \mathscr{F}_{L} + \mathscr{F}_g \right] \der\Omega,
\label{eq:Ch14FEcub2tet}
\end{equation}
where the bulk $\mathscr{F}_b$, the shear  $\mathscr{F}_s$, the Landau $\mathscr{F}_L$ and the Ginzburg  $\mathscr{F}_g$ energy contributions are defined as
\begin{align}
\label{eq:Ch14homog_free}
&\mathscr{F}_b= \frac{a_1}{2} e_1^2, &  &\mathscr{F}_s =  \frac{a_2}{2} \left(e_4^2 + e_5^2 +e_6^2\right), \nonumber\\
&\mathscr{F}_L = a_3 \tau  \left(e_2^2 + e_3^2\right) + a_4 e_3 \left(e_3^2 - 3 e_2^2\right)  + a_5 (e_2^2 + e_3^2)^2, &  &\mathscr{F}_g =  \frac{k_g}{2} \left( |\grad e_2|^2 + |\grad e_3|^2 \right).
\end{align}
Here, $ a_i $, $i\in\{1,\dots,5\}$ and $ k_g $ are material parameters, $ \tau $ is the dimensionless temperature defined as $ \tau = \displaystyle (\theta - \theta_m)/(\theta_0 - \theta_m) $, where $ \theta_0 $ and $ \theta_m $ are the material properties specifying the transformation start and end temperatures, and $|\cdot|$ denotes the Euclidean norm of a vector. The $\mathscr{F}_b$ and  $\mathscr{F}_s$ energy components represent non-OP contributions due to the bulk  and shear energies, respectively. The $\mathscr{F}_L$ corresponds to the OP contribution and represents the non-convex 2--3--4 polynomial Landau functional. The Landau functional is dependent on the temperature coefficient $\tau$. The $\mathscr{F}_L$ is plotted for two different $\tau$ values corresponding to the martensite, and the austenite phases in Fig. \ref{fig:Ch14CubicToTetragonal}(b), and Fig. \ref{fig:Ch14CubicToTetragonal}(c), respectively. 

The kinetic energy $\mathscr{K}$, the energy associated with the external body forces $\mathscr{B}$, and the Raleigh dissipation $\mathscr{R}$ are defined, respectively, as
\begin{equation}
\mathscr{K}[\dot{\vect u}]=\int_{\Omega}\frac{\rho}{2}|\dot{\vect u}|^2 \der\Omega,\quad 
\mathscr{B}[\vect u]=-\int_{\Omega} \vect f\cdot\vect u \der\Omega,\quad
\mathscr{R}[\dot{\vect u}]=\int_{\Omega}\frac{\eta}{2} |\dot{\vect e}|^2\  \der\Omega,
\end{equation}
where a dot over a function denotes partial differentiation with respect to time, $\rho$ is the density, $ \vect f $ is the body load vector, $ \eta $ is the dissipation coefficient, and $\vect e=\{e_i\}_{i=1,\dots,6}$. The potential energy of the system $\mathscr{U}$ is defined as $\mathscr{U}[{\vect u}]=\mathscr{F}[{\vect u}]+\mathscr{B}[{\vect u}]$, while the Lagrangian takes on the form $\mathscr{L}[\vect u,\dot{\vect u}]=\mathscr{K}[\dot{\vect u}]-\mathscr{U}[{\vect u}]$. Let us define the Hamiltonian of the system $\mathscr{H}$ as 
\begin{equation}
\label{Ch14hamiltonian}
\mathscr{H}[\vect u,\dot{\vect u}]=\int_0^t\mathscr{L}[\vect u,\dot{\vect u}]\der {\it t},
\end{equation}
where $[0,t]$ is the time interval of interest. Following a variational approach, the governing equation of motion has the form
\begin{equation}
\label{Ch14eqs_motion}
\frac{\partial}{\partial t}\left(\frac{\delta\mathscr{L}}{\delta\dot{\vect u}}\right)-\frac{\delta\mathscr{L}}{\delta \vect u}=-\frac{\delta\mathscr{R}}{\delta \dot{\vect u}},
\end{equation}
where the operators $\delta(\cdot)/\delta\vect u$ and $\delta(\cdot)/\delta\dot{\vect u}$ denote the variational derivatives with respect to $\vect u$ and $\dot{\vect u}$ respectively. Note that setting  the right-hand side of Eq. \eqref{Ch14eqs_motion} to zero, yields the Lagrange general equation  of motion. However, we let the system relax according to the Lagrange equation  of motion using the Raleigh dissipation $\mathscr{R}$.

After analytic manipulations, Eq. \eqref{Ch14eqs_motion} may be written as
\begin{equation}
\label{Ch14eq_momentum}
\rho\frac{\partial^2\vect u}{\partial t^2}=\divergence\vect\sigma+\eta\divergence\vect\sigma'+\divergence\vect\sigma_w+\vect f.
\end{equation}

The symmetric stress tensor $\pmb{\sigma} = \{ \sigma_{ij}\} $ components are defined in Appendix \ref{app:Ch14StressDefinations}, specifically for the cubic-to-tetragonal PTs. The equations for $\sigma_{ij}$ in terms of the strains are strongly non-linear. The cubic non-linearity allows us to capture the hysteretic phase transformation properties like shape memory effect and pseduoelasticity as a function of temperature. The stress expressions define material constitutive laws of the phase transformation. 


The symmetric dissipation stress tensor $\pmb{\sigma}^{\prime}= \{ \sigma_{ij}^{\prime}\} $
is a linear function of the strain rates $\dot{e}_i$, $i=1,\dots,6$ defined as
\begin{equation}
\left\{\begin{matrix}
\sigma'_{11} \\
\sigma'_{22} \\
\sigma'_{33} \\
\sigma'_{23} \\
\sigma'_{13} \\
\sigma'_{12} \\
\end{matrix}\right\} = \left[\begin{array}{c|c}
\mathbb{D}_3^T \hspace{.05cm}& \hspace{.05cm}\mathbb{O}_3 \\ [.1cm]
\hline \\[-.3cm]
\mathbb{O}_3 & \frac{1}{2}\mathbb{I}_3 
\end{array}\right]  \left\{\begin{matrix}
\dot{e}_{1} \\
\dot{e}_{2} \\
\dot{e}_{3} \\
\dot{e}_{4} \\
\dot{e}_{5} \\
\dot{e}_{6} \\
\end{matrix}\right\}.
\end{equation}
The fourth-order differential terms in Eq. (\ref{Ch14eq_momentum}) coming from $\divergence\vect\sigma_w$ represent domain walls between different phases in a domain.  The additional stress components corresponding to the gradient terms in the potential energy take on the form
\begin{equation}
\vect\sigma_w=\frac{k_g}{3}\lap\left(\grad^T\vect u-3\grad_d\vect u\right),
\end{equation}
where $\grad^T\vect u$ denotes the transpose of the displacement gradient (i.e., $\grad^T\vect u=\{u_{j,i}\}$). We also use the notation $\grad_d\vect u=\hbox{\rm diag}(u_{1,1},u_{2,2},u_{3,3})$, where $\hbox{\rm diag}(a,b,c)$ is a $3\times3$ diagonal matrix whose diagonal entries starting in the upper left corner are $a,b,c$.

The governing equation of the thermal field is obtained from the conservation law for internal energy $\iota$ \cite{Melnik2002} as 
\begin{equation}
\rho \frac{\partial\iota}{\partial t} - \pmb{\sigma}^T:\nabla \dot{\vect u} + \nabla 
\cdot \pmb{q} = g,
\label{eq:Ch14internal energy}
\end{equation}
where $\pmb{q}= -\kappa \nabla \theta$ is the Fourier heat flux vector, $\kappa$ is the heat conductance coefficient of the material, and $g$ is a thermal loading. The internal energy is connected with the potential energy constructed above via the Helmholtz free energy $ \Psi $ as
\begin{eqnarray}
\iota &=& \Psi(\theta, \epsilon) - \theta \frac{\partial \Psi(\theta, 
\epsilon)}{\partial \theta}, \\
\Psi(\theta, \epsilon) &=& \mathscr{L}[\vect u,\dot{\vect u}] - C_v 
\theta\hspace*{0.05cm} \ln\hspace*{0.05cm}(\theta),
\label{eq:Ch14thermalenergy}
\end{eqnarray}
where $C_v$ is the specific heat of the material. On substituting the above relationships in Eq. (\ref{eq:Ch14internal energy}), the governing equation of the thermal field is formulated as

\begin{equation}
\label{Ch14eq_thermal}
C_v \frac{\partial \theta}{\partial t} = \kappa\lap\theta + \Xi \theta\left(\divergence\vect u\divergence\dot{\vect u}-3\hbox{\rm tr}(\nabla_d\vect u\nabla_d\dot{\vect u})\right) + g,
\end{equation}
where the operator $\hbox{\rm tr}(\cdot)$ denotes the trace of a square matrix, and we have assumed $\kappa$ to be constant. The second term on the right-hand side of Eq. (\ref{Ch14eq_thermal}) is a non-linear term, which couples temperature, deformation gradient (strain), and rate of the deformation gradient (strain rate). Hence, Eqs. (\ref{Ch14eq_momentum}) and Eq. (\ref{Ch14eq_thermal}) describe the  thermo-mechanical physics of SMAs by creating the two-way coupling via $\theta$, $ \nabla\vect u$ and $\nabla \dot{\vect u}$.

In what follows, we use the component-wise version of the governing equations \eqref{Ch14eq_momentum} and \eqref{Ch14eq_thermal}. Let us define the tensor $\vect \mu=\{\mu_{ij}\}$, such that $\vect\mu=\frac{k_g}{3}\left(\grad^T\vect u-3\nabla_d\vect u\right)$, and $\vect \sigma_w=\lap\vect\mu$. Thus, the governing equations may be written in the component form as
\begin{subequations}
\label{eq:Ch14comp112}
\begin{eqnarray}
& & \rho \ddot{u}_{i}=\sigma_{ij,j}+\eta\sigma'_{ij,j}+\mu_{ij,kkj}+f_i, \\
& & C_v\dot{\theta}=\kappa\theta_{,ii}+ \Xi\theta\left(u_{i,i}\dot{u}_{j,j}-3u_{i,i}\dot{u}_{i,i}\right) + g,
\end{eqnarray}
\end{subequations}
where repeated indices indicate summation. The governing thermo-mechanical Eqs. \eqref{eq:Ch14comp112} are converted into the dimensionless form as:
\begin{subequations}
\label{eq:Ch14dimensionless}
\begin{eqnarray}
&\rho \ddot{\bar{u}}_{i}  & =\bar{\sigma}_{ij,j}+ \bar{\eta} \bar{\sigma}^{\prime}_{ij,j}+\bar{\mu}_{ij,kkj}+\bar{f}_i, \\
&\bar{C}_v \dot{\bar{\theta}} & =\bar{\kappa} \bar{\theta}_{,ii}+ \bar{\Xi} \bar{\theta} \left(\bar{u}_{i,i} \dot{\bar{u}}_{j,j}-3 \bar{u}_{i,i} \dot{\bar{u}}_{i,i}\right) + \bar{g}.
\end{eqnarray}
\end{subequations}
by using the following change of variables:
\begin{eqnarray}
e_i = e_c \bar{e_i}, \qquad
u_i = e_c \delta \bar{u_i}, \qquad
x = \delta \bar{x}, \qquad
\mathscr{F} = \mathscr{F}_c \bar{\mathscr{F}}, \qquad
t = t_c \bar{t}, \qquad
\theta = \theta_c \bar{\theta},
\label{eq:Ch14rescaleconst}
\end{eqnarray}
with the scaling constants defined as
\begin{eqnarray}
\delta = \sqrt{\frac{k_g}{a_{0}}}, \qquad \bar{a}_1 = \frac{a_1}{a_{0}}, 
\qquad \bar{a}_2 = \frac{a_2}{a_{0}}, \qquad \bar{a}_4 = 2, \qquad 
\bar{a}_5 = 1, \qquad \mathscr{F}_c = \delta^2 e_c^2 a_{0}, \nonumber \\
\bar{\eta} = \frac{\eta}{a_0} 
\sqrt{\frac{a_0}{\rho \delta^2}}, \qquad 
\bar{C_v} = \frac{\rho C_v \tau}{t_c}, \qquad
\bar{\kappa} = \frac{\kappa \tau}{\delta^2 \bar{C_v}}, \qquad 
\bar{\Xi} = -\frac{2}{3} \frac{a_0 e_c^2}{t_c \tau \bar{C_v}}.
\label{eq:Ch14rescaleconstval1}
\end{eqnarray}

The rescaled free energy takes the form:
\begin{eqnarray}
\bar{\mathscr{F}} &=& \displaystyle 
\frac{\bar{a}_1}{2} \left( \bar{e}_1 \right)^2 
+ \frac{\bar{a}_2}{2} (\bar{e}_4^2 + \bar{e}_5^2 
+\bar{e}_6^2) + \bar{a}_3 \tau  (\bar{e}_2^2 + \bar{e}_3^2) + \bar{a}_4 
\bar{e}_3 (\bar{e}_3^2 - 3 \bar{e}_2^2) \nonumber \\
&& + \bar{a}_5 (\bar{e}_2^2 + \bar{e}_3^2)^2 + \frac{\bar{k_g}}{2} 
\left[ |\nabla \bar{e}_2|^2 + |\nabla \bar{e}_3|^2 \right].
\label{eq:Ch14rescaleFE}
\end{eqnarray}
In the rest of the paper, we drop the bar symbol over the dimensionless variables for simplicity.
\section{Numerical Formulation} \label{sec:Ch14NumericalFormulation}

The governing equations \eqref{eq:Ch14dimensionless} are highly nonlinear to account for the hysteretic behavior with thermo-mechanical coupling and fourth-order differential terms in the general 3D formulation. These complexities present a number of numerical challenges. We have developed an IGA framework that allows the straightforward solution to the fourth-order equations. It also allows the use of coarser meshes, larger time steps along with geometrical flexibility and accuracy \cite{Hughes}. The IGA numerical implementation of the Eqs. \eqref{eq:Ch14dimensionless} are described in detail in \cite{dhote2014CMAME}.

Our computational method for the structural equations is based on the following principle of virtual work
\begin{equation}
\delta W(\delta u_i,u_i)=\delta W^{\hbox{\rm\scriptsize int}} - \delta W^{\hbox{\rm\scriptsize ext}} + \delta W^{\hbox{\rm\scriptsize kin}}=0, \qquad \forall \delta u_i,
\end{equation}
where the $\delta u_i$'s are the virtual displacements and 
\begin{subequations}
\begin{eqnarray}
\delta W^{\hbox{\rm\scriptsize int}} & = &\int_{\Omega}\delta\epsilon_{ij}\left(\sigma_{ij}+\eta\sigma'_{ij}\right)\hbox{\rm d}\Omega -\int_{\Omega}\delta u_{i,jk}\mu_{ij,k}\der\Omega, \\
\delta W^{\hbox{\rm\scriptsize ext}} & = &\int_{\Omega}\delta u_if_i\der\Omega,  \\
\delta W^{\hbox{\rm\scriptsize kin}} & = &\int_{\Omega}\delta u_i\rho\ddot{u}_i\der\Omega.
\end{eqnarray}
\end{subequations}
Here $\delta\vect\epsilon$ denotes the strain tensor associated with the virtual displacements $\delta\vect u$. 

For the thermal physics, we use the variational principle,
\begin{equation}
\int_\Omega \delta\theta\left(c_v\dot{\theta} -\Xi\theta(u_{i,i}\dot{u}_{j,j} -3u_{i,i}\dot{u}_{i,i}) \right) \hbox{\rm d}\Omega + \int_\Omega\delta\theta_{,i}\kappa \theta_{,i}\der\Omega=0, \qquad \forall\delta\theta.
\end{equation}

Note that for the integrals in $\delta W^{\hbox{\rm\scriptsize int}}$ to be well defined, the strains must be globally smooth, which cannot be easily achieved using the classical finite element method. To overcome this difficulty, we propose a numerical method based on the isogeometric analysis, a methodology in computational mechanics that permits globally continuous representations of strains.

To completely define our algorithm, we need to specify discrete versions of the displacements and temperature. We use the following expressions
\begin{equation}
\label{eq:Ch14grp1}
u_i^h(\vect x,t) = \sum\limits_{A=1}^{n_b} u_i^A(t) N^A(\vect x);\;\;
\theta^h(\vect x,t) = \sum\limits_{A=1}^{n_b} \theta^A(t) N^A(\vect x), 
\end{equation}
where the superscript $h$ indicates a discrete field and $n_b$ is the dimension of the discrete space. We note that identical approximations are used for the virtual counterparts of $u_i^h$ and $\theta^h$. The $N^A$'s are B-spline (or NURBS) basis  functions. Using the rich basis functions, the IGA allows a straightforward implementation of the fourth order equations. 

\begin{figure}[h!]
\centering
\subfigure[]
{
\includegraphics[width=0.35\linewidth]{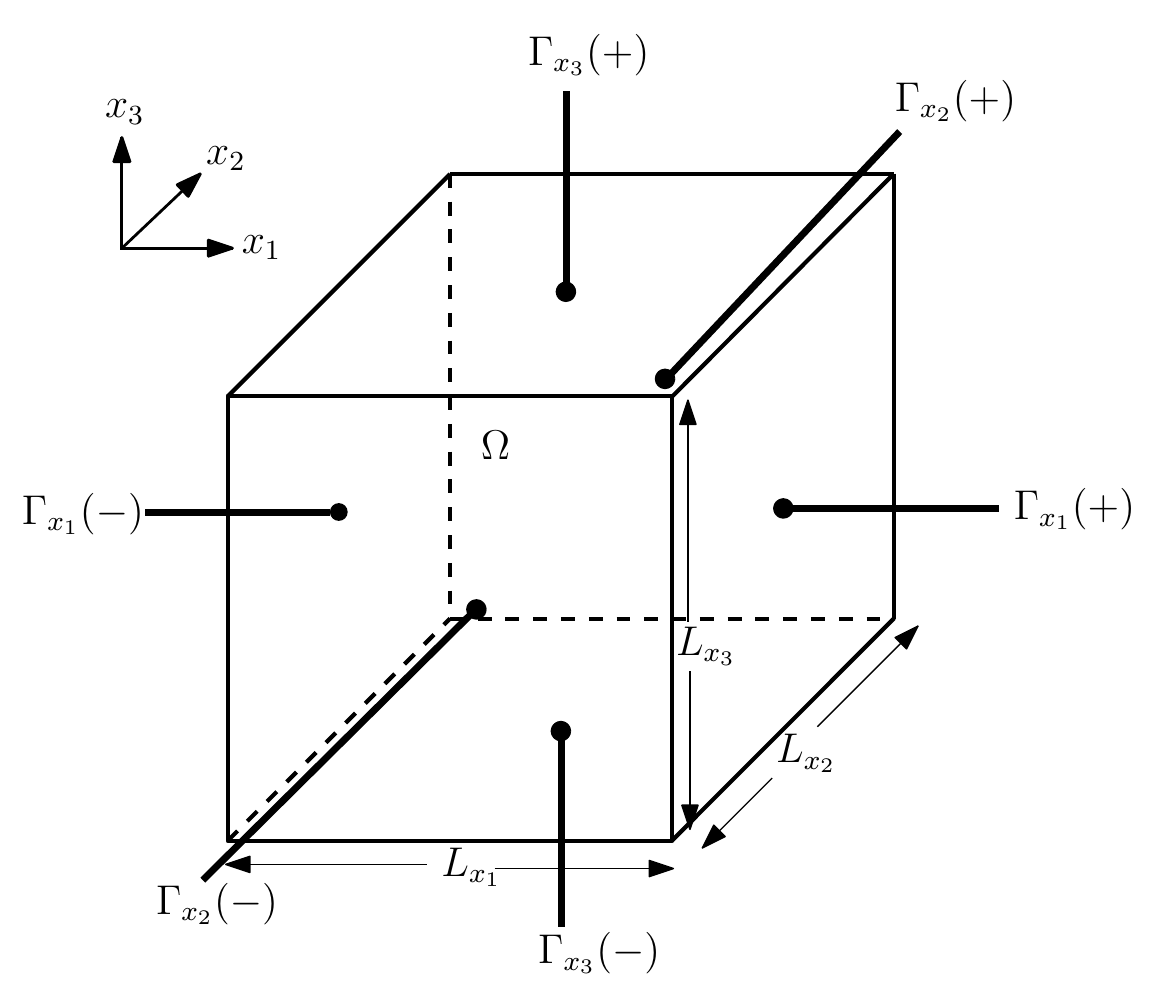}
\label{fig:Ch14SchematicRectangularPrism}
}
\caption{Schematic of the rectangular prism (domain $\Omega$ and nomenclature of boundaries $ \Gamma $).}
\label{fig:Ch14SchematicDomains}
\end{figure}

We illustrated the details about the IGA numerical implementation of the developed model, the geometrical flexibility, accuracy and robustness of the new numerical formulation in \cite{dhote2014CMAME}. In this paper, we focus on thermo-mechanical behavior of rectangular prism SMA nanostructures subjected to temperature- and stress- induced loadings, with our special attention to the effects of aspect ratio and boundary conditions. The nomenclature of domains and boundaries is presented in Fig. \ref{fig:Ch14SchematicDomains}. The simulations have been conducted on Fe$_{70}$Pd$_{30}$  material \cite{Ahluwalia2006}, whose properties have been summarized in Table \ref{tab:Ch14MatProperties}.  

\begin{table}[htbp]
  \centering
  \caption{Fe$_{70}$Pd$_{30}$ material constants}
    \begin{tabular}{ccccccc}
$a_1$ & $a_2 $ & $a_3 $ & $a_4 $ & $a_5 $ & $ \eta $ \\ \hline
192.3 GPa & 280 GPa & 19.7 GPa & 2.59$\times$10$^3$ GPa & 8.52$\times$10$^4$ Gpa & 0.25 N-s m$^2$   \\ \hline
$k_g $ & $\theta_m$ & $\theta_0$ & $C_v $ & $\kappa$ & $\rho$ \\ \hline
3.15$\times$10$^{-8}$ N & 270 K & 295 K & 350 J kg$^{-1}$ K$^{-1}$ & 78 W m$^{-1}$ K$^{-1}$ & 10000 kg m$^{-3}$ \\
\hline \\
    \end{tabular}%
  \label{tab:Ch14MatProperties}%
\end{table}%
 
\section{Temperature Induced Transformations} \label{sec:Ch14tempinduced}

SMAs evolve into complex microstructures below the transition temperature. In the following subsections, we study how the geometric aspect ratio and boundary conditions affect the microstructure morphology. 

\subsection{Influence of Geometric Aspect Ratio} \label{sec:Ch14ThreeDImp}
\begin{figure}[h!]
\centering
\subfigure[Thin film]
{
\includegraphics[trim = 0mm 20mm 0mm 0mm, clip, width=0.28\textwidth]{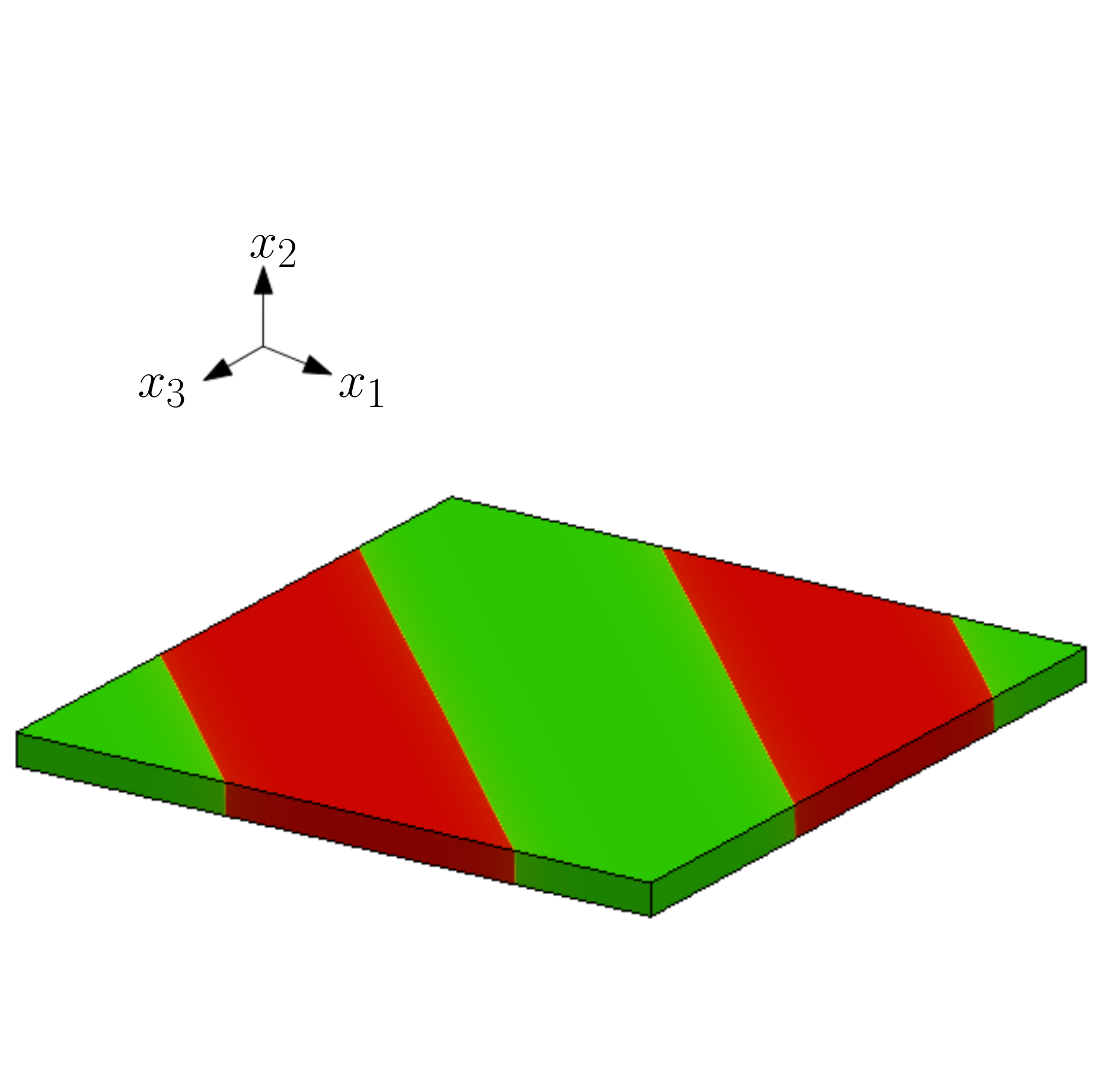}
\label{fig:Ch14ThreeDIMPThinFilm}
}
\subfigure[Slab]
{
\includegraphics[trim = 0mm 0mm 0mm 0mm, clip, width=0.28\textwidth]{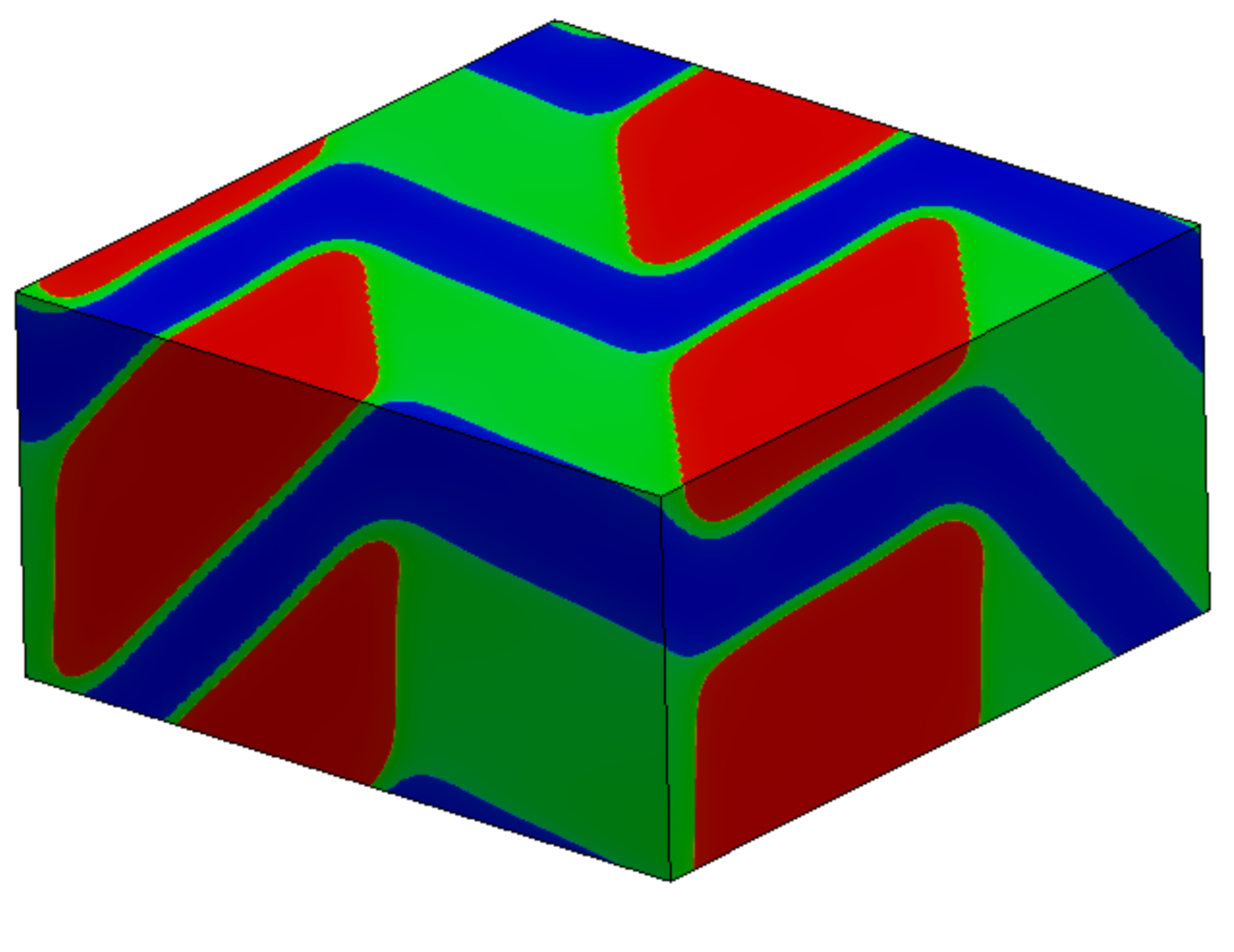}
\label{fig:Ch14ThreeDIMPSlab}
}
\subfigure[Cube]
{
\includegraphics[trim = 0mm 0mm 0mm 0mm, clip, width=0.28\linewidth]{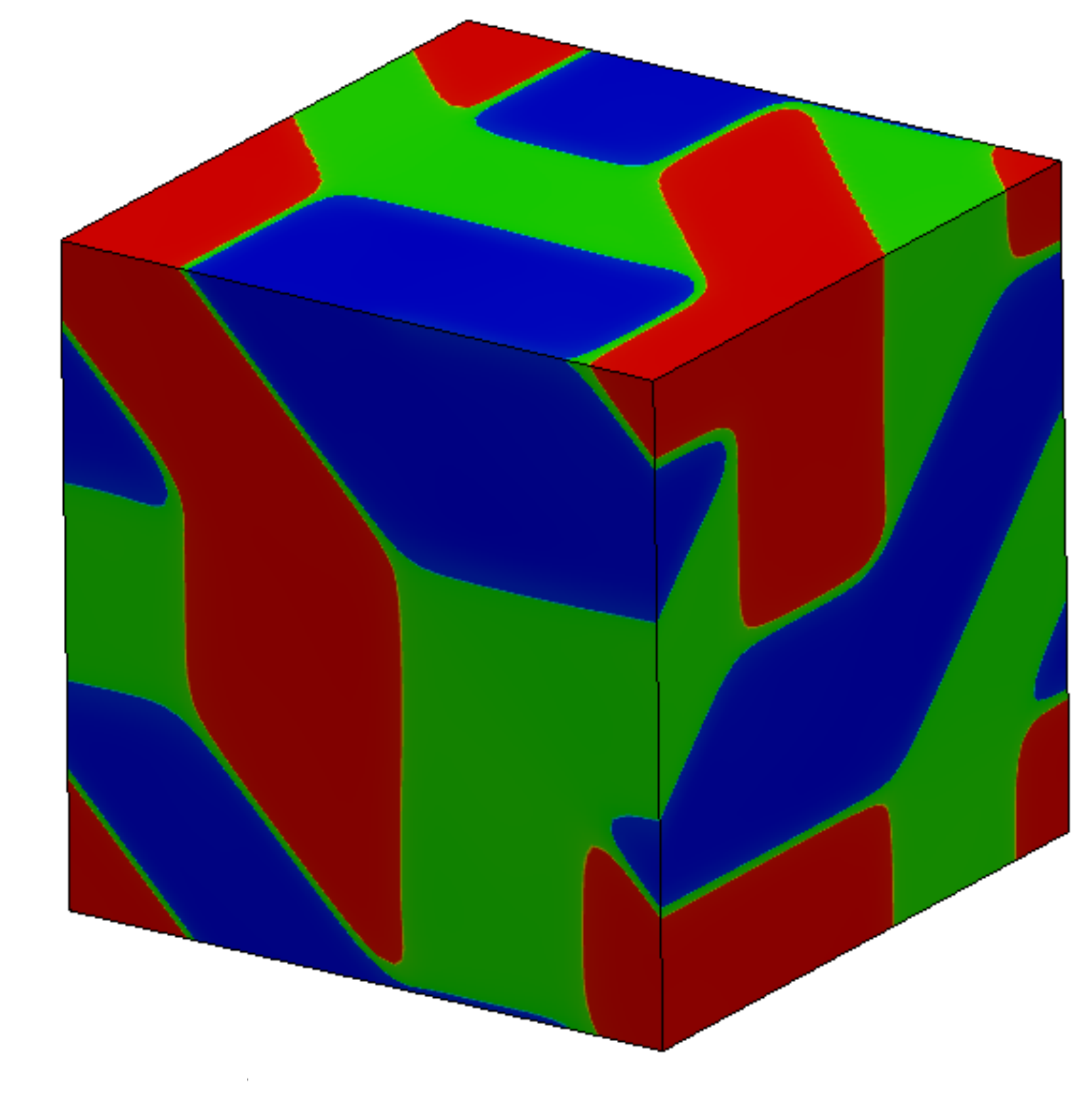}
\label{fig:Ch14ThreeDIMPCube}
}
\caption{(Color online) Self-accommodated microstructures in the (a) thin film, (b) slab  and (c) cube SMA domains (red, blue, and green colors represent M$_1$, M$_2$, and M$_3$ variants, respectively).}
\label{fig:Ch14Importance3D}
\end{figure}

Here, the simulations have been conducted on rectangular prism  SMA specimens with $ L_{x_1}$=$L_{x_3}$=32 nm and three different thickness ratios $ L_{x_2}/L_{x_1} =$ 1/32, 1/2, 1 representing a thin-film, a slab, and a cube domain, respectively. The specimens are quenched to temperature corresponding to $ \tau  = -1.2$ and allowed to evolve until the energy and microstructures are stabilized. Fully periodic boundary conditions have been applied to the displacement vector $\vect u$ in all the simulations in Section \ref{sec:Ch14ThreeDImp}. The stabilized martensitic microstructures at $ t $ = 0.504 ns for these three domains are plotted in Fig. \ref{fig:Ch14Importance3D}. The red, blue, and green colors represent M$_1$, M$_2$, M$_3$ martensitic variants, respectively. 

\begin{figure}[h!]
\centering
\subfigure[]
{
\includegraphics[width=0.31\textwidth]{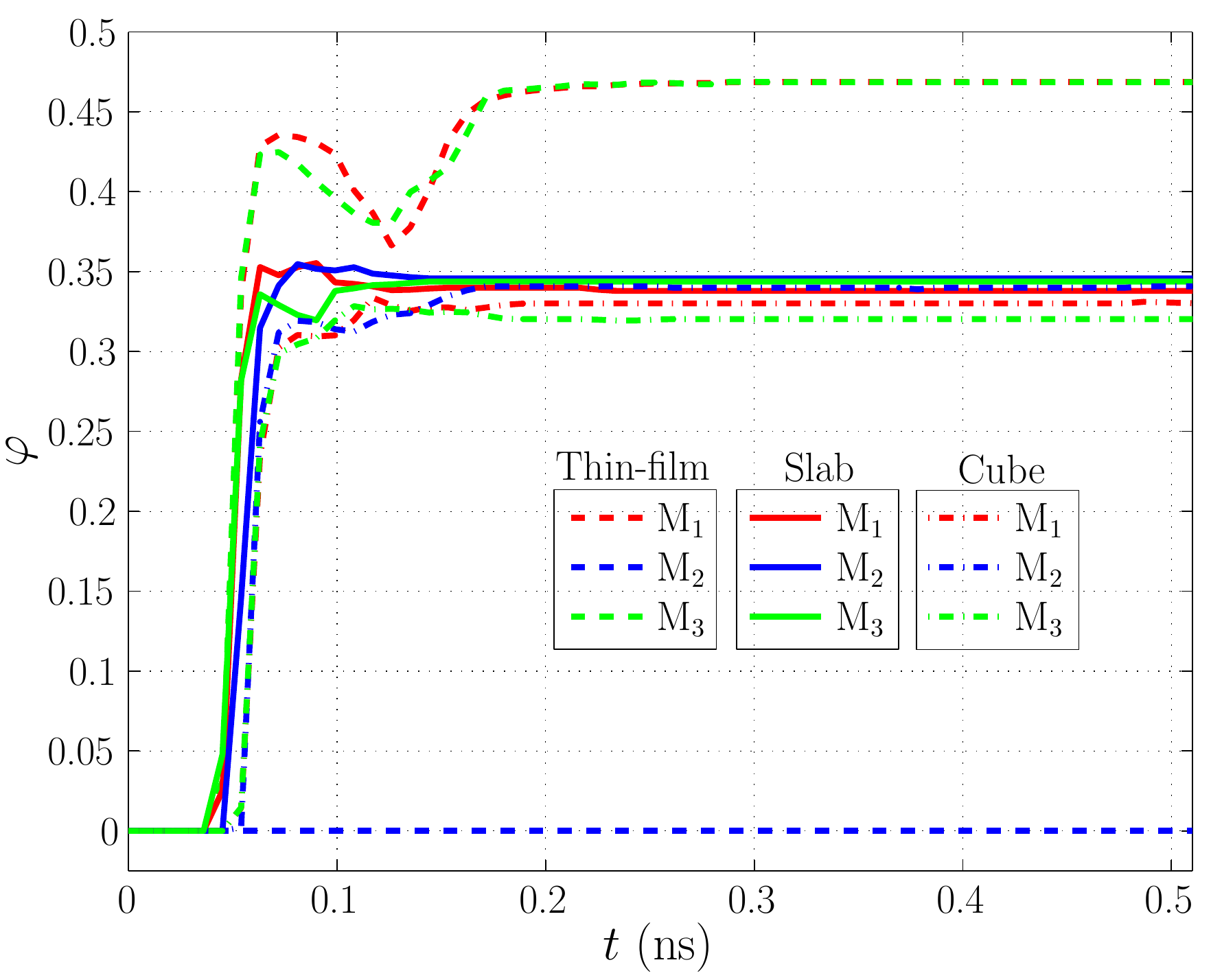}
\label{fig:Ch14ThreeDIMPCompareCSTFTSPhaseFraction}
}
\subfigure[]
{
\includegraphics[width=0.3\linewidth]{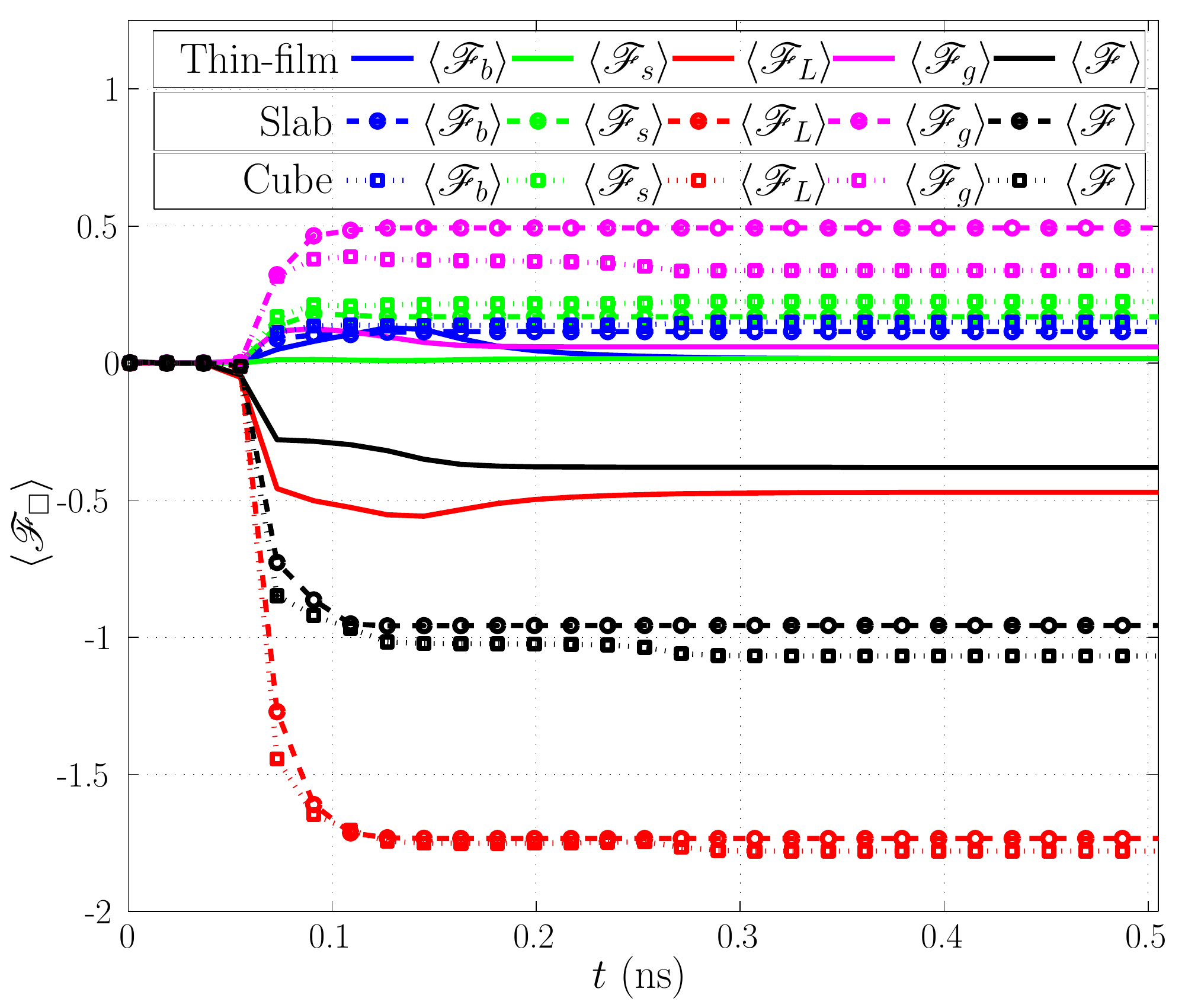}
\label{fig:Ch14ThreeDIMPCompareCSTFEnergyEvolution}
}
\subfigure[]
{
\includegraphics[width=0.31\textwidth]{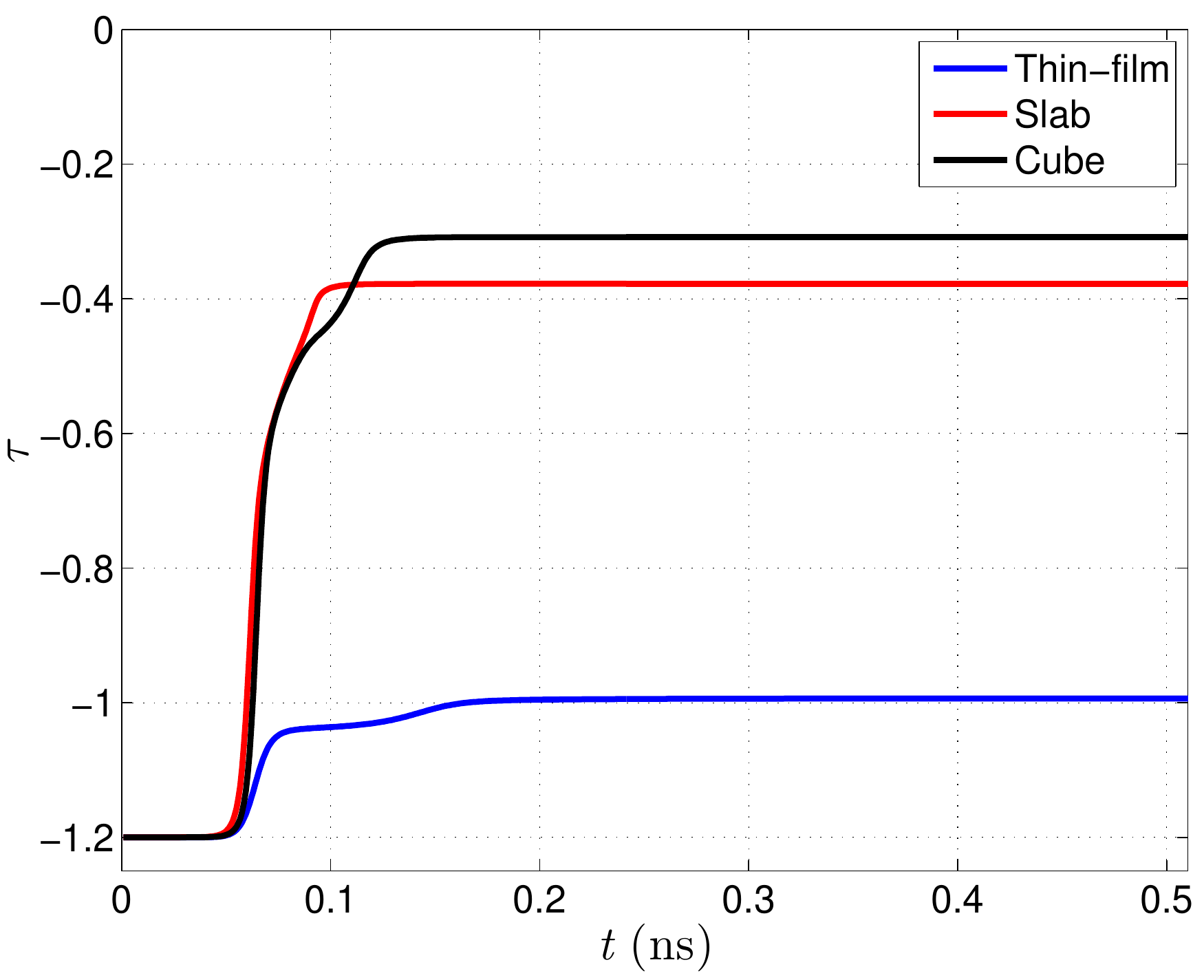}
\label{fig:Ch14ThreeDIMPCompareCSTFTimeVsTemp}
}	
\caption{(Color online) Evolution of (a) phase fraction $\varphi$ (red, green and blue colors refer to M$_1$, M$_2$, and M$_3$ variants, respectively), (b) energy density and (c) average temperature coefficient $ \tau $ with time.}
\end{figure}

The microstructures in the SMA thin-film domain self-accommodate to a twin morphology to minimize the energy as shown in Fig. \ref{fig:Ch14Importance3D}(a). Due to the thin-film geometry, the M$_2$ variant is not allowed and not favored. The microstructure reveals only two variants \mOne and \mThree with a complete suppression of the out-of-plane variant M$_2$. The microstructure is a sequence of \mOne and \mThree variant bands, which are evolved into approximately equal proportions as observed in Fig. \ref{fig:Ch14ThreeDIMPCompareCSTFTSPhaseFraction} (refer to the dashed lines). The twin boundaries are aligned along the \pozo plane. The stabilized microstructure reproduces the morphology previously reported in the 2D model \cite{dhote2013AST,Dhote2012}, the crystallographic theory \cite{Sapriel1975}, and experiments \cite{Kaushik}, thus verifying our numerical implementation and validating the results. The suppression of the out-of-plane variant in a thin-film has also been reported experimentally \cite{ma2012freestanding}. 

For energetic analysis, the energy density components $\langle \mathscr{F}_{\Box} \rangle$ (in the dimensionless units) are plotted with time. Fig. \ref{fig:Ch14ThreeDIMPCompareCSTFEnergyEvolution} reveals the $\langle \mathscr{F}_{\Box} \rangle$ evolution  for a thin-film specimen (refer to the solid lines). The auto-catalytic nucleation starts at different locations in the domain at approximately 0.05 ns, the domain gets deformed, leading to internal strain, which in-turn, transcends as the long-range elastic interactions causing the increase in $\langle \mathscr{F}_g \rangle$. Due to the strained state of the domain, the $\langle \mathscr{F}_b \rangle$ and $\langle \mathscr{F}_s \rangle$ increase. As the microstructure evolves, the $\langle \mathscr{F} \rangle$ is minimized with stabilization of energy components. The competition between $\langle \mathscr{F}_g \rangle$, $\langle \mathscr{F}_b \rangle$ and $\langle \mathscr{F}_s \rangle$ leads to the formation of the twinned microstructure.

The stabilized microstructure in the SMA slab domain is shown in Fig. \ref{fig:Ch14Importance3D}(b). The morphology reveals the primary bands of \mOne and \mThree variants, as observed in Fig. \ref{fig:Ch14Importance3D}(a), however, they are penetrated by the secondary band of the \mTwo variant. The three variants self-accommodate to form Chevron or herringbone patterns. The  evolution to the approximately equal proportion of three martensitic variants is shown in Fig. \ref{fig:Ch14ThreeDIMPCompareCSTFTSPhaseFraction} (refer to the solid line). The observation of equal phase fractions of martensitic variants in the domains has been reported in the literature \cite{cui8strain}. The width ratios of M$_1$:M$_2$ and M$_3$:M$_2$ bands are observed to be approximately 2:1. Similar width ratios have been observed numerically \cite{wang1997three,jacobs2003simulations} and experimentally in ferroelectric and ferroelastic ceramics \cite{arlt1990twinning,pertsev1992theory}. The step-like morphology of M$_2$ has also been reported experimentally \cite{boullay2003nano}. The energetic analysis reveals that the nucleation and energy stabilization occur faster in the slab, than in the thin-film domain. The $\langle \mathscr{F}_g \rangle$ dominates and is higher than the $\langle \mathscr{F}_b \rangle$ and $\langle \mathscr{F}_s \rangle$.

The microstructure morphology in the cube domain is distinct compared to the thin-film and slab domains as shown in Fig. \ref{fig:Ch14Importance3D}(c). The bands of \mOne and \mThree variants are not prominent as observed in Figs. \ref{fig:Ch14Importance3D}(a) and (b). Due to the periodic boundary conditions and elastic strain accommodation, the domain walls are aligned along the \pooz planes and the microstructure forms polytwinned domains. Such accommodations of martensites have been reported in \cite{jacobs2003simulations}. The three variants evolve in an equal phase fraction $\varphi$ as shown in Fig. \ref{fig:Ch14ThreeDIMPCompareCSTFTSPhaseFraction} (refer to the dash-dot lines). 

Interestingly, the $\langle \mathscr{F}_g \rangle$ in the SMA cube specimen is smaller than in the slab specimen. The periodic boundary condition forces the slab domain to create more domain walls to maintain the equal proportion of martensitic variants in less volume (as compared to the cube domain) leading to a higher gradient energy. 

The evolution of the average temperature coefficient $ \tau $ for the three domains is  plotted in Fig. \ref{fig:Ch14ThreeDIMPCompareCSTFTimeVsTemp}. A temperature increase is observed during microstructure evolution due to  the thermo-mechanical coupling and insulated boundary conditions. The evolved temperature decreases with higher aspect ratios, due to the formation of fewer twins. This suggests that a better understanding the thermo-mechanical coupling is especially important in 3D, as different microstructure morphologies are evolved based on the aspect ratios, that cannot be captured by the 2D models \cite{Bouville2008,dhote2014isogeometric}.

Although all the above three simulations use periodic boundary conditions, it can be concluded that the aspect ratio of the domain plays an important role in the microstructure evolution. A lower aspect ratio promotes an equal proportion of martensitic variants, while a higher aspect ratio leads to suppression of the out-of-plane variant. Thus, by tuning the geometric aspect ratio, the domain patterns (microstructure) can be modified, which in turn can control preferential deformations for building special applications using SMAs \cite{bhattacharya2005material}.

\subsection{Influence of Boundary Conditions} 

In this section, we study the influence of boundary conditions on the microstructure morphology of a 80 nm side SMA cube specimen. We start with random initial conditions in the displacement vector $\vect u$. In all the simulations, the SMA specimen is quenched to temperature corresponding to $\tau = -1.2$ and allowed to evolve for a sufficiently long time until the energy and microstructure are stabilized. Three different boundary conditions on $\Gamma$ have been used for the simulations: 
(i) fully constrained (FC) case with $\vect u = 0$, 
(ii) normally constrained (NC) case with $\vect u\cdot\vect n= 0$ and
(iii) fully periodic (FP) case.

\begin{figure}[h!]
\centering
\subfigure[Fully Constrained]
{
\includegraphics[width=0.23\textwidth]{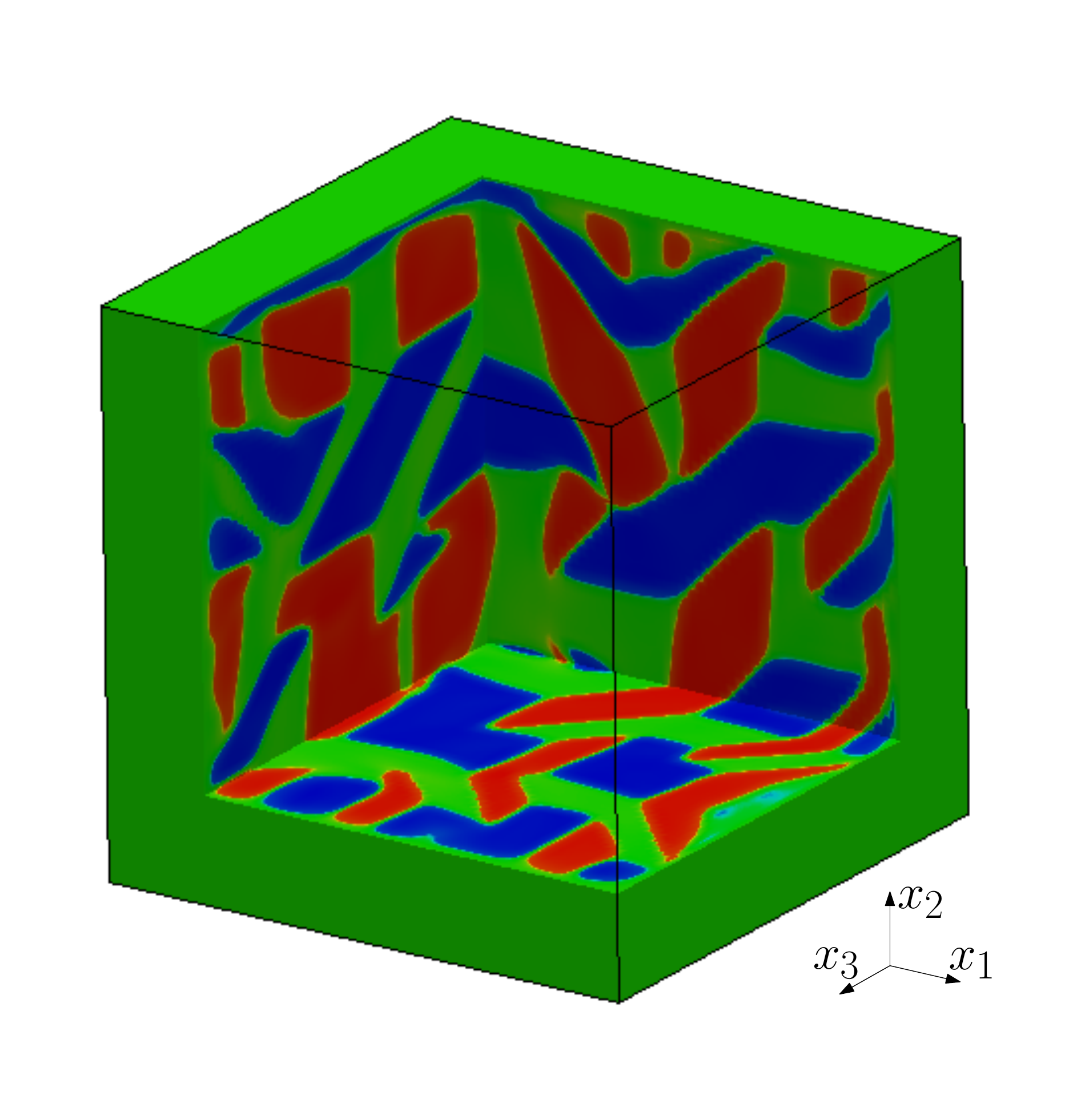}
\label{fig:Ch14DiffBCFullConstraint}
}
\subfigure[Normally Constraint]
{
\includegraphics[width=0.23\textwidth]{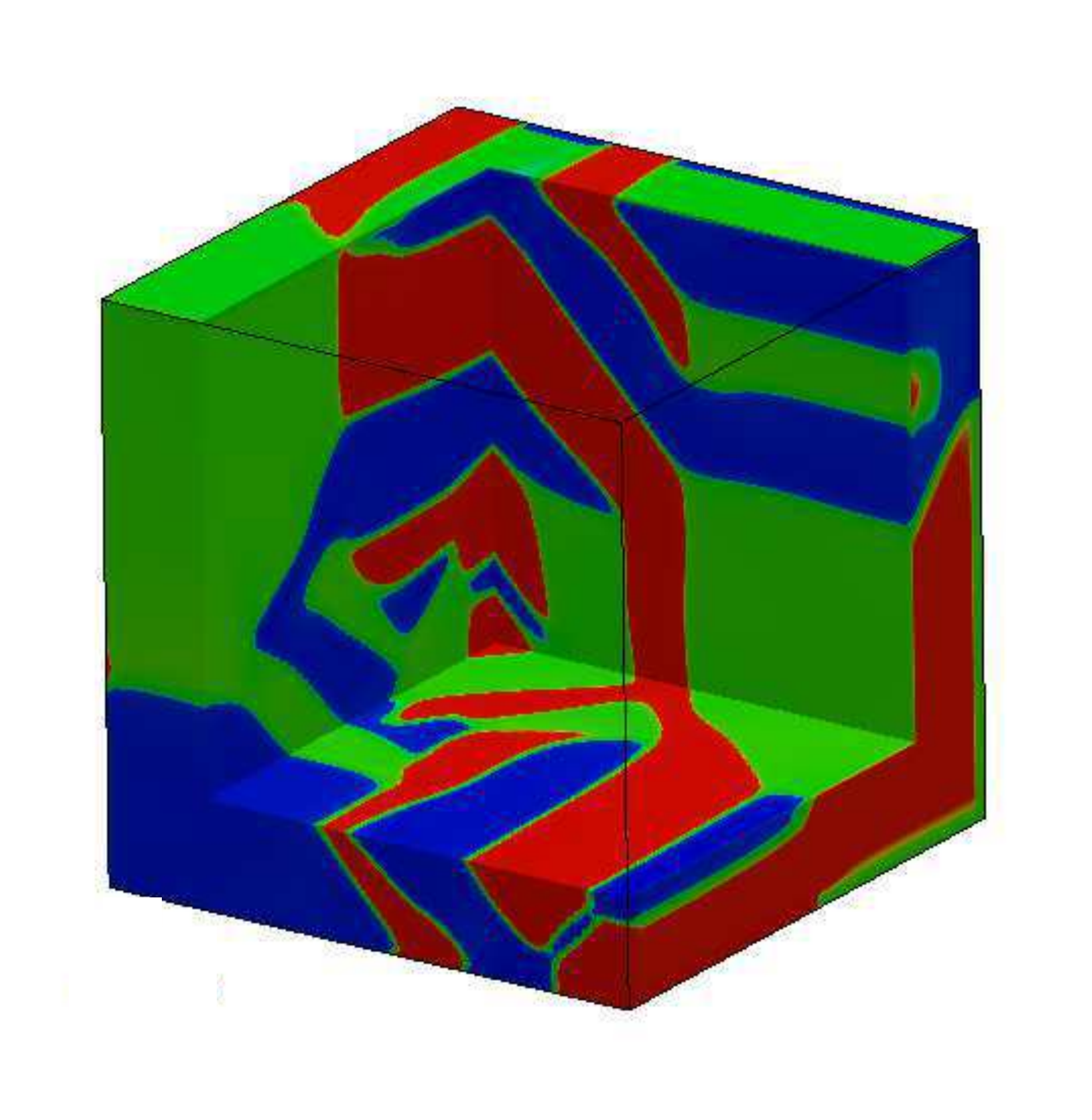}
\label{fig:Ch14DiffBCNormalizedConstraint}
}
\subfigure[Fully Periodic]
{
\includegraphics[width=0.23\linewidth]{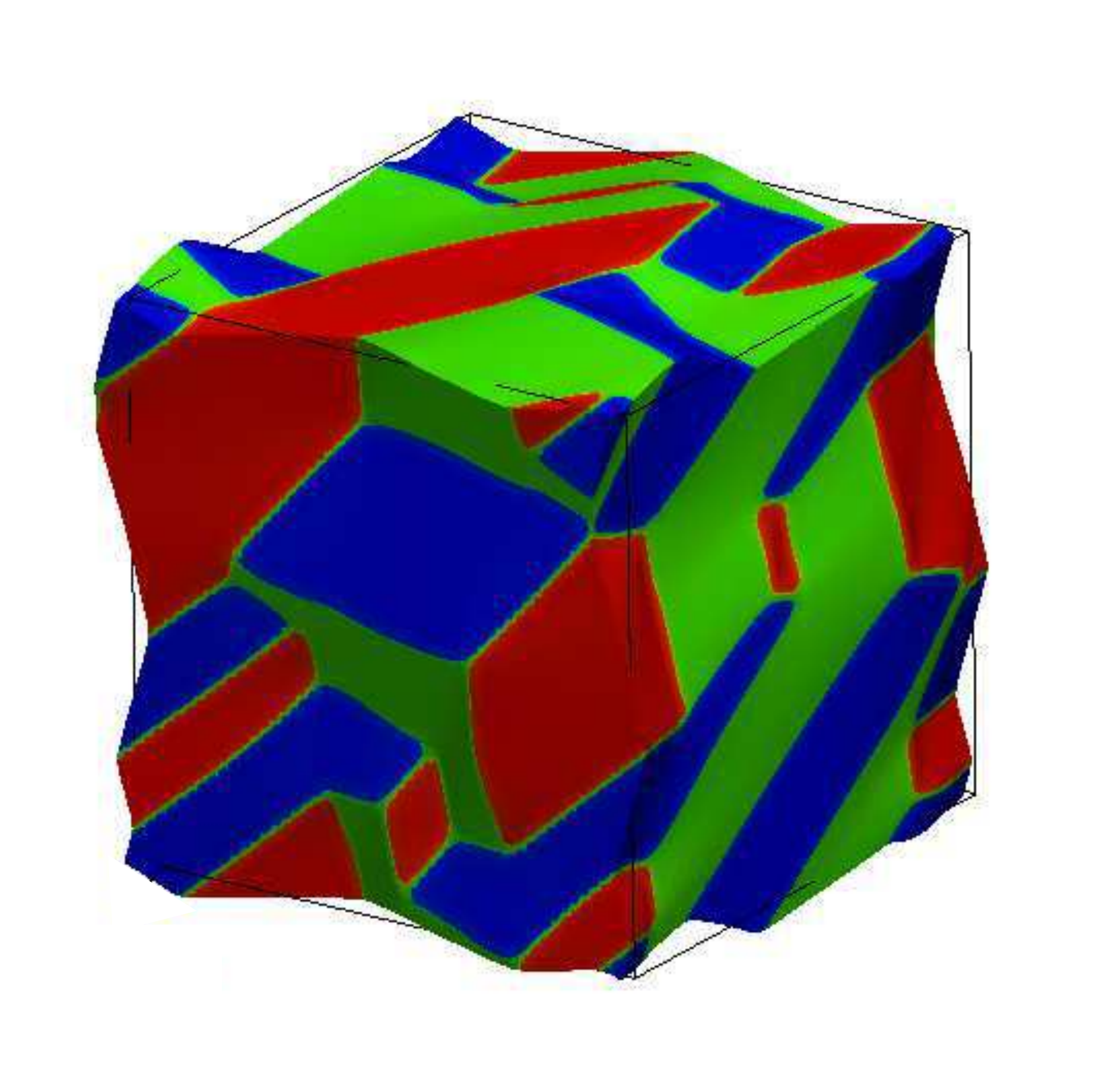}
\label{fig:Ch14DiffBCFullPeriodic}
}
\caption{(Color online) Self-accommodated microstructure in a 80 nm side SMA cube under (a) fully constrained, (b) normally constrained and (c) fully periodic boundary conditions (red, blue, and green colors represent M$_1$, M$_2$, and M$_3$ variants, respectively).}
\label{fig:Ch14DiffBCMicrostructures}
\end{figure}

Fig. \ref{fig:Ch14DiffBCMicrostructures} presents the stabilized microstructure morphology at $t$ = 1.5 ns in the SMA cube domain with different boundary conditions. The sensitivity of the microstructure morphology evolution to the boundary conditions during temperature induced transformations is evident. The cube domain does not change the shape in the case of FC and NC boundary conditions. However, it can  change the shape in the FP case. 

\begin{figure}[h!]
\centering
\subfigure[]
{
\includegraphics[width=0.30\textwidth]{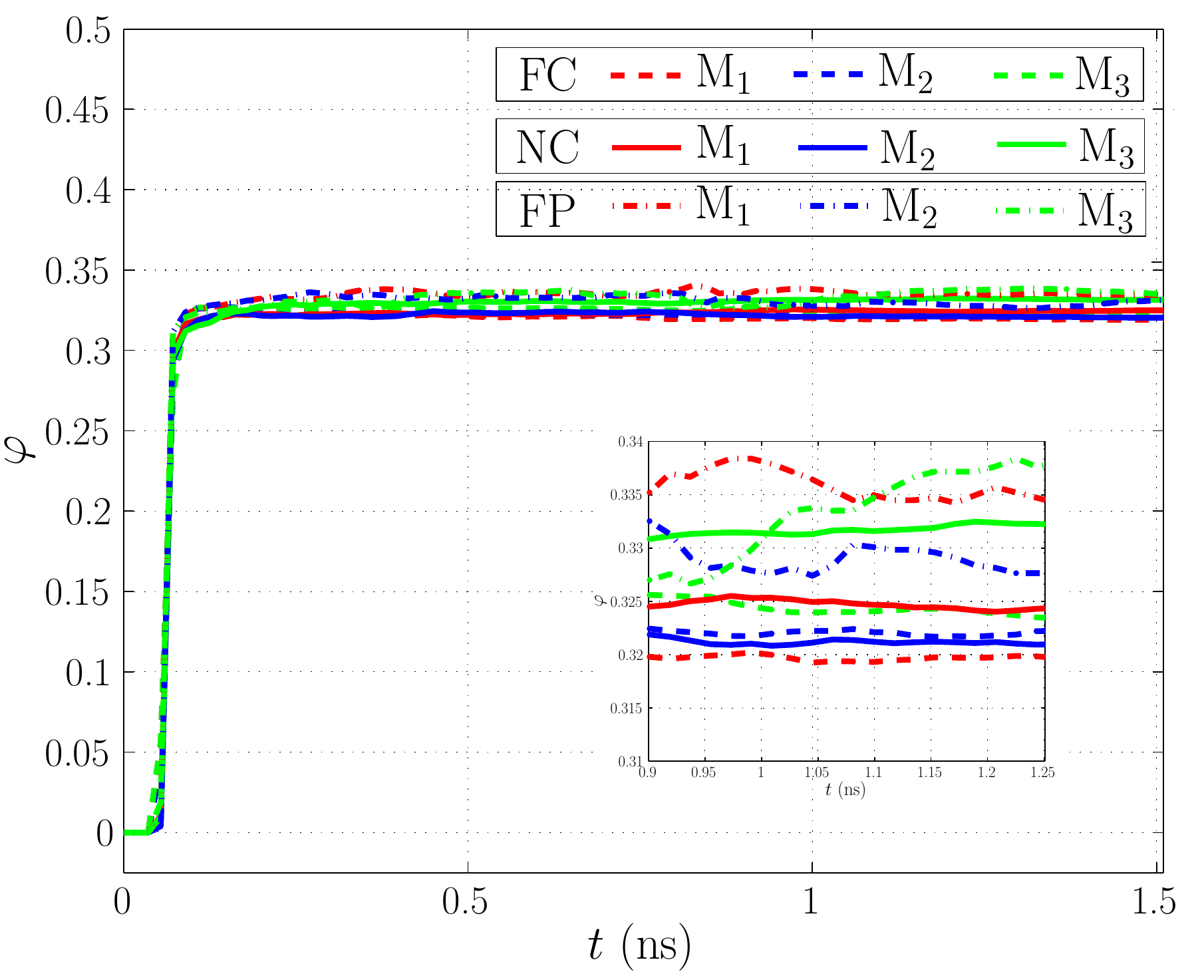}
\label{fig:Ch14DiffBCPhaseFraction}
}
\subfigure[]
{
\includegraphics[width=0.335\linewidth]{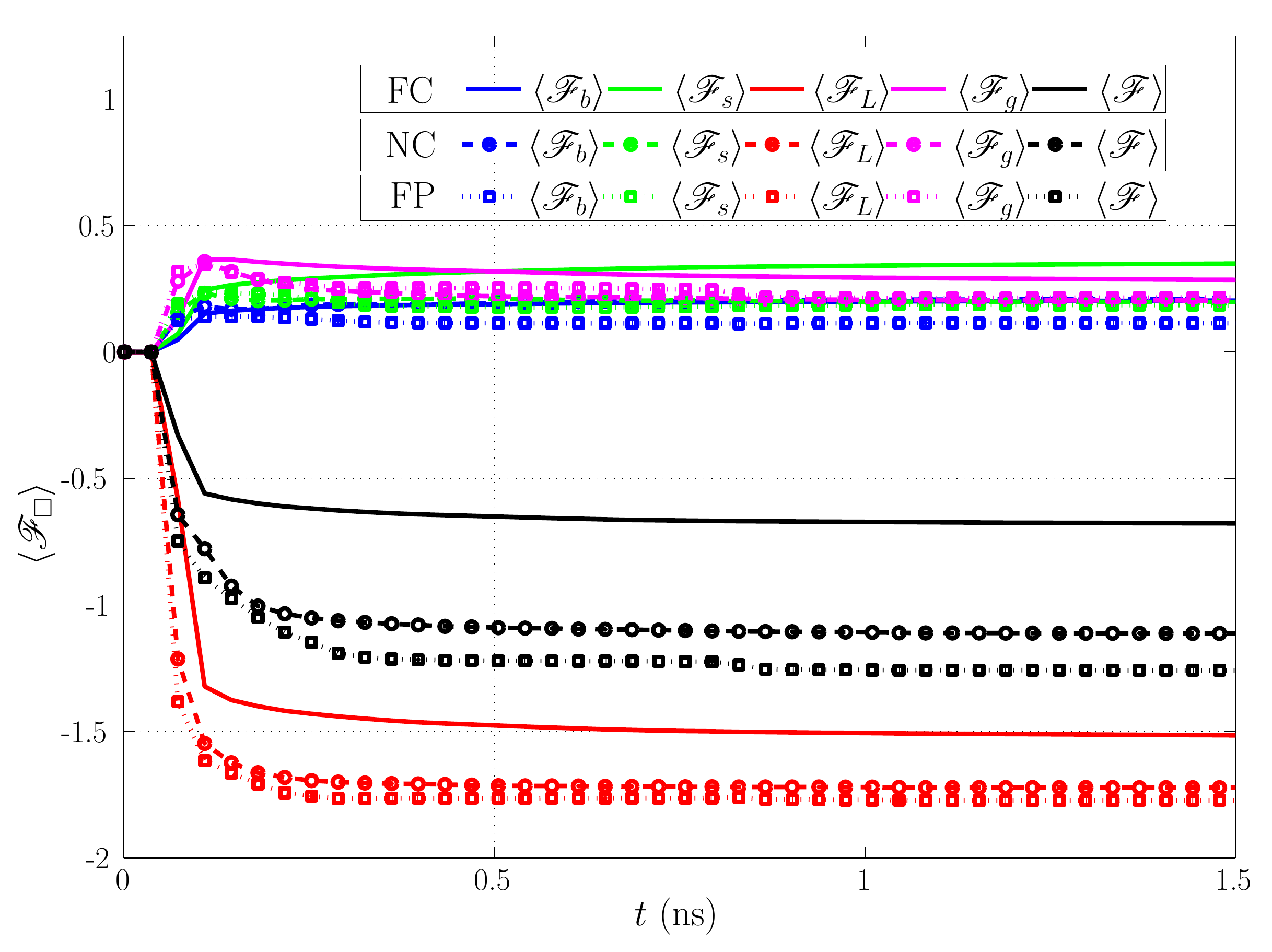}
\label{fig:Ch14DiffBCEnergyEvolution}
}
\subfigure[]
{
\includegraphics[width=0.30\textwidth]{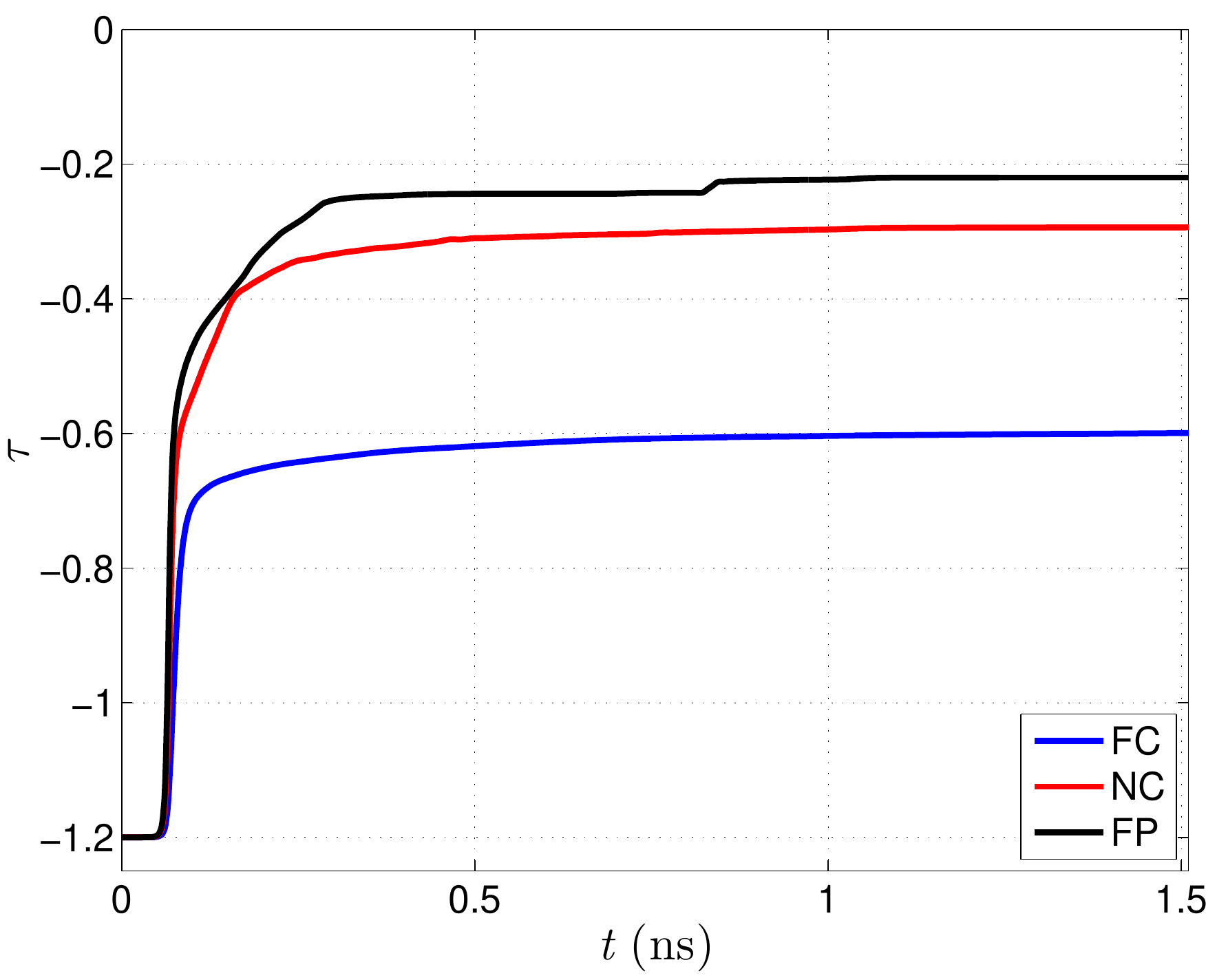}
\label{fig:Ch14DiffBCTimeVsTemp}
}	
\caption{(Color online)  Evolution of (a) phase fraction $\varphi$ (red, green and blue colors refer to M$_1$, M$_2$, and M$_3$ variants, respectively), (b) energy density $\langle\mathscr{F}_{\Box}\rangle$, and (c) average temperature coefficient $ \tau $ with time. FC, NC and FP represent fully constrained, normally constrained and fully periodic boundary conditions, respectively.}
\label{fig:Ch14DiffBCEvolution}
\end{figure}

Examining the FC boundary condition case in Fig. \ref{fig:Ch14DiffBCMicrostructures}(a) (a portion of the domain is not shown for viewing the variants inside the constrained surfaces), it is observed that the martensite variants are self-accommodated to form a complex microstructure morphology. The nucleation starts homogeneously, and the self-accommodation and stabilization of martensitic variants take place quickly into smaller lamella aligned along \pooz planes. The mechanical constraints on surfaces $\Gamma$ forcefully introduce additional domain walls to maintain the size and shape of the cube domain. This leads to higher $\langle \mathscr{F}_g \rangle$ in the case of FC boundary conditions. The higher $\langle \mathscr{F}_g \rangle$ and additional domain walls lead to indistinguishable martensitic variant regions in the domain. Thus, the FC boundary condition case is governed by short-range effects. The $\langle \mathscr{F}_s \rangle$ dominates over the $\langle \mathscr{F}_g \rangle$ due to mechanical constraints. The $\langle \mathscr{F} \rangle$ is highest among all the cases. The martensitic variants evolve into equal proportions as shown in Fig. \ref{fig:Ch14DiffBCEvolution}(b).

The morphology for the NC boundary condition case is shown in Fig. \ref{fig:Ch14DiffBCMicrostructures}(b). Although the size and shape of the cube domain are constant, the martensitic variants are self-accommodated in a plate morphology, in contrast with the FC boundary condition case. The nucleation starts homogeneously. The $\langle \mathscr{F}_g \rangle$ is smaller compared to the FC boundary condition case because the domain walls can move to self-accommodate as the tangential displacement is allowed at the surface nodes. The bigger martensitic domains are a result of the long-range elastic interactions. Due to the system's ability for self-adjustment of the martensitic domain, the SMA specimen minimizes to a lower $\langle \mathscr{F} \rangle$  as compared to the FC boundary condition case.

The FP boundary condition morphology is shown in Fig. \ref{fig:Ch14DiffBCMicrostructures}(c). Due to the system's ability to change shape, the martensitic variants are self-accommodated in equal proportions. The $\langle \mathscr{F}_g \rangle$ dominates over the $\langle \mathscr{F}_s \rangle$. The nucleation is homogeneous and domain walls are aligned along \pooz planes. Due to the flexibility of the system to change shape, the $\langle \mathscr{F} \rangle$ is lowest in the three boundary condition cases. The fully periodic system imposes an artificial constraint on the boundary to satisfy the continuity in degrees of freedom across the opposite surfaces of the cube. 

The effect of different boundary conditions is also observed on the evolution of $\tau$, as shown in Fig. \ref{fig:Ch14DiffBCEvolution}(c). The boundary conditions and elastic interactions in the domain also affect the thermal behavior. The temperature gets stabilized quickly in the FC case, while it takes a comparatively long time for the FP boundary condition case. 

\section{Stress Induced Transformations} \label{sec:Ch14TT}

SMAs exhibit phase transformations and hysteretic properties during stress induced loadings. Here, the simulations have been conducted on SMA nanowires to understand the thermo-mechanical behavior under dynamic loading conditions. The current study focuses on the behavior of prismatic specimens. The study of microstructure morphology on complex 3D geometries like tube, tubular torus and tubular spring using the developed model and isogeometric formulation have been reported in \cite{dhote2014CMAME,dhote2014Letters}. The shape-memory effect and pseudoelastic behaviors of SMA specimens have been studied in the following subsections. 

\subsection{Shape Memory Effect} \label{sec:Ch14TTSME}
\begin{figure}[h!]
\centering
\subfigure[\textit{t} = 0 ns]
{
\includegraphics[trim=0mm 0mm 0mm 0mm,clip, width=0.3\linewidth]{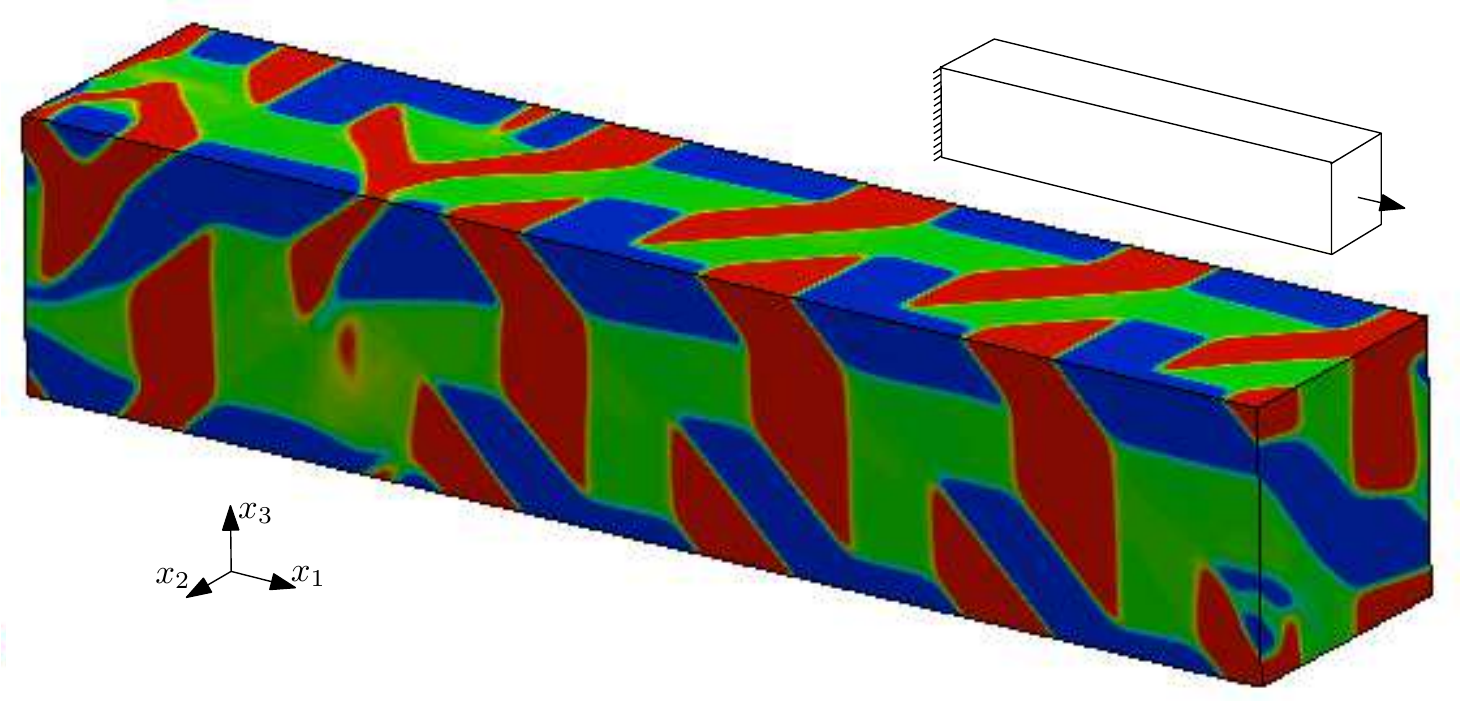}
}
\subfigure[\textit{t} = 0.017 ns]
{
\includegraphics[trim=0mm 0mm 0mm 0mm,clip, width=0.3\textwidth]{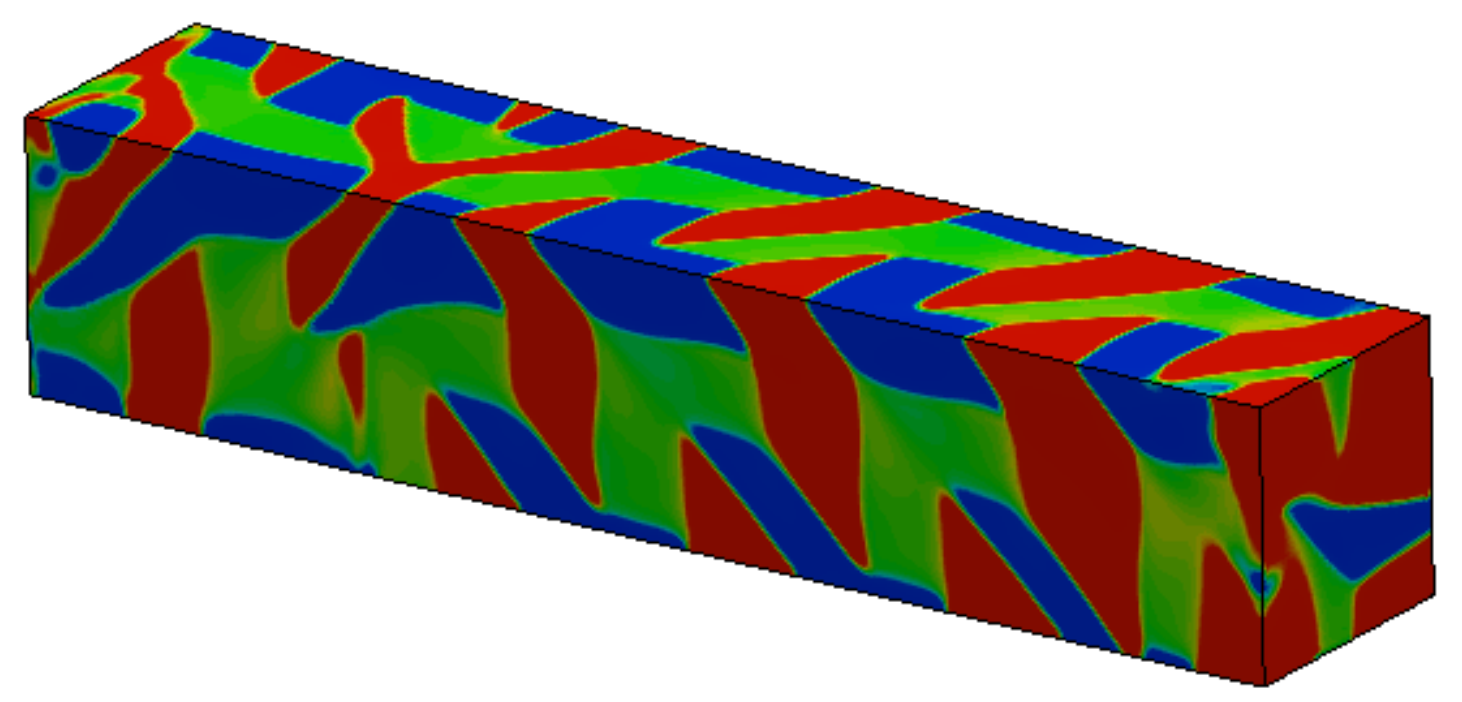}
}
\subfigure[\textit{t} = 0.033 ns]
{
\includegraphics[trim=0mm 0mm 0mm 0mm,clip, width=0.3\textwidth]{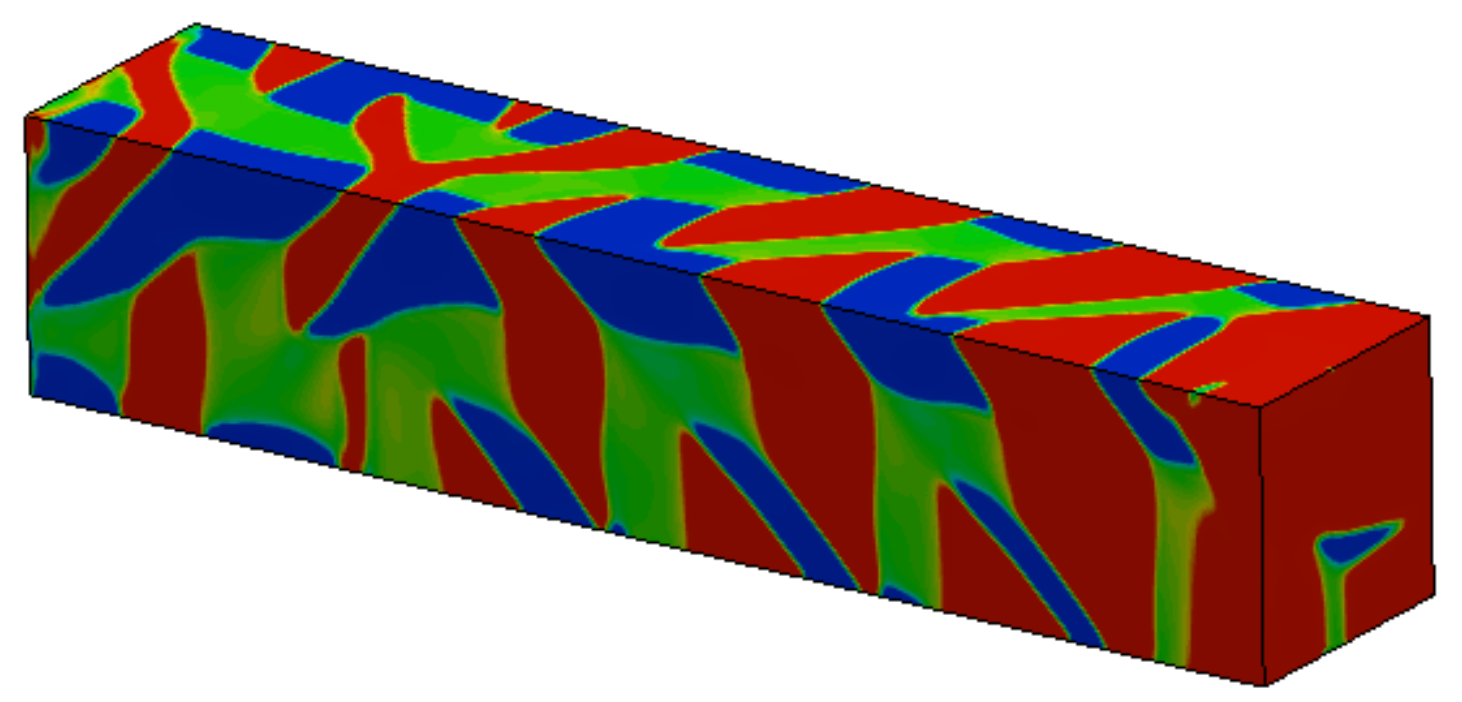}
}
\subfigure[\textit{t} = 0.05 ns]
{
\includegraphics[trim=0mm 0mm 0mm 0mm,clip, width=0.3\linewidth]{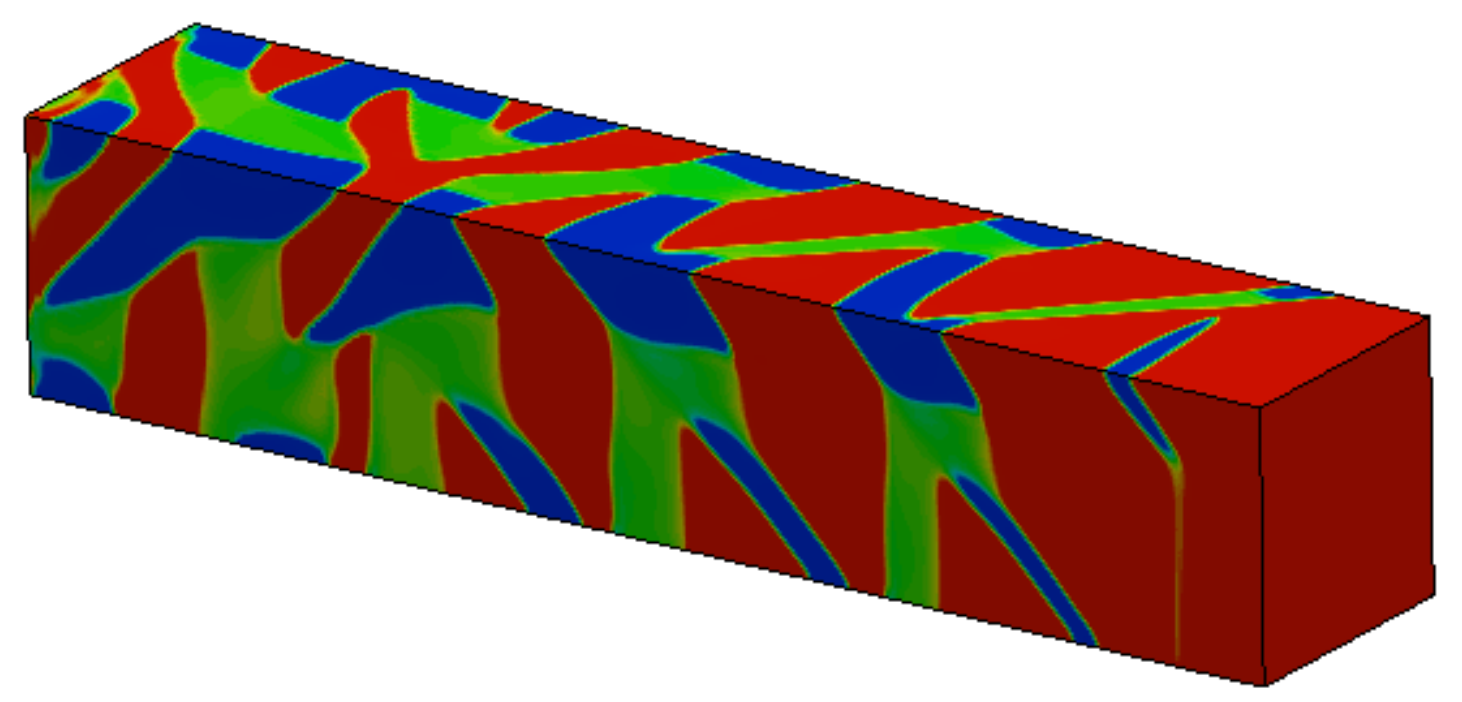}
}
\subfigure[\textit{t} = 0.067 ns]
{
\includegraphics[trim=0mm 0mm 0mm 0mm,clip, width=0.3\textwidth]{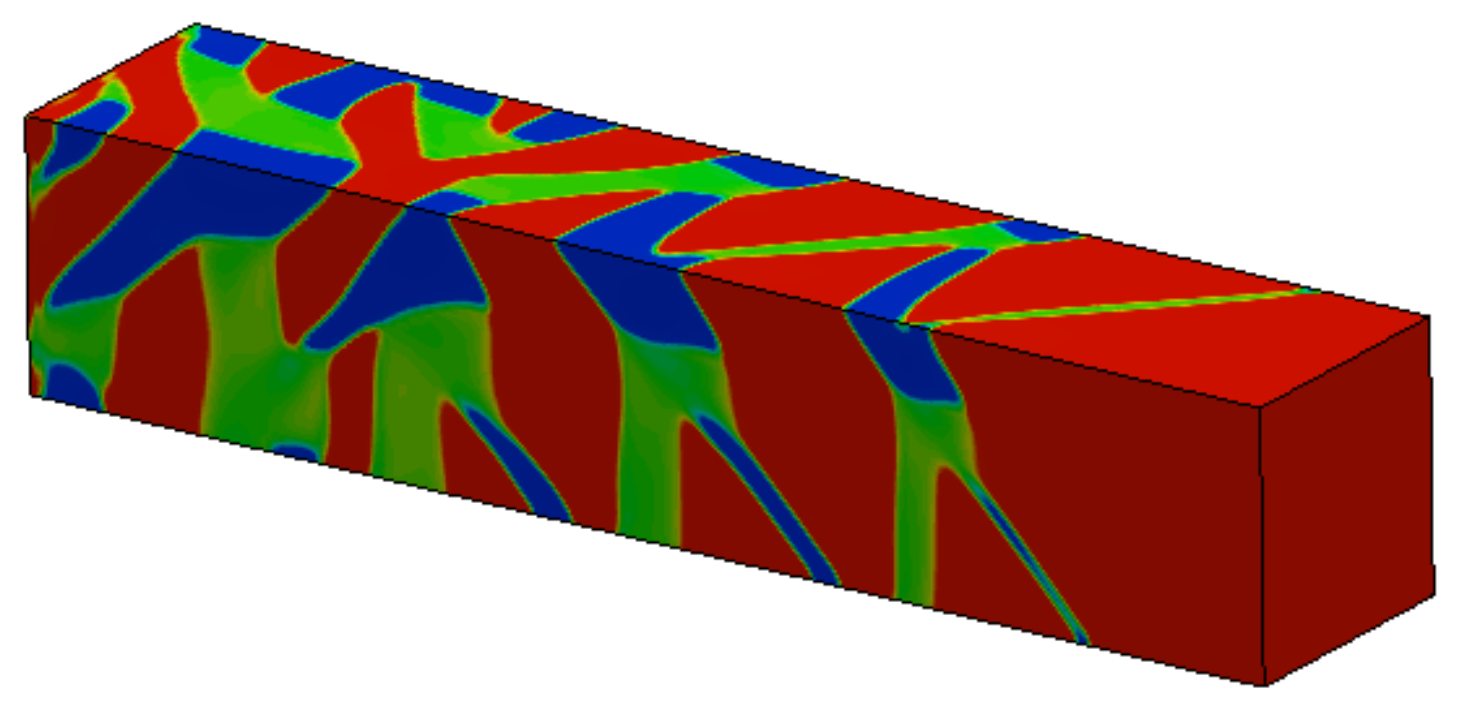}
}
\subfigure[\textit{t} = 0.083 ns]
{
\includegraphics[trim=0mm 0mm 0mm 0mm,clip, width=0.3\textwidth]{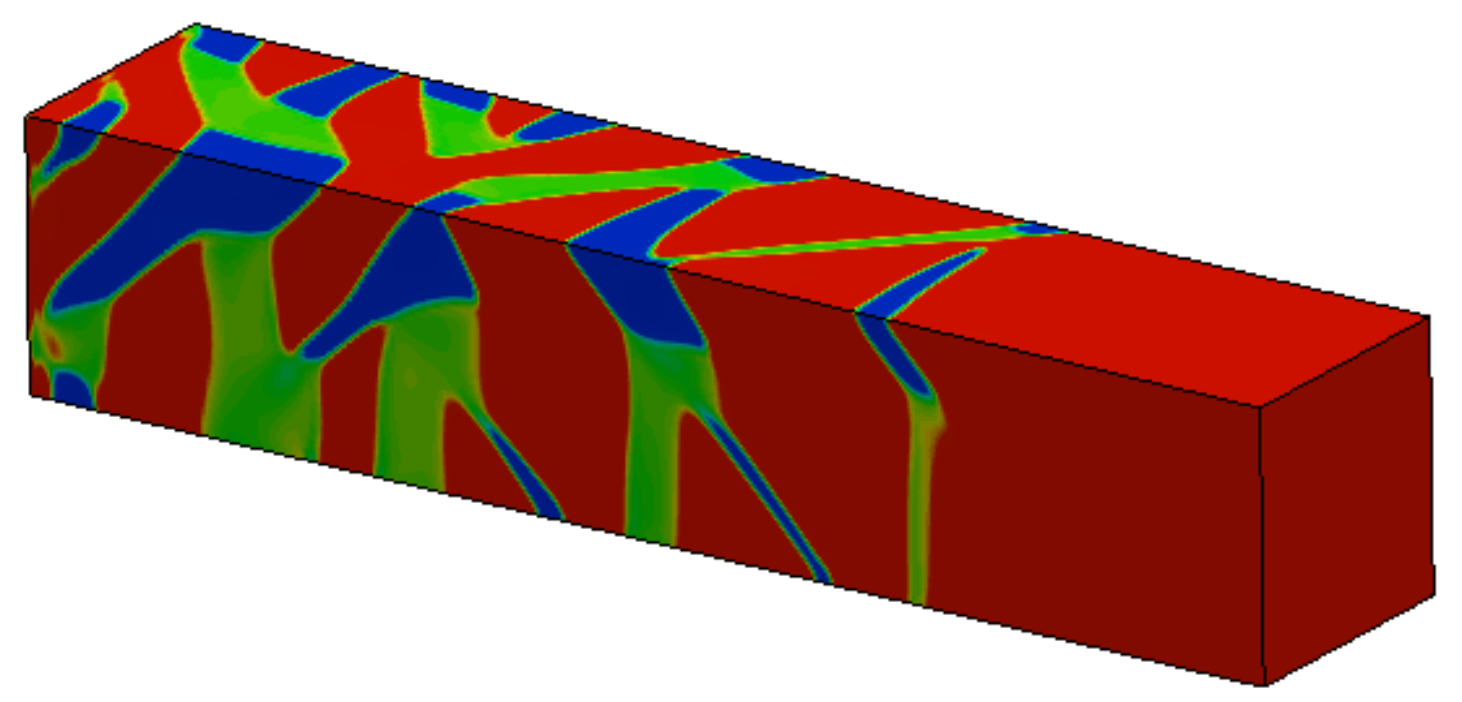}
}
\subfigure[\textit{t} = 0.1 ns]
{
\includegraphics[trim=0mm 0mm 0mm 0mm,clip, width=0.3\linewidth]{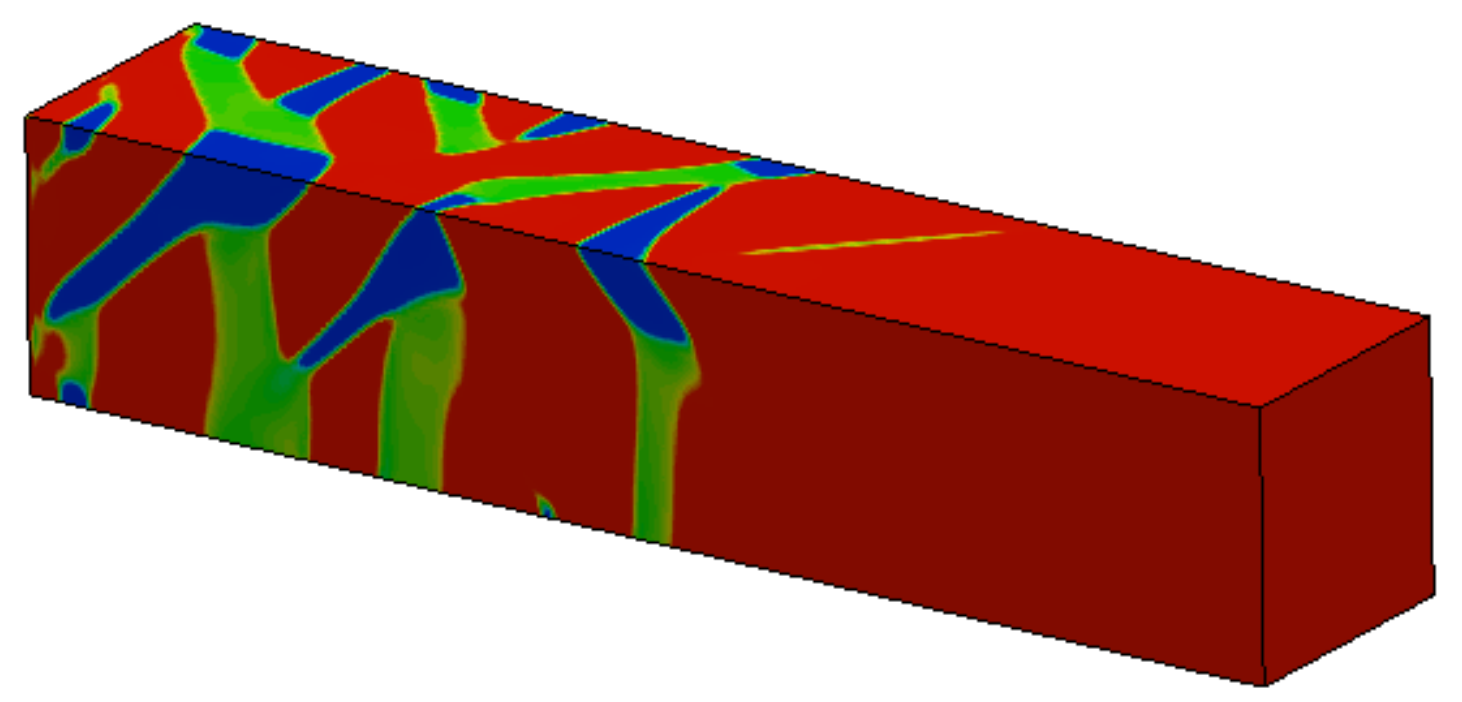}
}
\subfigure[\textit{t} = 0.117 ns]
{
\includegraphics[trim=0mm 0mm 0mm 0mm,clip, width=0.3\textwidth]{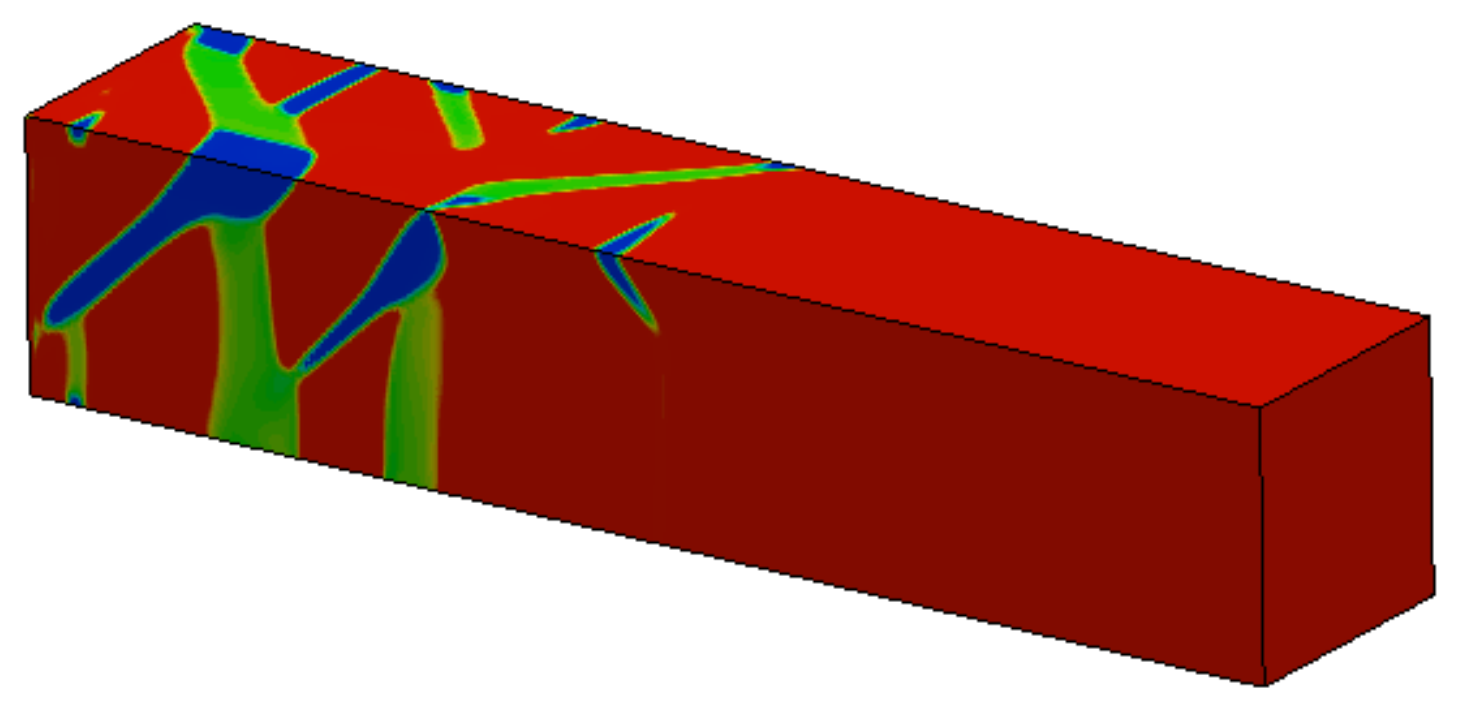}
}
\subfigure[\textit{t} = 0.3 ns]
{
\includegraphics[trim=0mm 0mm 0mm 0mm,clip, width=0.3\textwidth]{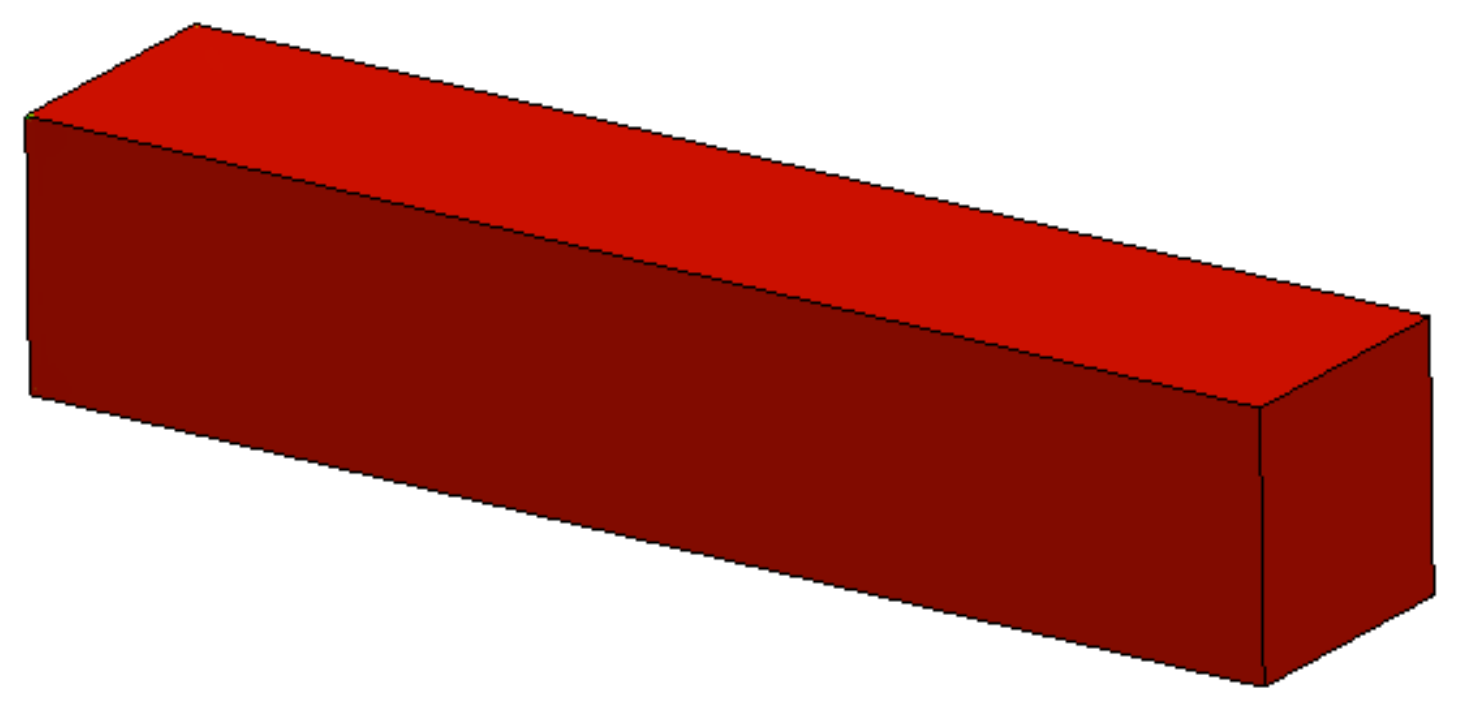}
}
\caption{(Color online) SME: microstructure morphology evolution in a 200$\times$40$\times$40 nm nanowire (red, blue, and green colors represent M$_1$, M$_2$, and M$_3$ variants, respectively).  }
\label{fig:Ch14SMETTmicrostructure}
\end{figure}

\begin{figure}[h!]
\centering
\subfigure[]
{
\includegraphics[width=0.31\textwidth]{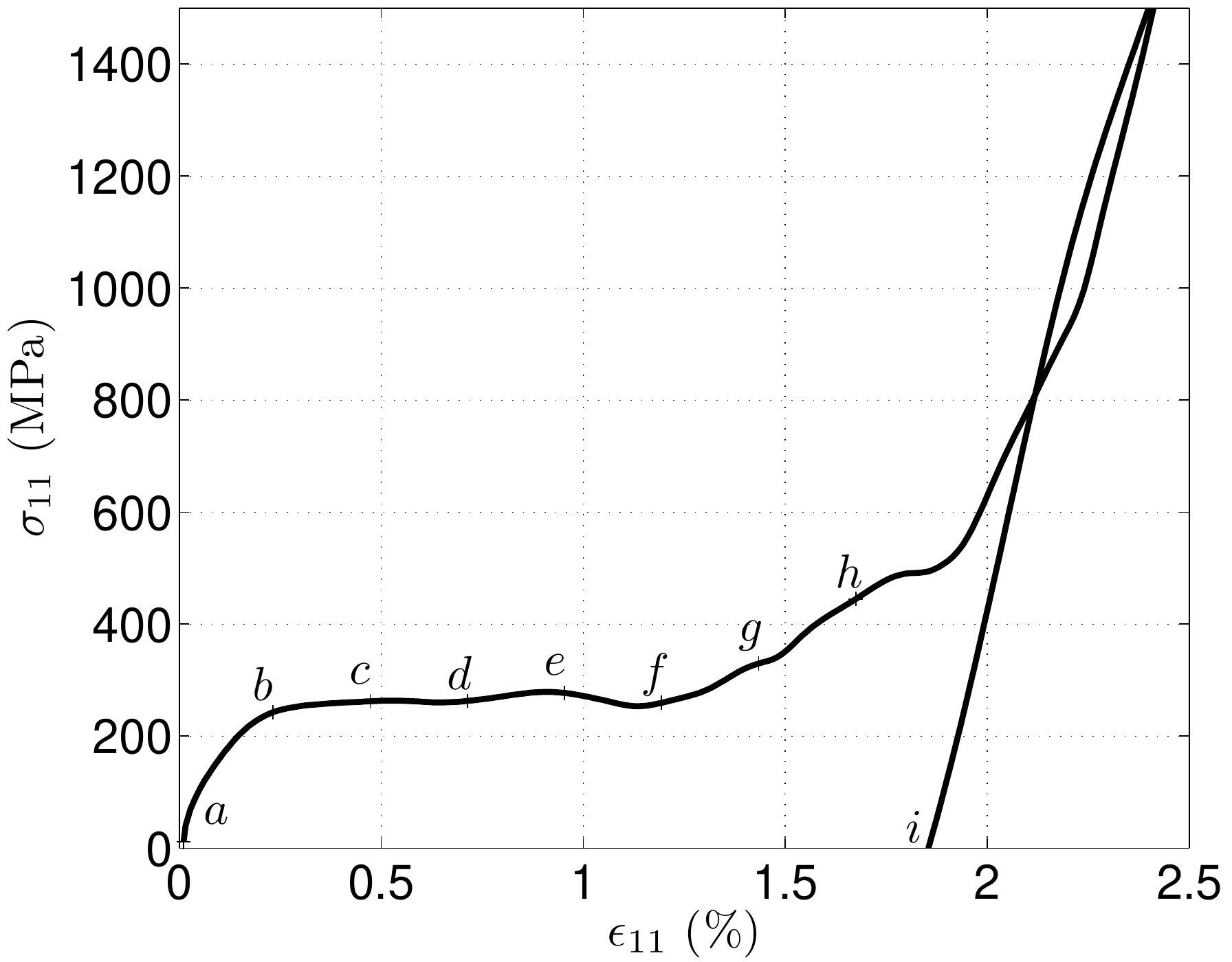}
\label{fig:Ch14SMETTSSxx}
}
\subfigure[]
{
\includegraphics[width=0.31\linewidth]{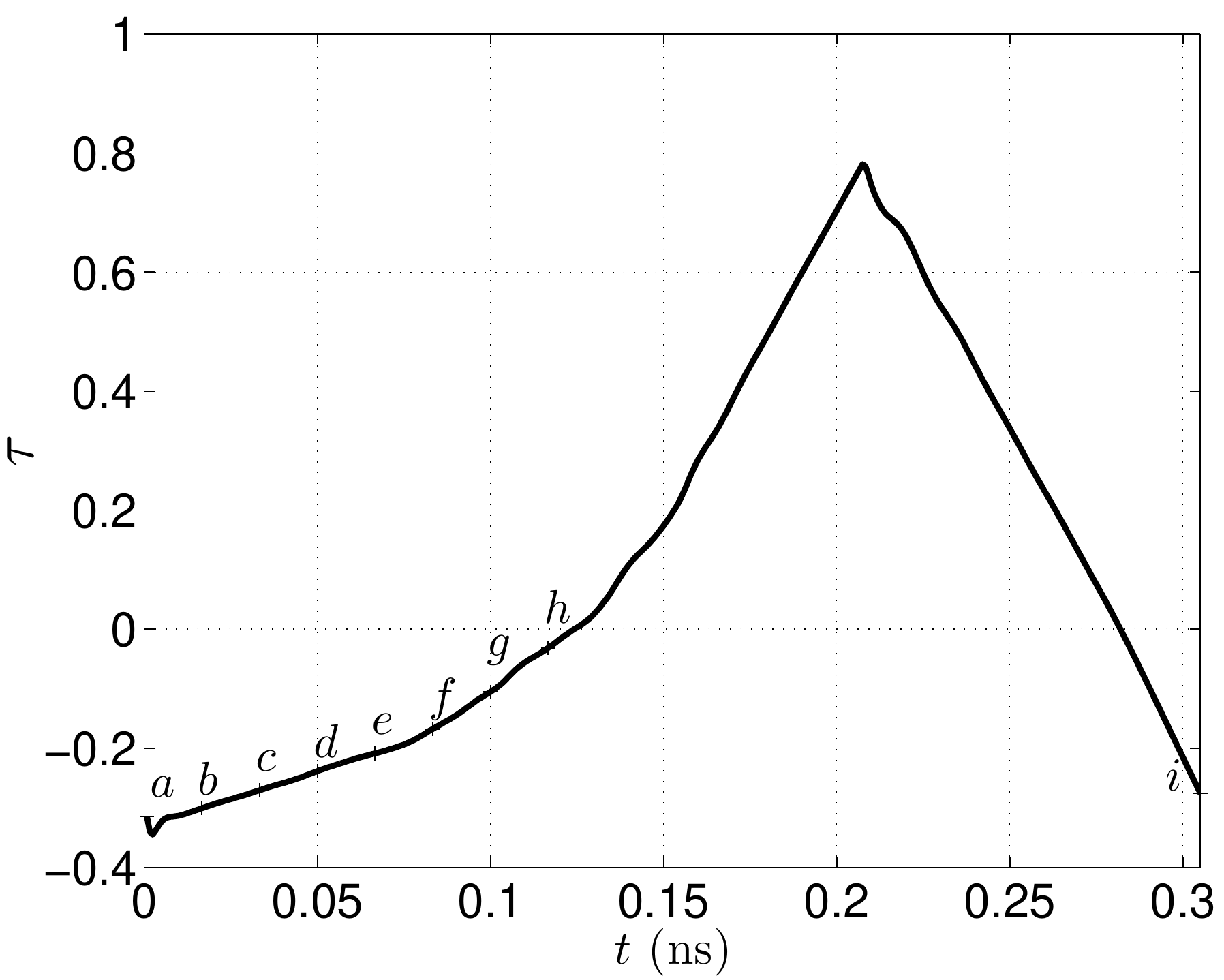}
\label{fig:Ch14SMETTTimeVsTemp}
}
\subfigure[]
{
\includegraphics[width=0.31\textwidth]{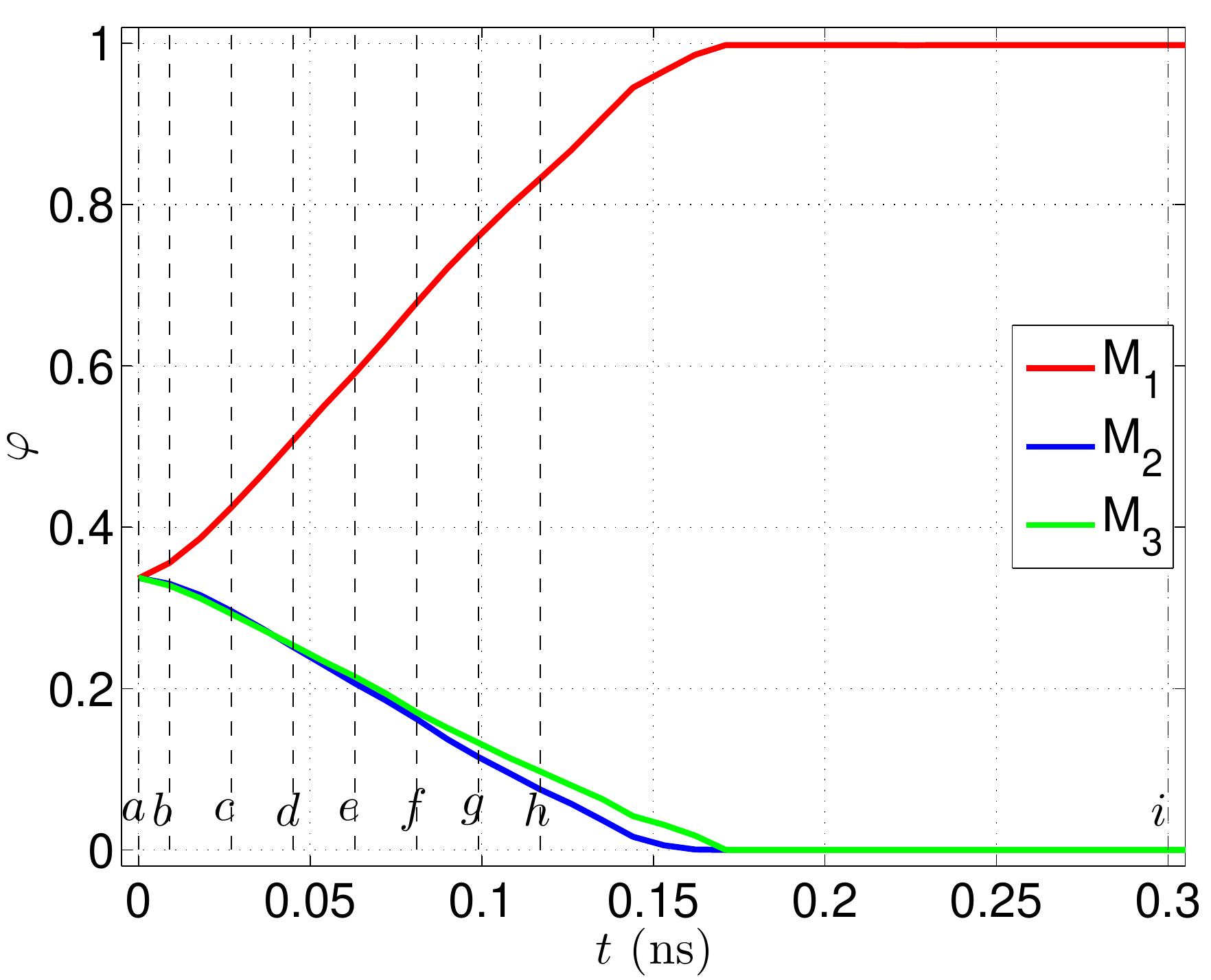}
\label{fig:Ch14SMETTTimeVsPhaseFraction}
}	
\caption{(Color online) SME: (a) the axial stress-strain ($\sigma_{11}$--$\epsilon_{11}$) relation, and time evolution of (b) average $ \tau $ and (c) phase fraction $\varphi$.}
\label{fig:Ch14SMETTAvgProperties}
\end{figure}

SMAs reveal shape-memory effect (SME) hysteretic behavior below the transition temperature.  A twinned martensite phase is transformed into the detwinned martensite phase under mechanical loading. The twinned microstructures are evolved in a rectangular prism nanowire of dimension (\lx$\times$\ly$\times$\lz) 200$\times$40$\times$40 nm. The simulation is conducted in two steps. First, the self-accommodated martensitic variants are evolved in the SMA specimen starting from initial random conditions of the displacement vector $\vect u$. The specimen is quenched to the temperature corresponding to $ \tau  = -1.2$ and allowed to evolve until microstructure and energy are stabilized. The three martensitic variants evolve into approximately equal proportions as seen at time $ t $ = 0 from the phase fraction $\varphi$ plot in Fig. \ref{fig:Ch14SMETTTimeVsPhaseFraction}. On energy stabilization, the  accommodated microstructure morphology is shown in Fig. \ref{fig:Ch14SMETTmicrostructure}(a). During the microstructure evolution, a temperature increase is observed due to the thermo-mechanical coupling owing to the movements of domain walls and the adiabatic conditions. The evolved microstructure is taken as an initial condition to the tensile test in the second step. 

The tensile test is conducted by constraining all displacements ($\vect u=0$) on the surface $ \Gamma_{x_1}(-) $ and axially loading and unloading the opposite surface $ \Gamma_{x_1}(+) $ in the out-of-material plane direction using a ramp-based displacement loading and unloading equivalent to the axial strain rate $\dot{\epsilon}_{11}$. The stress-free boundary conditions have been applied on the transverse surfaces $\Gamma_{x_i}(\pm)\vert_{i=2,3}$. 

The SME behavior of the nanowire is studied at axial strain rate $\dot{\epsilon}_{11}$ = 14.4$\times$10$^7$/s by achieving \mbox{3 $\%$} axial strain in 0.208 ns. Figs. \ref{fig:Ch14SMETTmicrostructure}(b-i) show  time snapshots of the microstructure evolution during axial loading and unloading. The phase fraction evolution and thermo-mechanical behavior in SMA specimen are plotted in Fig. \ref{fig:Ch14SMETTAvgProperties}. As the tensile test starts from a point $a$, the specimen is elastically loaded until it reaches  point $b$. Because of the loading and boundary condition, phase transformations start at the ends of the nanowire.  As the load progresses, the martensite variants start to coalesce to form distinct bands. The \mTwo and \mThree variants start converting into the favorable \mOne variant by the process of detwinning between points $b$--$h$. The motion, merging and vanishing of \mTwo and \mThree domains and growth of \mOne domain is apparent. During the loading, the domain walls move along the \pooz planes. During the detwinning process, the axial stress is nearly constant between points $b$--$f$. The SMA specimen undergoes phase transformations (M$_2$ $\rightarrow$\mOne and \mThree$\rightarrow$M$_1$) and elastic loading simultaneously between points $f$--$h$. After all the \mTwo and \mThree are converted to the \mOne variant, the specimen is loaded elastically. During unloading, the \mOne variant is elastically unloaded, till the stress in the domain is zero at the point $i$. The remnant axial strain is approximately 1.8 \% at the end of unloading. The effect of thermo-mechanical coupling is observed during the evolution of the average $\tau$ as shown in Fig. \ref{fig:Ch14SMETTAvgProperties}(b). The increase and decrease of temperature are a  result of exothermic and endothermic processes during loading and unloading of the nanowire. Similar behaviors of temperature increase and decrease have been observed experimentally during the dynamic loading-unloading of SMA specimens  \cite{shaw1995thermomechanical,Gadaj2002,Pieczyska2004,Pieczyska2010}.

As observed in Figs. \ref{fig:Ch14SMETTmicrostructure}--\ref{fig:Ch14SMETTAvgProperties}, the dynamic loading of SMAs involves growth, merging, elimination of martensitic variants, domain wall movements and temperature changes. It is now imperative to investigate how the domain size and strain rate affect  these mechanisms and properties. In the following sections, the simulations focus on these aspects.

\subsubsection{SME: Aspect Ratio Study} \label{sec:Ch14TTSMEAspectRatio}
To investigate the influence of aspect ratios on the SME behavior, the simulations have been conducted on nanowires of four dimensions: (i) 200$\times$40$\times$40 nm, (ii) 160$\times$40$\times$40 nm, (iii) 80$\times$40$\times$40 nm and (iv) 160$\times$40$\times$80 nm. All the simulations have been conducted according to the two-step procedure mentioned in the last section. The axial strain rate $\dot{\epsilon}_{11}$ = 14.4$\times$10$^7$/s  is used. 

\begin{figure}[h]
\centering
\subfigure[]
{
\includegraphics[width=0.4\textwidth]{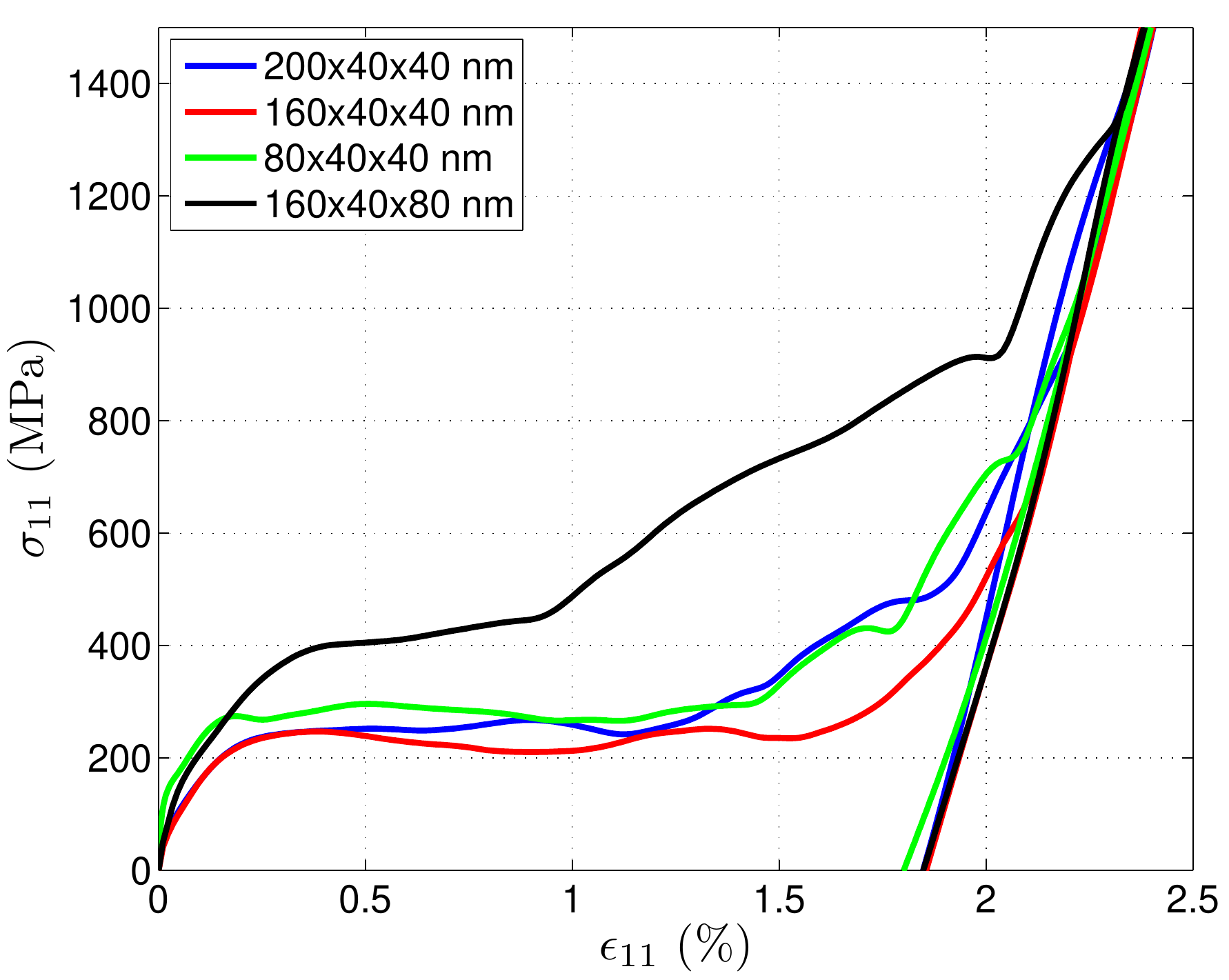}
\label{fig:Ch14SMEAspectRatioSSxx}
}
\subfigure[]
{
\includegraphics[width=0.4\linewidth]{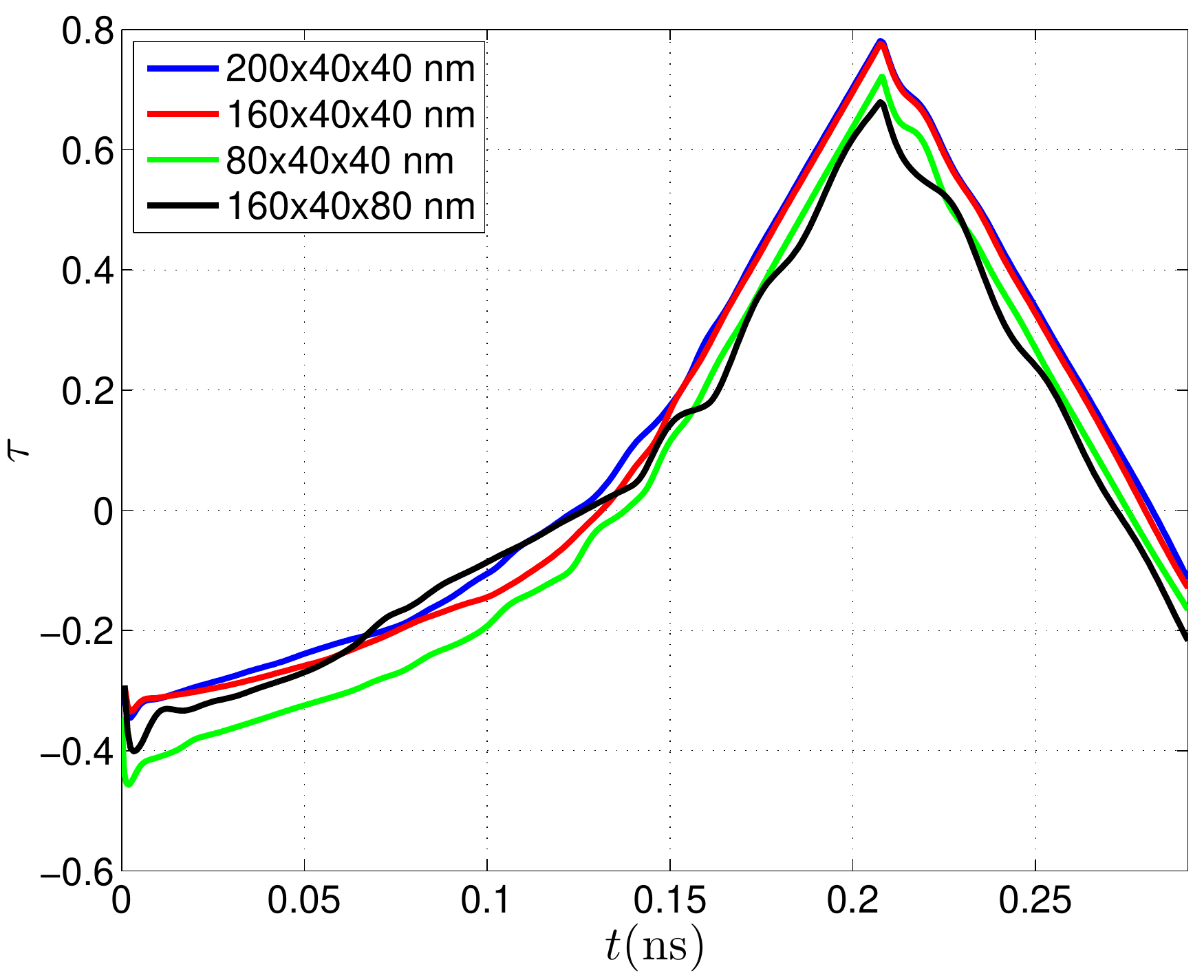}
\label{fig:Ch14SMEAspectRatioTimeVsTemp}
}
\caption{(Color online) SME aspect ratio study: (a) the average $\sigma_{11}$--$\epsilon_{11}$ behavior  and (b) average evolution of $ \tau $ with time.}
\label{fig:Ch14SMEAspectRatioAvgProperties}
\end{figure}

Fig. \ref{fig:Ch14SMEAspectRatioAvgProperties} presents the thermo-mechanical behavior on SMA nanowires of different aspect ratios. For the same lateral dimensions (\ly=\lz=40 nm), the shorter length nanowire behaves in a stiffer manner and phase transformations occur at approximately constant axial stress. This is because the deformation wave travels in a shorter domain, elastically loading the domain faster. The detwinning phase transformations, M$_2$ $\rightarrow$ M$_1$ and \mbox{M$_3$ $\rightarrow$ M$_1$}, occur at higher $\sigma_{11}$ value and have a distinct plateau. 

The nanowires with lower (160$\times$40$\times$40 nm) and higher (160$\times$40$\times$80 nm) aspect ratios behave differently during loading. The detwinning phase transformation of a lower aspect ratio nanowires occurs at nearly constant $\sigma_{11}$, before the elastic loading of M$_1$ after phase transformation is complete. In the case of high aspect ratio nanowires, the elastic loading and detwinning phase transformation occur simultaneously. The evolution of average $\tau$ for different aspect ratios shows similar trends, as presented in Fig. \ref{fig:Ch14PEAspectRatioTimeVsTemp}.

\subsubsection{SME: Strain Rate Study} \label{sec:Ch14TTSMEStrainRate}

The influence of strain rate on SME behavior is investigated on a SMA 160$\times$40$\times$40 nm nanowire. First, the SMA specimen is evolved to the self-accommodated microstructure morphology by quenching it to the temperature coefficient $\tau=-1.2$ and allowing the energy and microstructures to stabilize. The specimen with evolved microstructures is then taken as an initial condition to the tensile test. The specimen is subjected to loading and unloading with different strain rates. The seven strain rates used during the studies are 9$\times$ 10$^7/$s, 10.3 $\times$ 10$^7/$s, 12$\times$ 10$^7/$s, 14.4$\times$ 10$^7/$s, 18$\times$ 10$^7/$s, 24$\times$ 10$^7/$s, and 36$\times$ 10$^7/$s. 

\begin{figure}[h!]
\centering
\subfigure[]
{
\includegraphics[width=0.4\textwidth]{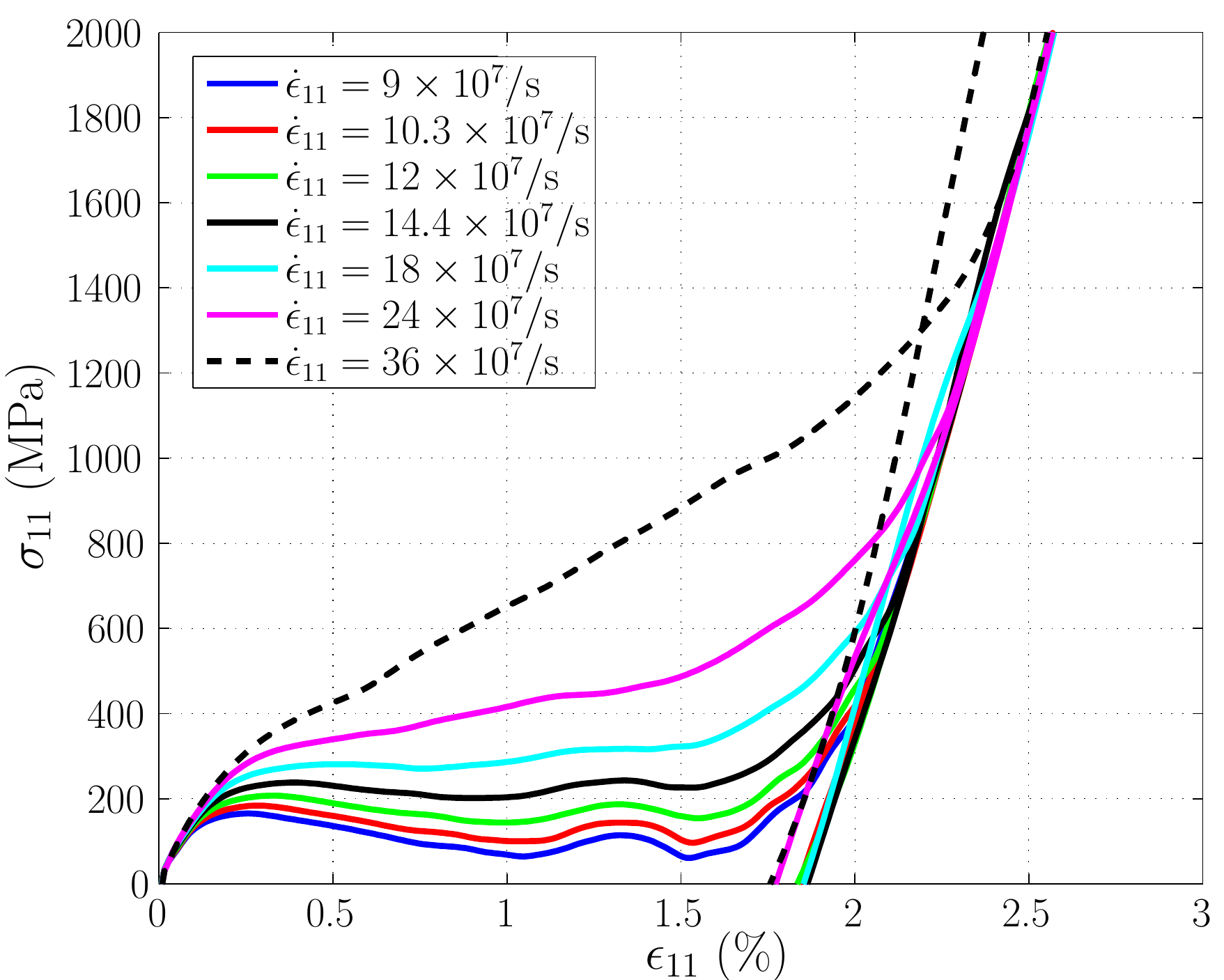}
\label{fig:Ch14SMEStrainRateSSxx}
}
\subfigure[]
{
\includegraphics[width=0.4\linewidth]{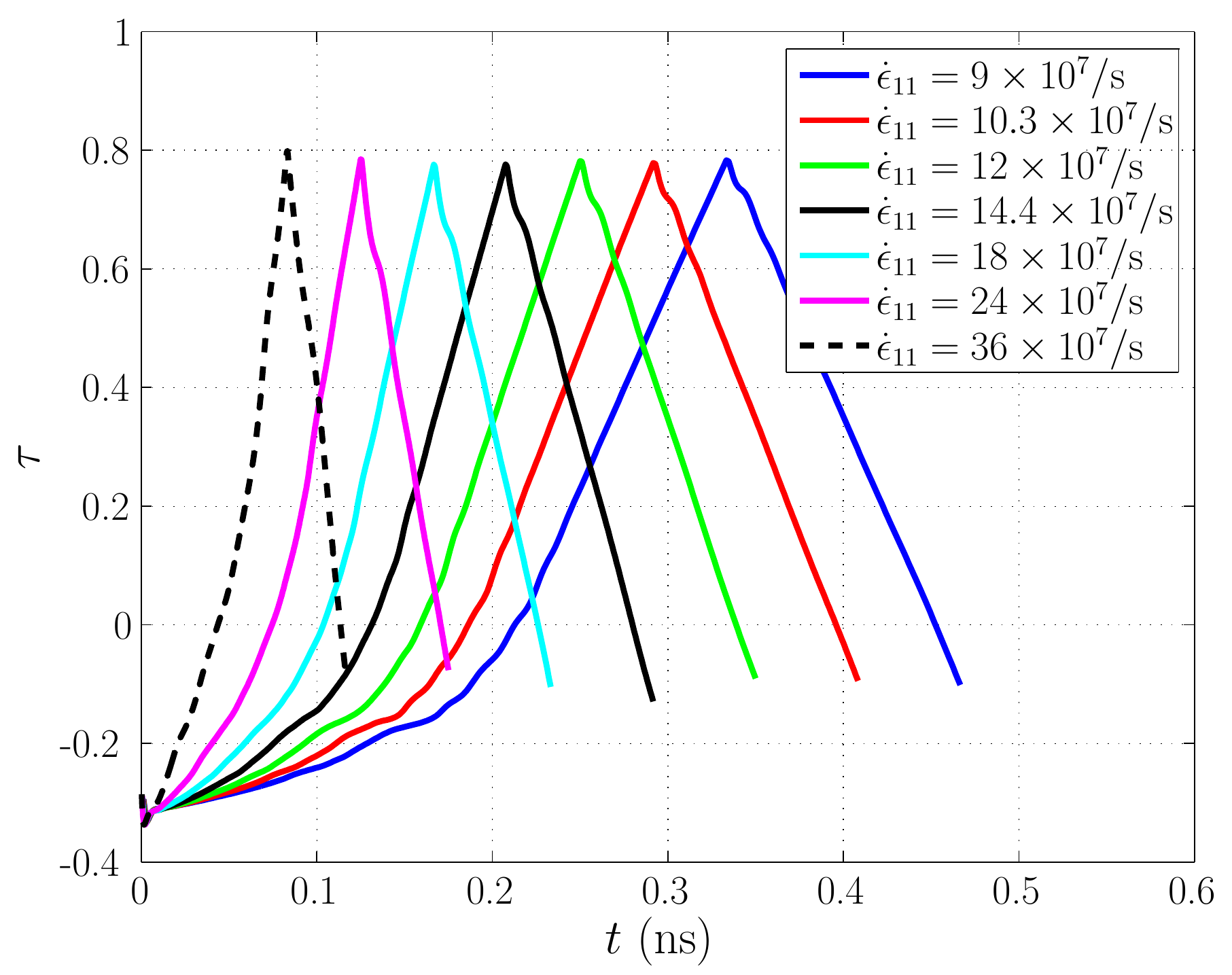}
\label{fig:Ch14SMEStrainRateTimeVsTemp}
}
\caption{(Color online) SME strain rate study: (a) average $\sigma_{11}$--$\epsilon_{11}$ behavior  and (b) average evolution of $ \tau $ with time.}
\label{fig:Ch14SMEStrainRateAvgProperties}
\end{figure}

The thermo-mechanical behavior of the SMA nanowire at different strain rates is plotted in Fig. \ref{fig:Ch14SMEStrainRateAvgProperties}. At lower strain rates between 9$\times$ 10$^7/$s to 14.4$\times$ 10$^7/$s, a drop in $\sigma_{11}$ is observed during phase transformations. This is because  during detwinning phase transformations (M$_2$ $\rightarrow$ M$_1$ and  M$_3$ $\rightarrow$ M$_1$), the wire is relaxed axially for a short time before loading again. The phase transformation occurs and is completed at higher $\sigma_{11}$ values and at higher strain rates. At the intermediate strain rate of 18$\times$ 10$^7/$s, the phase transformation takes place at approximately constant  $\sigma_{11}$. At higher strain rates like 24$\times$ 10$^7/$s and 36$\times$ 10$^7/$s, the phase transformation and elastic loading take place simultaneously, and the  phase transformation takes place at higher stress.  As the SMA specimen does not have enough time to respond to the loading, the  phase transformation no longer takes place at the constant stress value. The influence of strain rate is also evident on a temperature increase in a specimen as observed in Fig. \ref{fig:Ch14SMEStrainRateTimeVsTemp}. The temperature increases faster in a nanowire at high strain rates. Similarly, the strain rate sensitivity on the thermo-mechanical behavior of SMA specimens has been reported during dynamic loading  experiments  \cite{Gadaj2002,Pieczyska2004,Pieczyska2010}.

\subsection{Pseudoelastic Behavior} \label{sec:Ch14TTPE}

\begin{figure}[h!]
\centering
\subfigure[\textit{t} = 0 ns]
{
\includegraphics[trim=0mm 0mm 0mm 0mm,clip, width=0.3\linewidth]{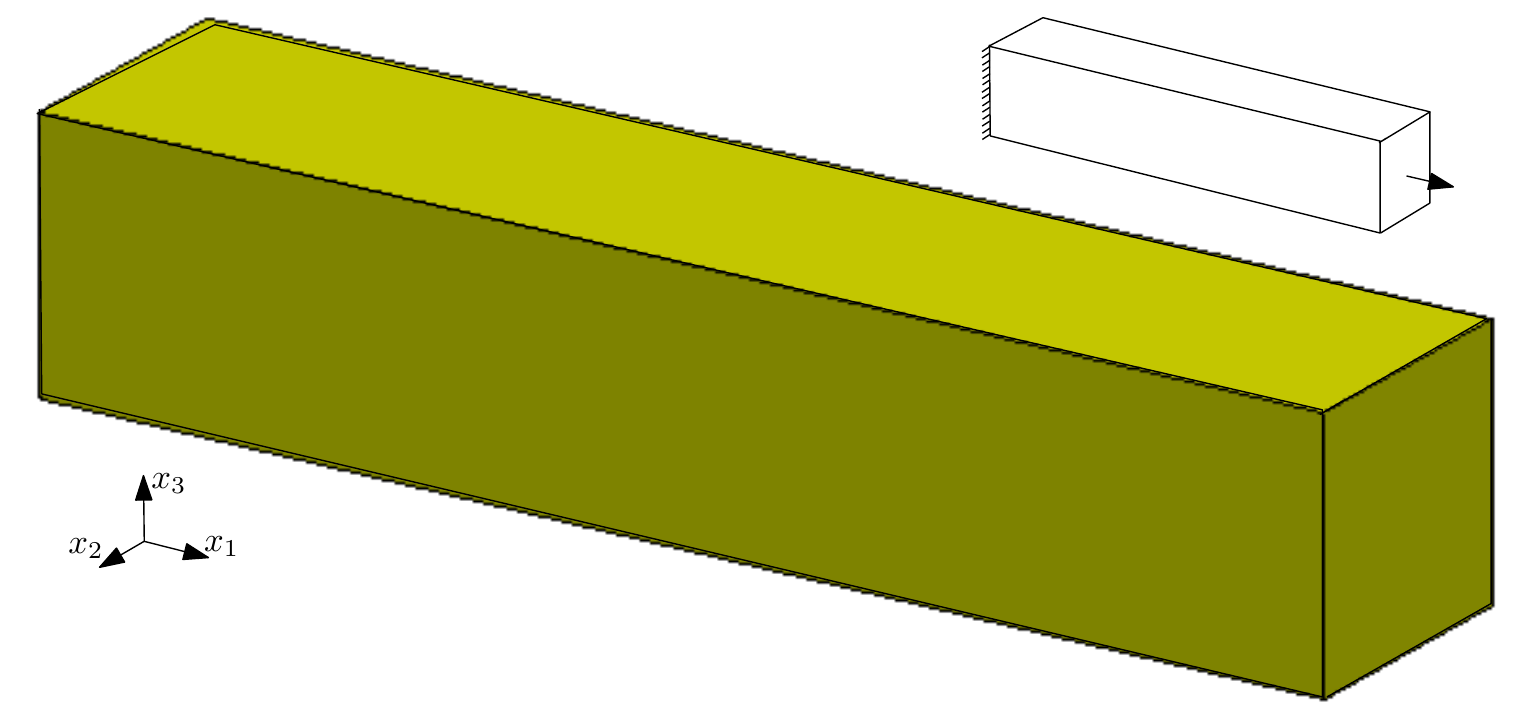}
}
\subfigure[\textit{t} = 0.067 ns]
{
\includegraphics[trim=0mm 0mm 0mm 0mm,clip, width=0.3\textwidth]{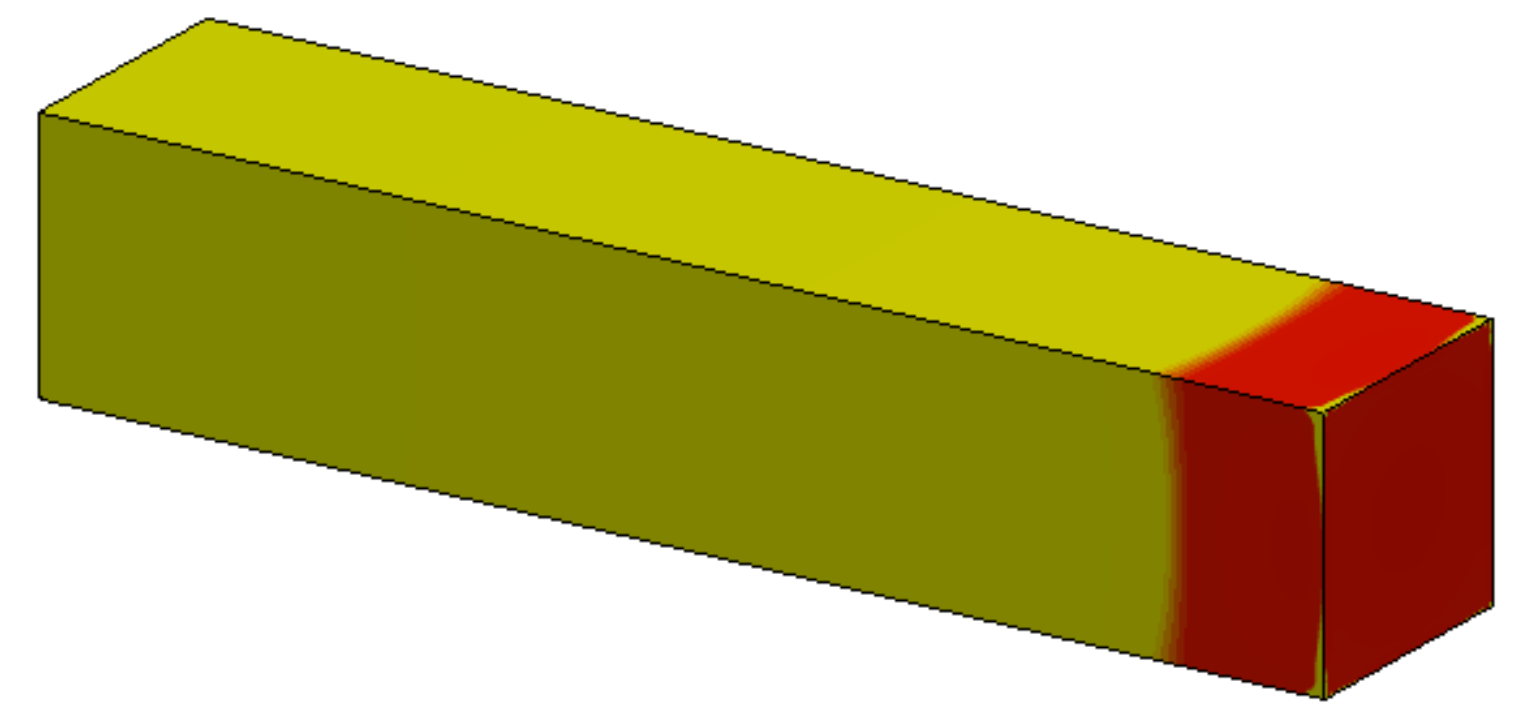}
}
\subfigure[\textit{t} = 0.083 ns]
{
\includegraphics[trim=0mm 0mm 0mm 0mm,clip, width=0.3\textwidth]{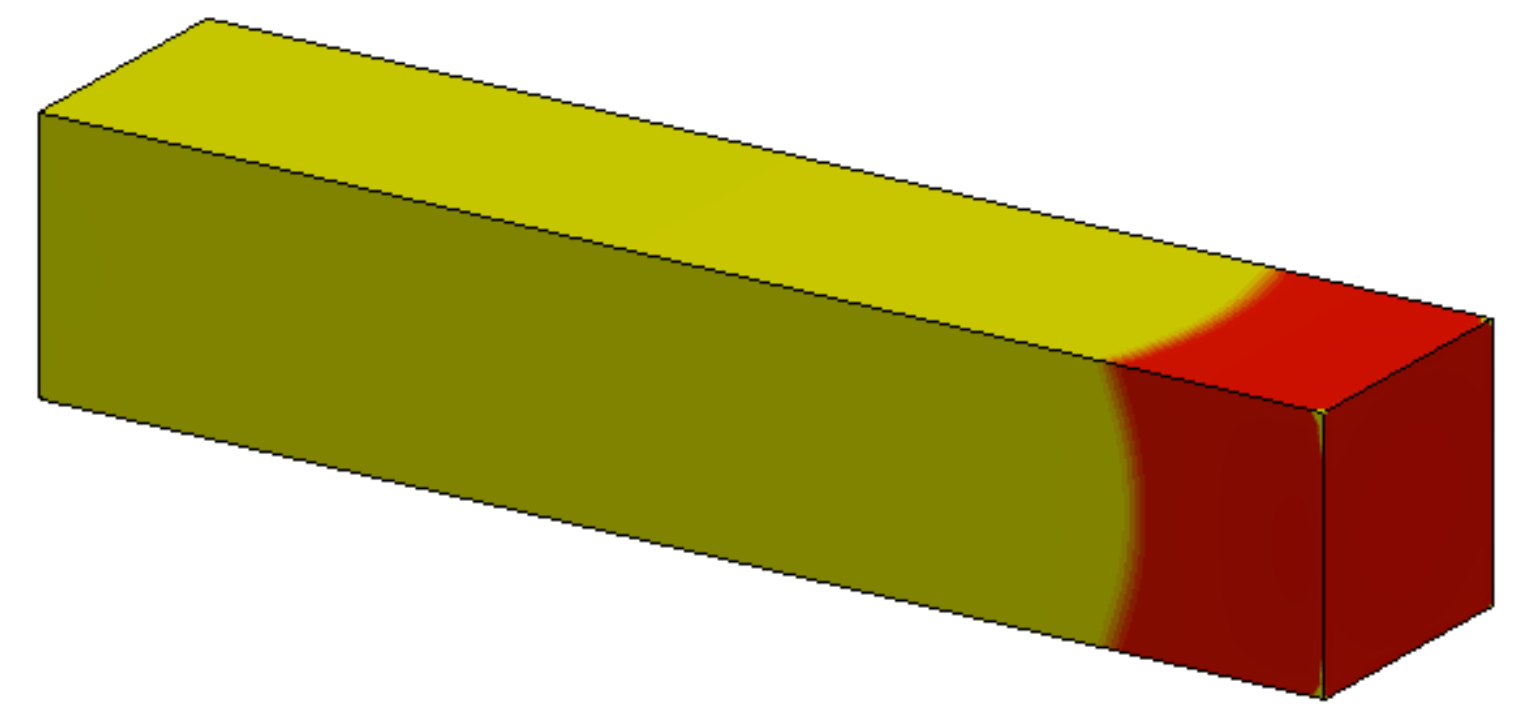}
}
\subfigure[\textit{t} = 0.117 ns]
{
\includegraphics[trim=0mm 0mm 0mm 0mm,clip, width=0.3\linewidth]{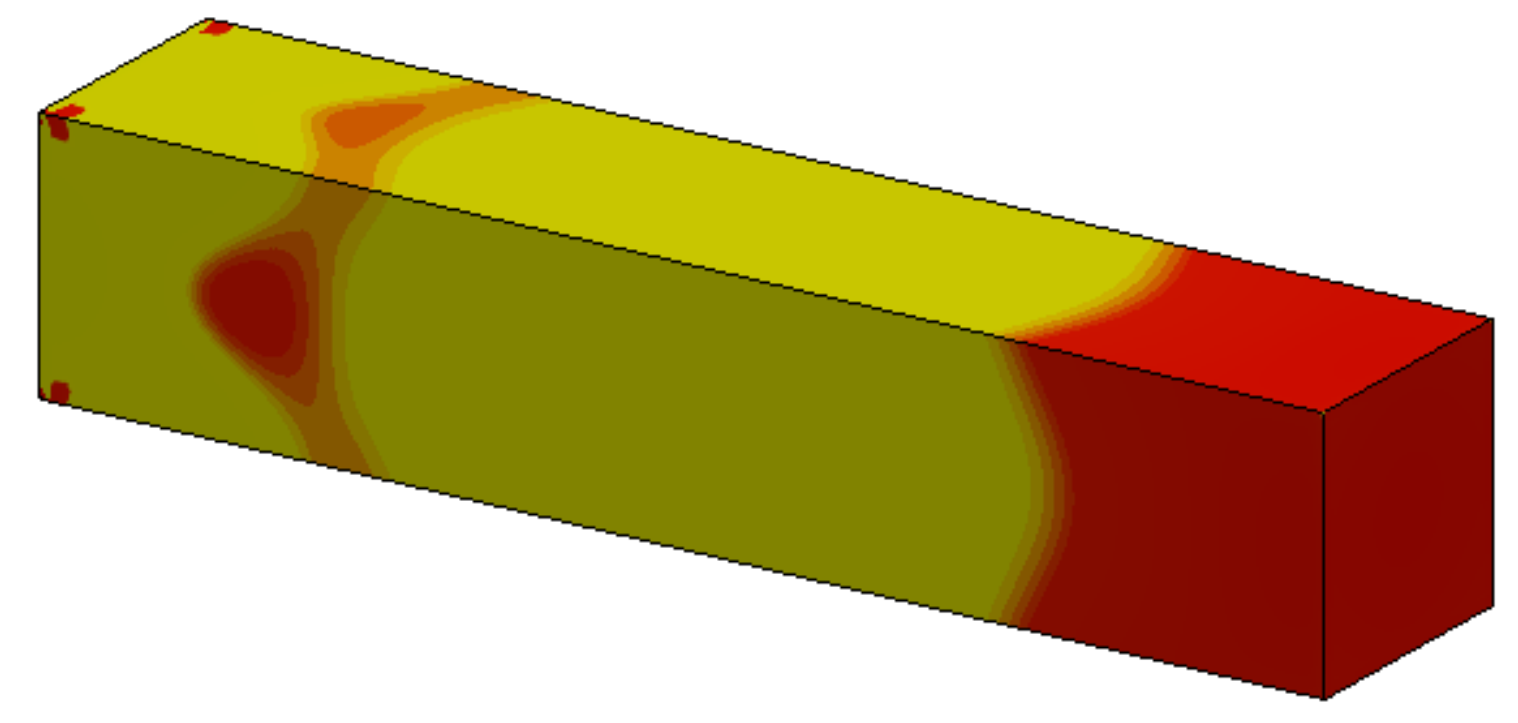}
}
\subfigure[\textit{t} = 0.133 ns]
{
\includegraphics[trim=0mm 0mm 0mm 0mm,clip, width=0.3\textwidth]{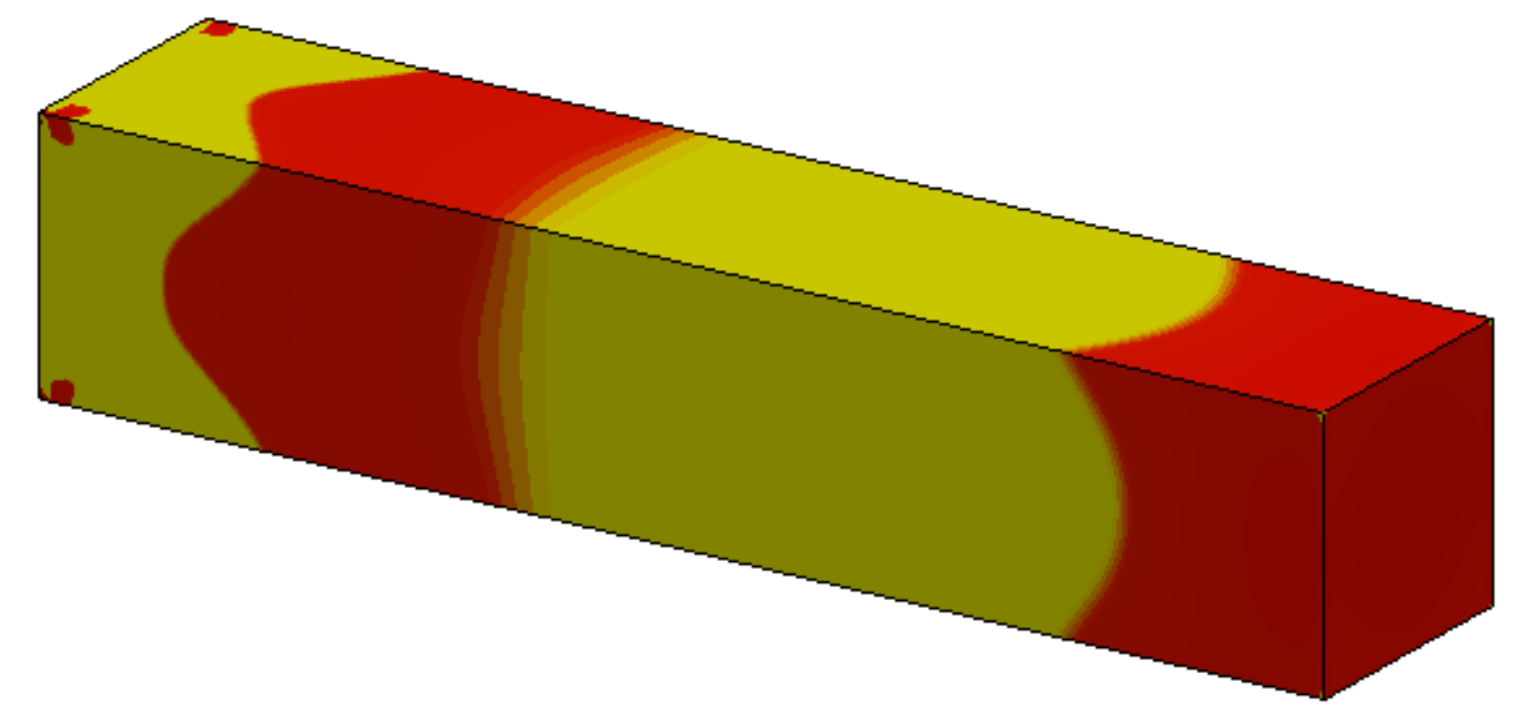}
}
\subfigure[\textit{t} = 0.15 ns]
{
\includegraphics[trim=0mm 0mm 0mm 0mm,clip, width=0.3\textwidth]{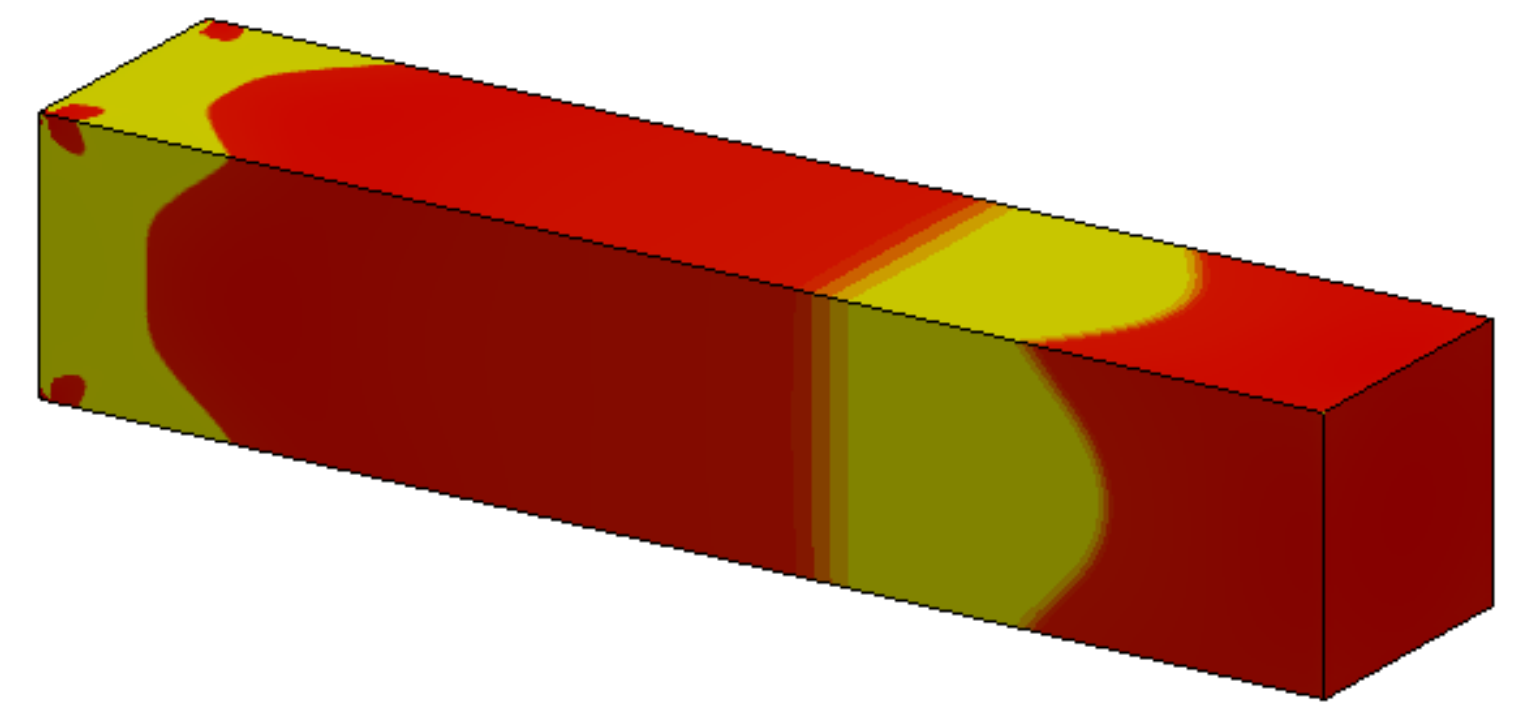}
}
\subfigure[\textit{t} = 0.167 ns]
{
\includegraphics[trim=0mm 0mm 0mm 0mm,clip, width=0.3\linewidth]{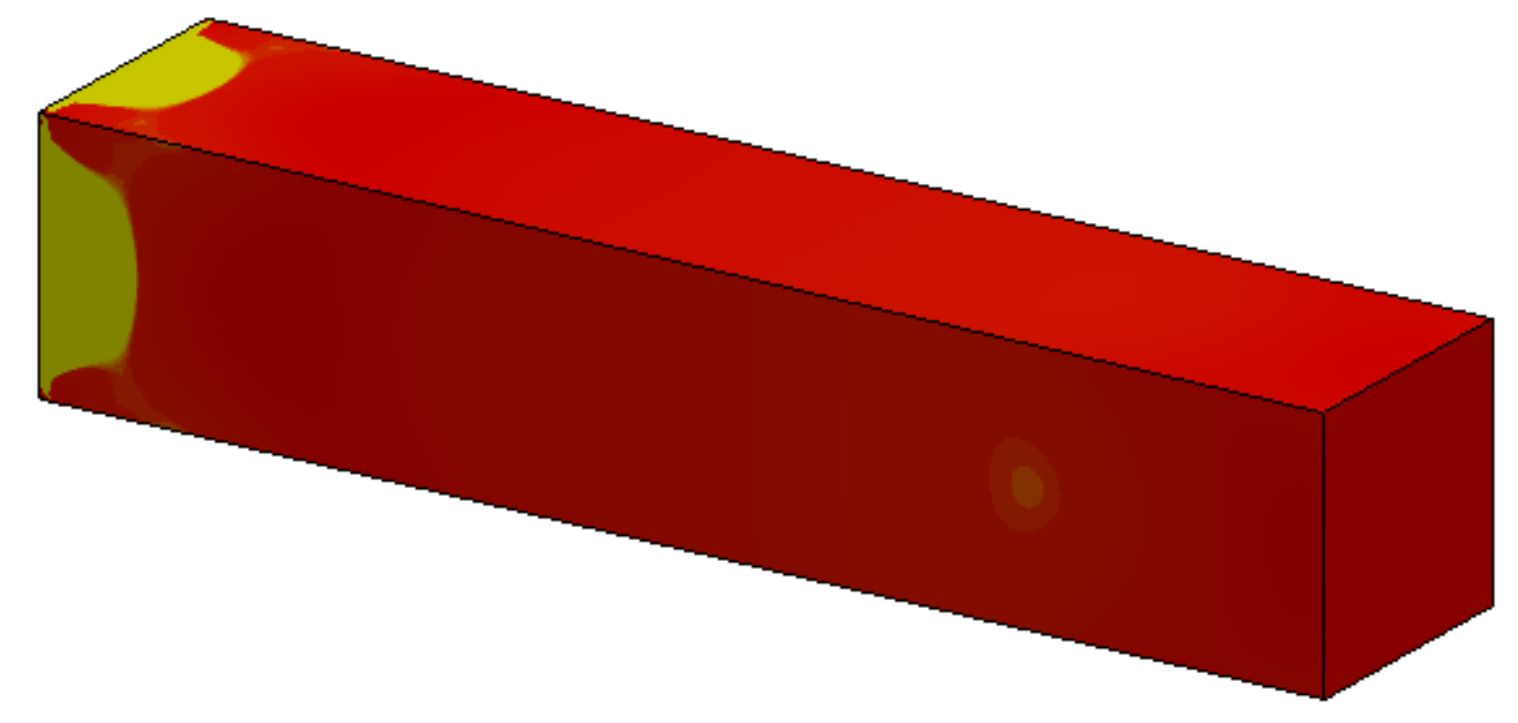}
}
\subfigure[\textit{t} = 0.25 ns]
{
\includegraphics[trim=0mm 0mm 0mm 0mm,clip, width=0.3\textwidth]{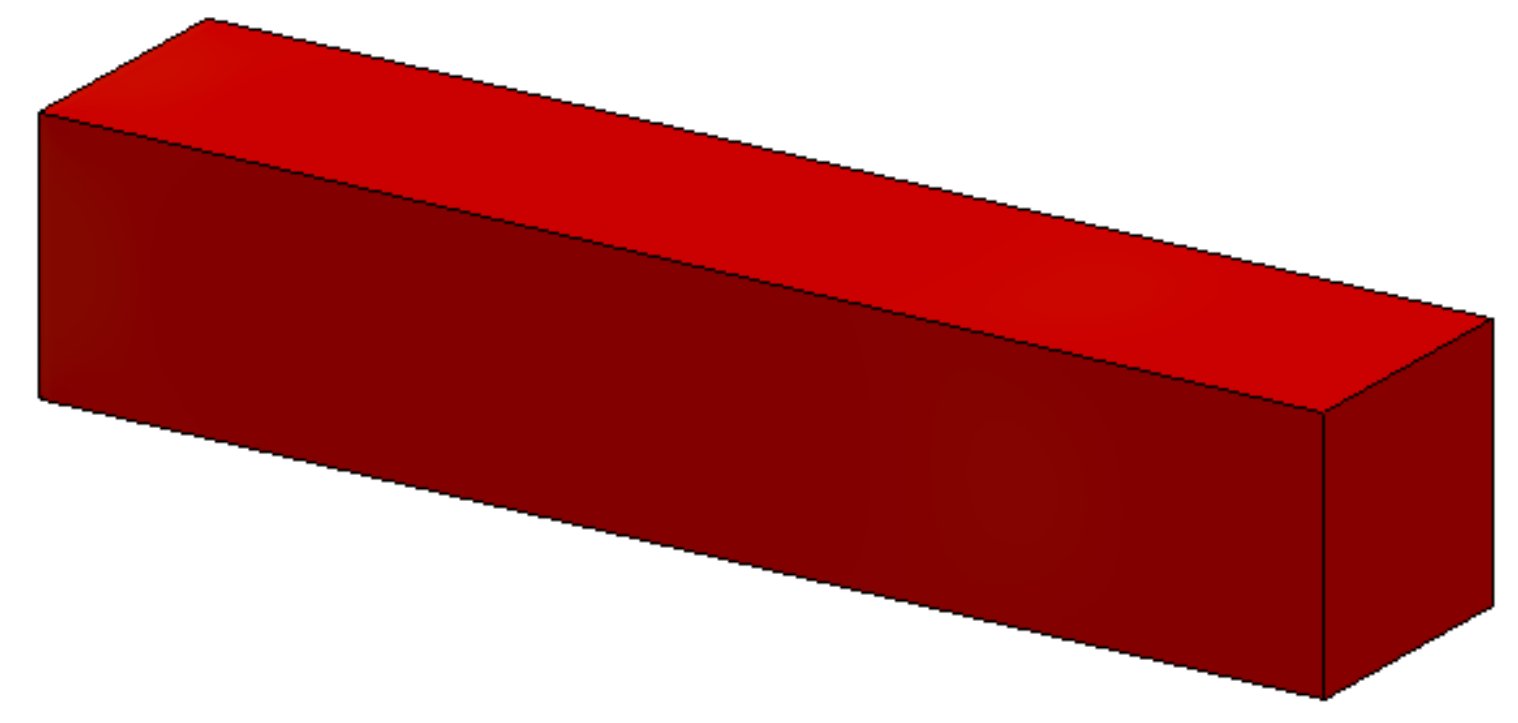}
}
\subfigure[\textit{t} = 0.833 ns]
{
\includegraphics[trim=0mm 0mm 0mm 0mm,clip, width=0.3\textwidth]{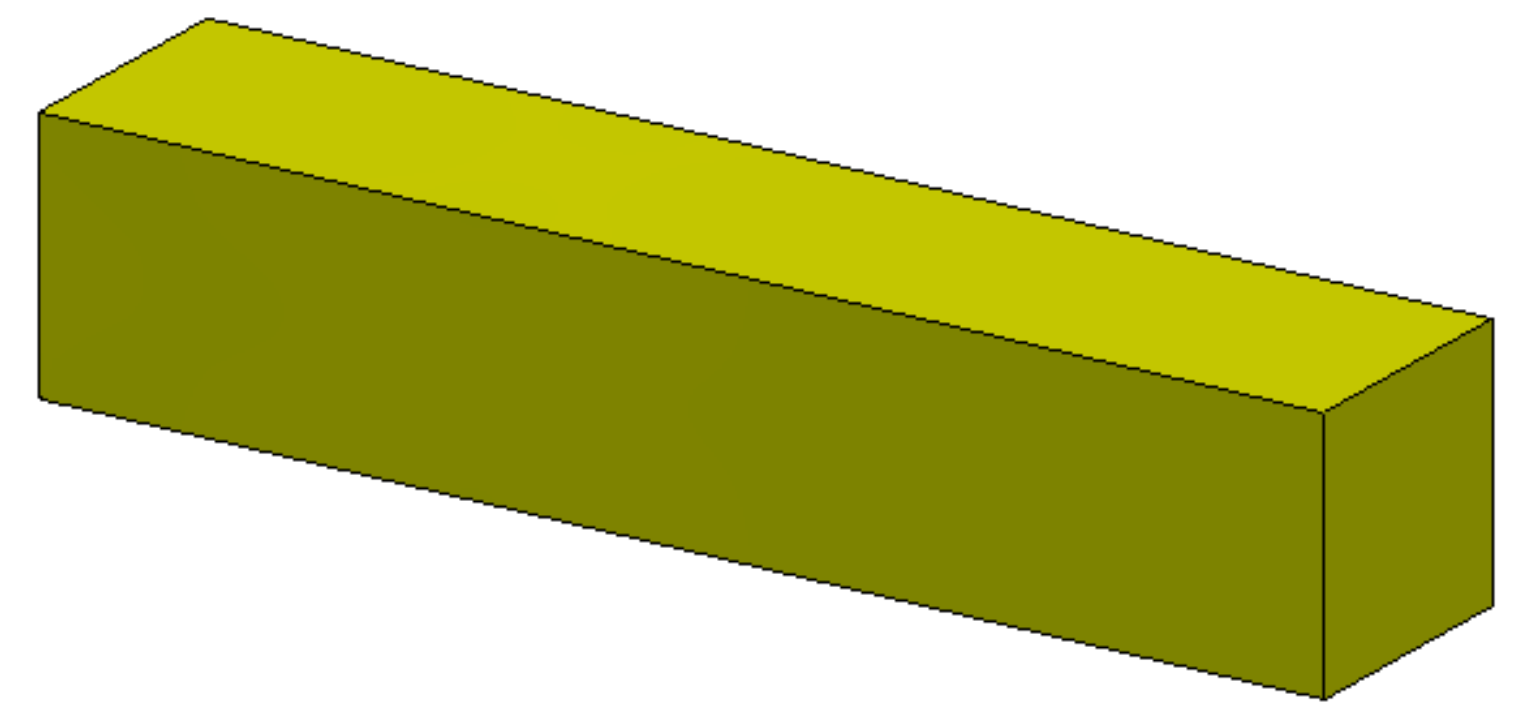}
}
\caption{(Color online) PE: microstructure morphology evolution in a 200$\times$40$\times$40 nm nanowire (red and yellow colors represent M$_1$ variant  and austenite phase). }
\label{fig:Ch14PEEvolution}
\end{figure}

SMAs exhibit pseudoelastic (PE) hysteretic behavior with complete recovery of strain above the transition temperature. We follow the two-step procedure mentioned in Section \ref{sec:Ch14TTSME}. The 200$\times$40$\times$40 nm SMA nanowire is evolved to the austenite phase with the temperature corresponding to $\tau$ = 1.12, starting with an initial random condition of displacement $\pmb{u}$. The evolved austenite phase, as shown in Fig. \ref{fig:Ch14PEEvolution}(a), is taken as an initial condition to the tensile test. In the following simulation, axial strain rate $\dot{\epsilon}_{11}$ = 3$\times$10$^7$/s is used. \\

\begin{figure}[h!]
\centering
\subfigure[]
{
\includegraphics[width=0.3\textwidth]{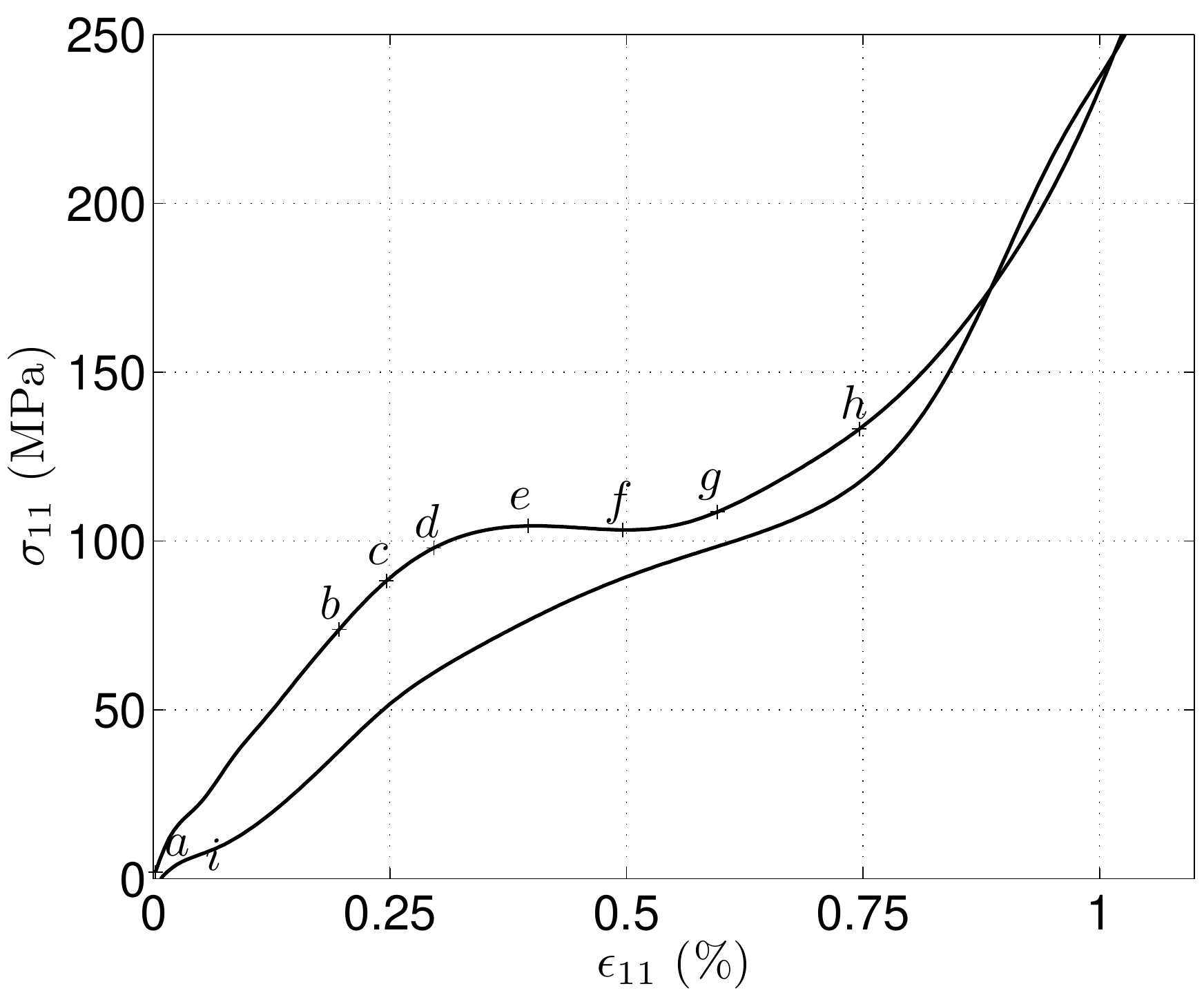}
\label{fig:Ch14PESSxx}
}
\subfigure[]
{
\includegraphics[width=0.3\linewidth]{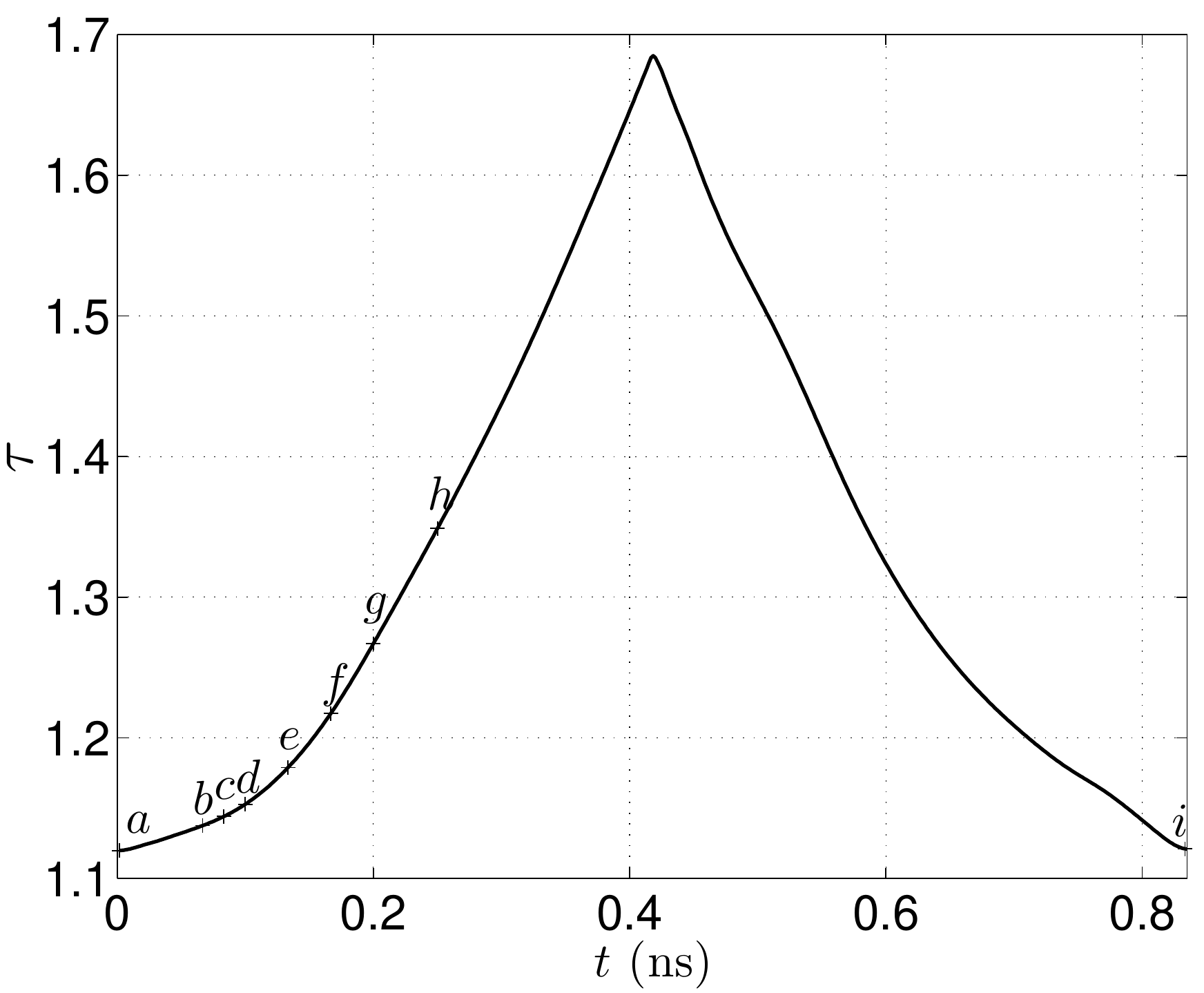}
\label{fig:Ch14PETimeVsTemp}
}
\subfigure[]
{
\includegraphics[width=0.3\linewidth]{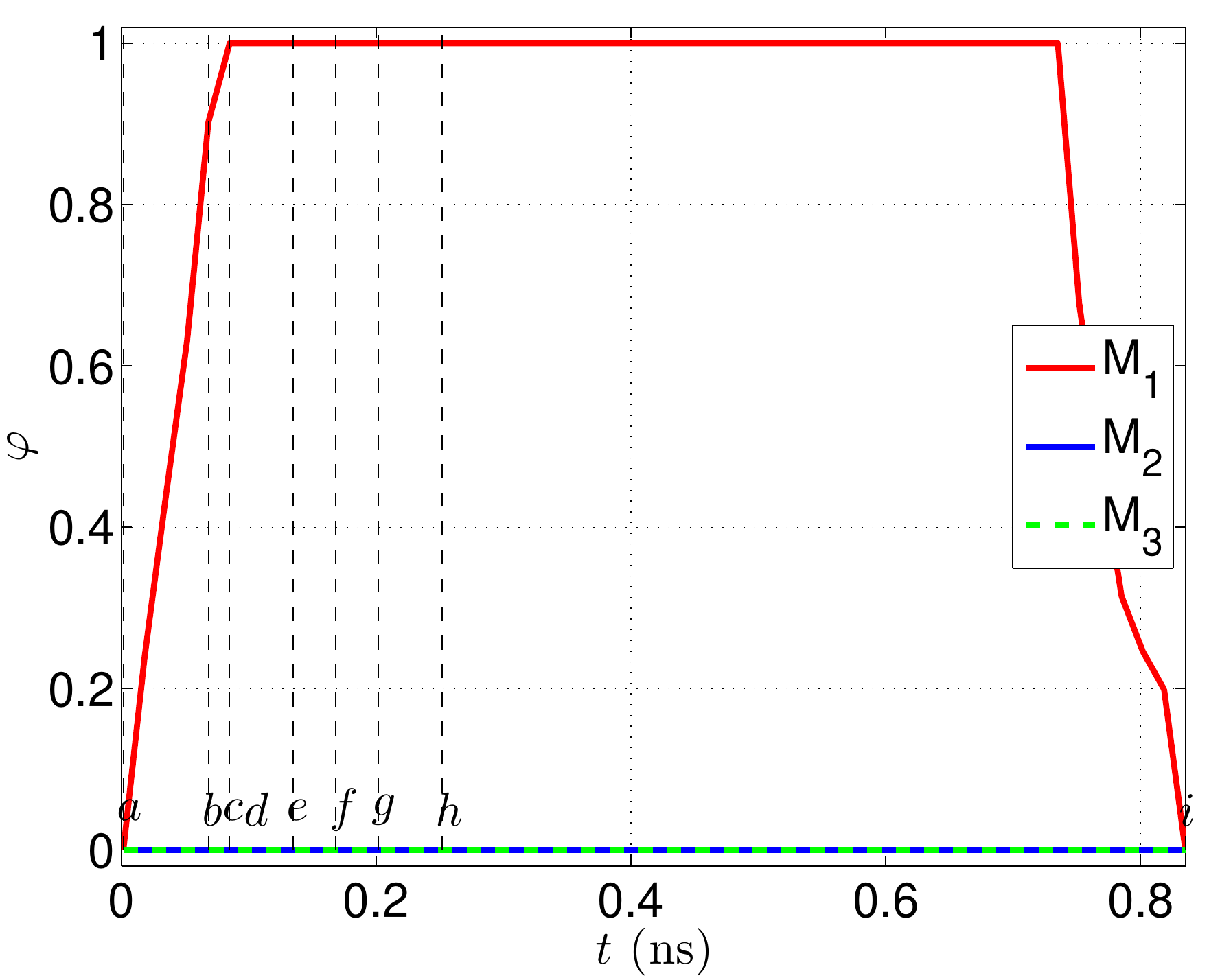}
\label{fig:Ch14PETimeVsPhaseFraction}
}
\caption{(Color online) PE: (a) axial stress-strain ($\sigma_{11}$--$\epsilon_{11}$) relation, and time evolution of (b) average $ \tau $ and (c) phase fraction $\varphi$.}
\label{fig:Ch14PETTAvgProperties}
\end{figure}

The time snapshots of microstructure morphology evolution, and thermo-mechanical properties of SMA specimen are presented in Figs. \ref{fig:Ch14PEEvolution}(b-i), and Fig. \ref{fig:Ch14PETTAvgProperties}, respectively. Initially, the specimen is loaded elastically before the phase transformations start. The phase transformation \austenite $\rightarrow$ \mOne starts near the surface $ \Gamma_{x_1}(+) $ as the loading progresses. The domain wall front moves towards the opposite end of the specimen with the phase transformation.  The phase transformation also nucleates near the $ \Gamma_{x_1}(-) $ end. The combined phase transformation and axial loading occur at the later stages of loading, as observed with a steep rise in the axial $\sigma_{11}$-$\epsilon_{11}$ curve. The whole SMA specimen is converted into the \mOne phase towards the end of loading as shown in Fig. \ref{fig:Ch14PEEvolution}(h).  During the unloading, the reverse phase transformation \mOne $\rightarrow$  \austenite takes place with a complete recovery of strain at the end of unloading as shown in Fig. \ref{fig:Ch14PEEvolution}(i). The influence of thermo-mechanical coupling on the average temperature coefficient $ \tau $ is evident from Fig. \ref{fig:Ch14PETTAvgProperties}(b). The increase and decrease of $\tau$ are a result of exothermic and endothermic processes during loading and unloading of a specimen. 

\subsubsection{PE: Aspect Ratio Study} \label{sec:Ch14TTPEAspectRatio}

To investigate the influence of aspect ratios on a PE regime, the simulations have been conducted on nanowires of the same four dimensions as described in Section \ref{sec:Ch14TTSMEAspectRatio}. The simulations have been conducted following the two-step procedure mentioned in the previous section, but now evolving the domains at the temperature corresponding to $\tau$ = 1.12, starting with an initial random condition of displacement $\pmb{u}$.  The axial strain rate $\dot{\epsilon}_{11}$ = 3$\times$10$^7$/s is used. 

\begin{figure}[h]
\centering
\subfigure[]
{
\includegraphics[width=0.35\textwidth]{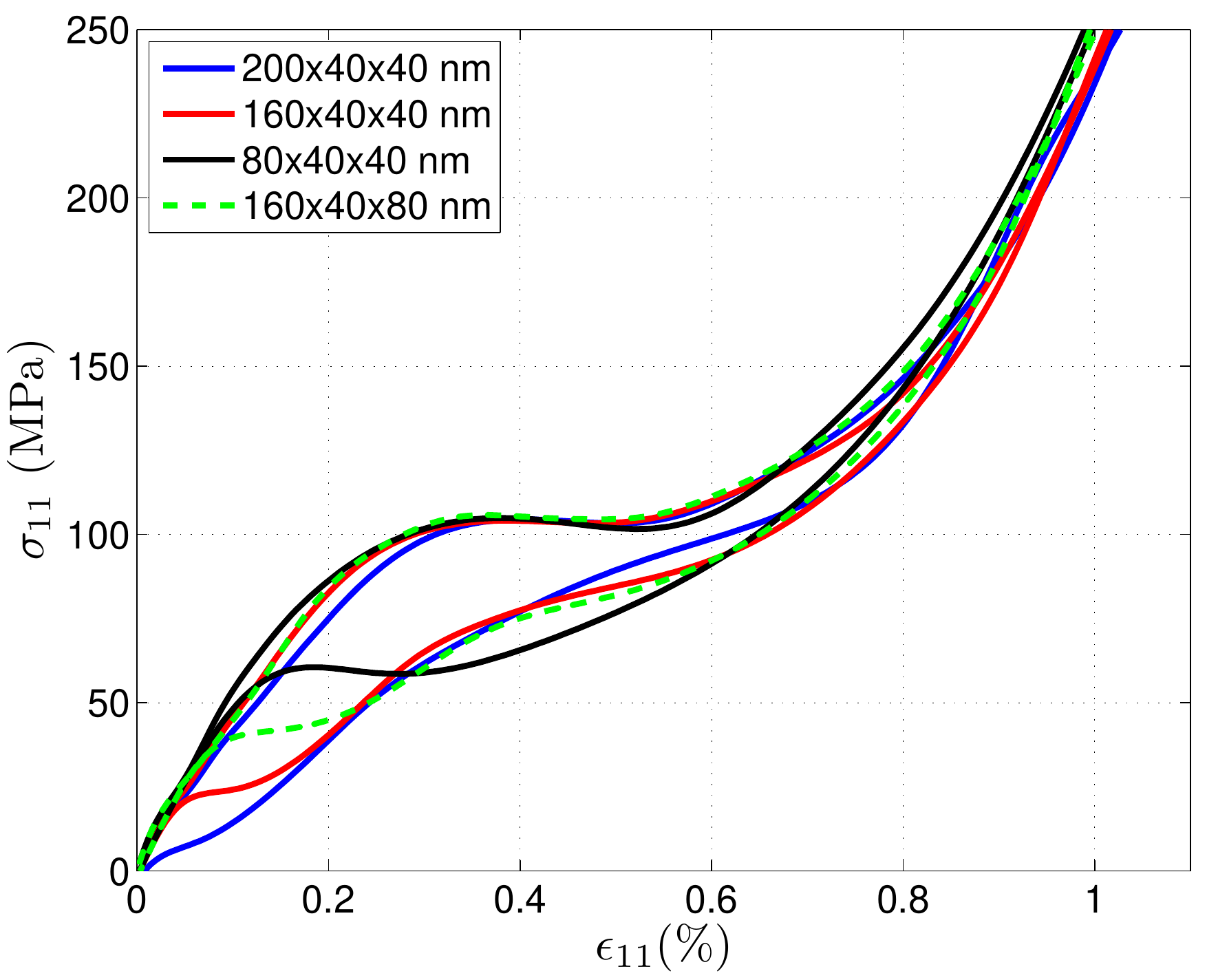}
\label{fig:Ch14PEAspectRatioSSxx}
}
\subfigure[]
{
\includegraphics[width=0.35\linewidth]{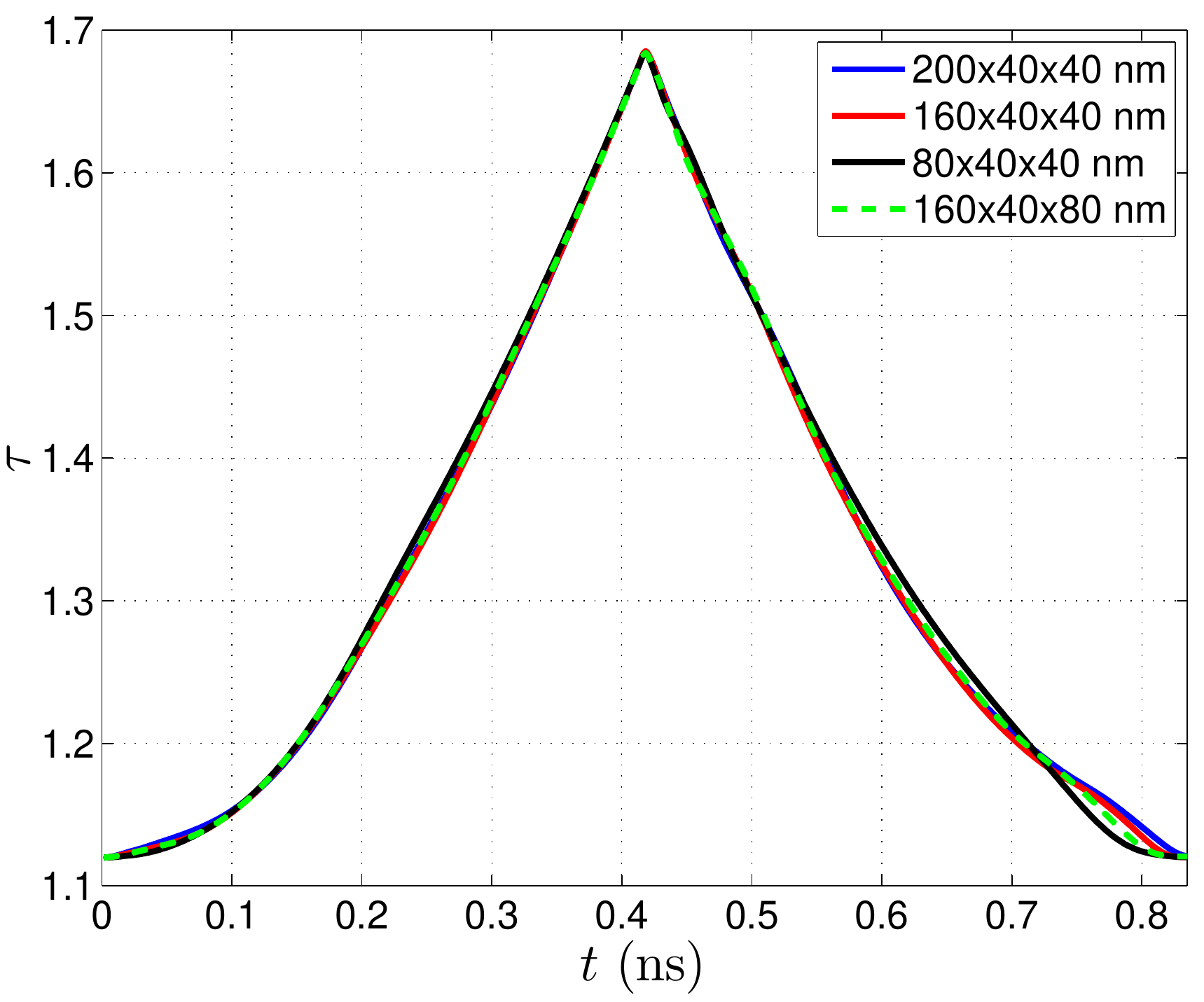}
\label{fig:Ch14PEAspectRatioTimeVsTemp}
}
\caption{(Color online) PE aspect ratio study: (a) average $\sigma_{11}$--$\epsilon_{11}$ behavior  and (b) average evolution of $ \tau $ with time.}
\label{fig:Ch14PEAspectRatiosAvgProperties}
\end{figure}

Fig. \ref{fig:Ch14PEAspectRatiosAvgProperties} illustrates the thermo-mechanical behavior of nanowires with different aspect ratios. For the same lateral dimensions (\ly=\lz=40 nm), the shorter length nanowire behaves in a stiffer manner during elastic loading.  For a particular axial strain, the elastic loading  of longer nanowires takes place at lower $\sigma_{11}$. Thus, the shorter nanowire behaves in a stiffer manner, as the loading deformation wave travels quicker. The phase transformation (\austenite $\rightarrow$M$_1$) occurs at nearly constant $\sigma_{11}$. During unloading, the phase transformation (\mOne $\rightarrow$A) completes early in shorter nanowires. Eventually, the nanowire returns to its original shape, without remnant strain in the system.

The nanowires with lower (160$\times$40$\times$40 nm) and higher (160$\times$40$\times$80 nm) aspect ratios behave similarly during loading. However, during unloading, the phase transformation is completed early in high-aspect ratio nanowires. The evolution of average $\tau$ for different aspect ratios shows similar trends, as presented in Fig. \ref{fig:Ch14PEAspectRatioTimeVsTemp}.

\subsubsection{PE: Strain Rate Study} \label{sec:Ch14TTPEStrainRate}

The strain rate influence on PE behavior is studied on the 160$\times$40$\times$40 nm SMA nanowire. The specimen is subjected to loading and unloading at three different strain rates: $3 \times 10^7/$s, $4 \times 10^7/$s and $6 \times 10^7/$s. 

\begin{figure}[h]
\centering
\subfigure[]
{
\includegraphics[width=0.35\textwidth]{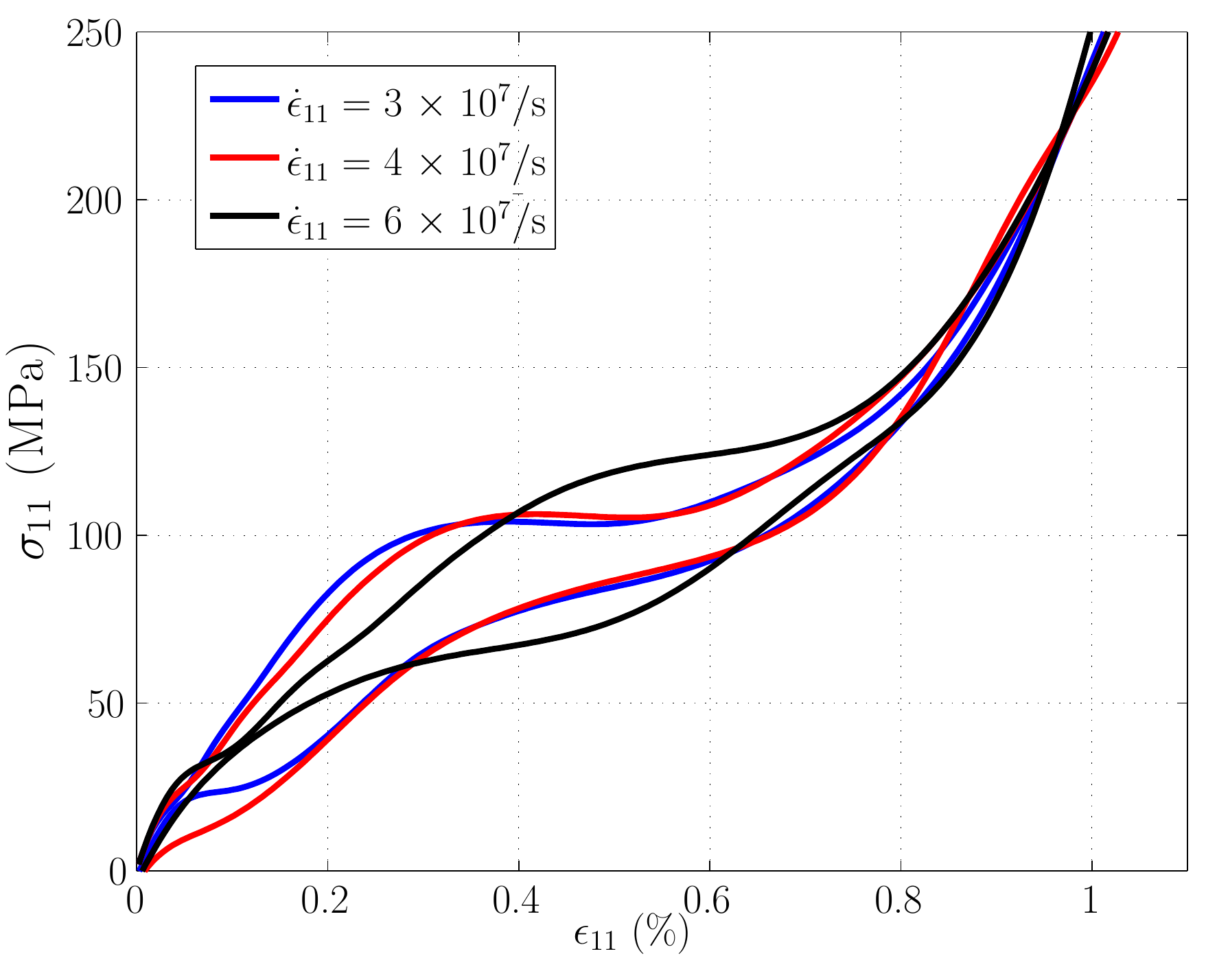}
\label{fig:Ch14PEStrainRateSSxx}
}
\subfigure[]
{
\includegraphics[width=0.35\linewidth]{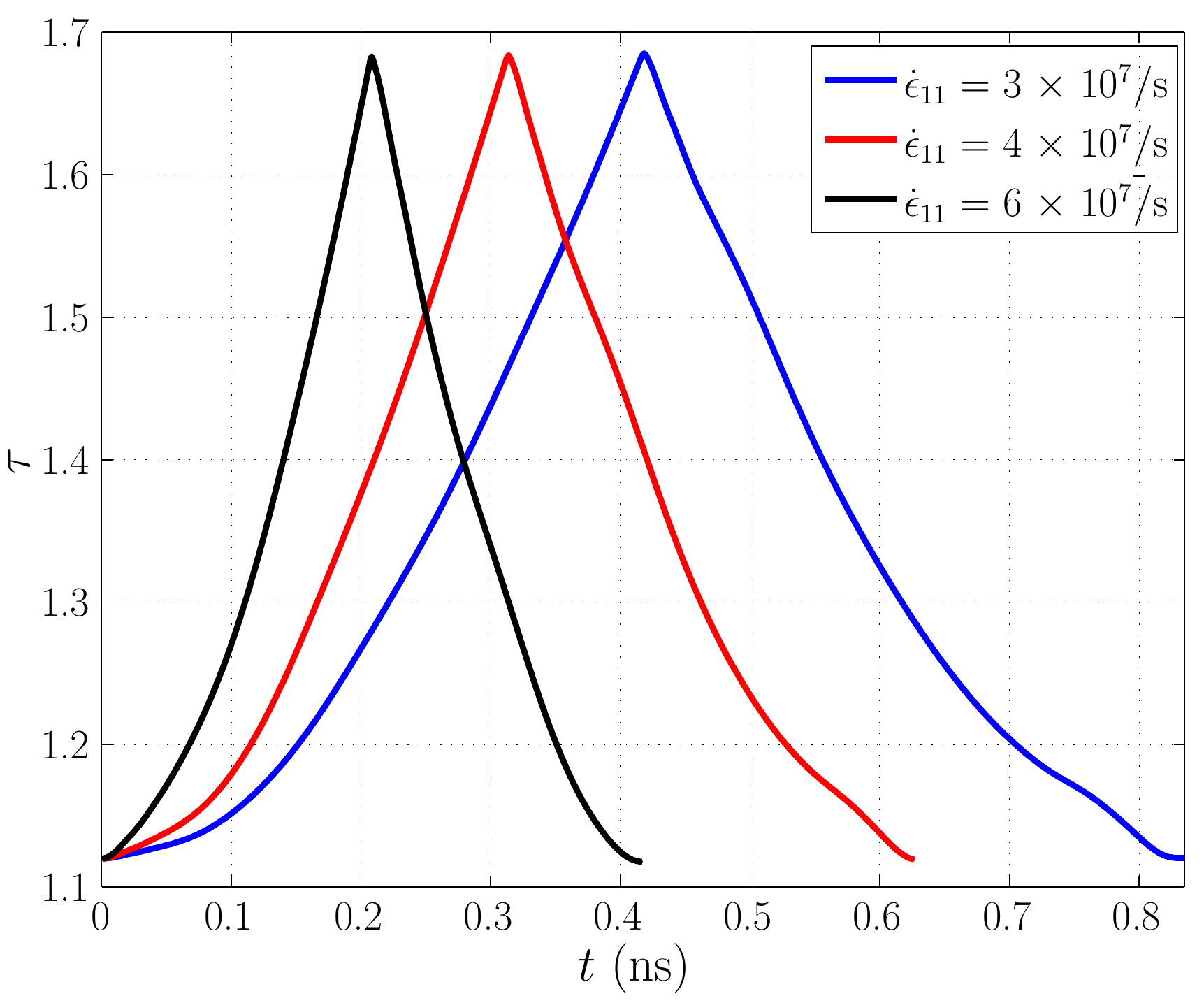}
\label{fig:Ch14PEStrainRateTimeVsTemp}
}
\caption{(Color online) PE strain rate study: (a) average $\sigma_{11}$--$\epsilon_{11}$ behavior  and (b) average evolution of $ \tau $ with time.}
\label{fig:Ch14PEStrainRateAvgProperties}
\end{figure}

The thermo-mechanical behavior of SMA nanowires at different strain rates is plotted in Fig. \ref{fig:Ch14PEStrainRateAvgProperties}. It is apparent from Fig. \ref{fig:Ch14PEStrainRateSSxx} that the elastic loading and phase transformation \mbox{(\austenite $\rightarrow$ \mOne)} processes are distinct at lower strain rates. However, at higher strain rates, the elastic loading and phase transformation occur simultaneously. The competition between loading rates and the evolution dynamics plays an important role. The influence of mechanical loading is observed in a specimen with temperature increase and decrease during loading, as seen from Fig. \ref{fig:Ch14PEStrainRateTimeVsTemp}.


\section{Conclusions} \label{sec:Ch14Conclusions}

\begin{itemize}
\item A new fully coupled dynamic thermo-mechanical 3D phase-field model has been developed for cubic-to-tetragonal phase transformations in SMAs.  The model has a bi-directional coupling via temperature, strain and strain rate. The governing equations have been numerically implemented in a variational form based on the isogeometric analysis, which allows for straightforward implementation of the fourth order differential terms using the rich NURBS basis functions.

\item The aspect ratio of the domain plays an important role in the microstructure morphology evolution during thermally induced transformations. The lower aspect ratio promotes the equal proportion of martensitic variants, while the higher aspect ratio leads to suppression of the out-of-plane variant.  The gradient energy in a slab is higher than in the cube specimen in order to maintain an equal phase fraction of the martensitic variants in the domain.
 
\item During thermally induced transformations, the boundary conditions have a sensitive influence on the microstructure morphology. The SMA specimen with fully constrained boundary conditions is stabilized in the shortest time, followed by normally constrained and then fully periodic boundary conditions. 
Energetically, the fully constrained boundary condition has the highest free energy, and the fully periodic condition has the lowest. The fully periodic boundary condition allows the formation of distinct self-accommodated domains due to long-range elastic interactions. 

\item The model reproduces the shape memory effect and pseudoelastic behavior in the dynamic tensile loading-unloading test. The influence of mechanical loading on temperature evolution is in agreement with the experiments reported in the literature. 

\item  The tensile test on rectangular prisms of different aspect ratios reveals the impact of geometry on the thermo-mechanical behavior of SMAs. For the same lateral dimensions, the shorter length nanowire behaves in a stiffer manner and phase transformations occur at approximately constant axial stress. In the higher aspect ratio specimen, during SME loading, the elastic loading and phase transformations occur simultaneously. In the case of PE, the loading behavior is similar, however the unloading phase transformation is completed earlier. 

\item The lower strain rate loading causes a phase transformation to take place at constant axial stress, with a distinct plateau. However, at higher strain rates, the elastic loading and phase transformation occur  simultaneously. The dynamics of phase transformations and loading rate, as well as the time scale of twin interface motion, play an important role. 
\end{itemize}

Our model  and numerical framework is a step forward in a better understanding of the dynamics of thermo-mechanical behavior of SMA nanostructures, which should assist in the development of novel SMA-based applications. The methodology developed in this paper  can be extended to the study of the phase transformations and thermo-mechanical behavior in other 3D structures as well as to the study of the influence of strain and phase transformations in electronic band structure calculations \cite{imada1998metal,prabhakar2010influence}. 

\section*{Acknowledgments}
RD, RM, and JZ have been supported by the NSERC and CRC program (RM), Canada. RM thanks colleagues at Mevlana University for their hospitality during his visit there and T\"{U}BITAK for its support.  HG was partially supported by the European Research Council through the FP7 Ideas Starting Grant program (project \# 307201) and by \emph{Conseller\'ia de Educaci\'on e Ordenaci\'on Universitaria} (\emph{Xunta de Galicia}). Their support is gratefully acknowledged. This work was made possible with the facilities of the Shared Hierarchical Academic Research Computing Network (SHARCNET: www.sharcnet.ca) and Compute/Calcul Canada.
\appendix
\section{Stress Components} \label{app:Ch14StressDefinations}
The stress tensor  components $ \{ \sigma_{ij} \}$ are
\begin{subequations}
\renewcommand{\theequation}{\theparentequation.\arabic{equation}}
\label{eq:Ch14stresscomponents}
\begin{eqnarray}
\sigma_{11} &=& \!\!\!\frac{a_1 e_1}{\sqrt{3}}
+ \frac{e_2}{\sqrt{2}} \left[ 2 \tau a_3 - 6 a_4 e_3 + 4 a_5 (e_2^2+e_3^2)  \right] + \frac{1}{\sqrt{6}} \left[ e_3(2 \tau a_3 +  4 a_5 (e_2^2+e_3^2)) + 3 a_4 (e_3^2-e_2^2)  \right], \nonumber \\
&& \\
\sigma_{12} &=& \!\!\!\sigma_{21} = \frac{1}{2} a_2 e_6, \\
\sigma_{13} &=& \!\!\!\sigma_{31} = \frac{1}{2} a_2 e_5, \\
\sigma_{22} &=& \!\!\!\frac{a_1 e_1}{\sqrt{3}} - \frac{e_2}{\sqrt{2}} \left[ 2 \tau a_3 - 6 a_4 e_3 + 4 a_5 (e_2^2+e_3^2)  \right] + \frac{1}{\sqrt{6}} \left[ e_3(2 \tau a_3 +  4 a_5 (e_2^2+e_3^2)) + 3 a_4 (e_3^2-e_2^2)  \right], \nonumber \\
&& \\
\sigma_{23} &=& \!\!\!\sigma_{32} = \frac{1}{2} a_2 e_4, \\
\sigma_{33} &=& \!\!\!\frac{1}{\sqrt{3}} a_1 e_1
- \frac{2}{\sqrt{6}} \left[ 2 \tau a_3 e_3 +  3 a_4 (e_3^2-e_2^2) + 4 a_5 e_3 (e_2^2+e_3^2)   \right].
\end{eqnarray}
\end{subequations}


\bibliographystyle{ieeetr}
\bibliography{IGA3DPhysicsarXiv}

\end{document}